\documentclass[twocolumn]{aastex631}
\usepackage{amsmath, amssymb}
\usepackage[T1]{fontenc}
\usepackage{apjfonts}
\usepackage{hyperref}
\usepackage{color,xcolor}
\usepackage{epsfig}
\usepackage{float}
\usepackage{amssymb}
\usepackage{graphicx}
\usepackage{amsmath}
\usepackage{multirow}
\usepackage{threeparttable}
\usepackage[caption=false,farskip=0pt,labelfont={bf}]{subfig}
\usepackage{threeparttable}
\usepackage{epstopdf}

\begin{document}

\title{Broadband multi-wavelength study of LHAASO detected AGN}

\author[0000-0002-3883-6669]{Ze-Rui Wang}
\affiliation{College of Physics and Electronic Engineering, Qilu Normal University, Jinan 250200, People's Republic of China}

\author[0000-0003-1721-151X]{Rui Xue}
\altaffiliation{Corresponding authors}
\affiliation{Department of Physics, Zhejiang Normal University, Jinhua 321004, People's Republic of China; \textcolor{blue}{ruixue@zjnu.edu.cn}}

\author[0000-0002-6809-9575]{Dingrong Xiong}
\altaffiliation{Corresponding authors}
\affiliation{Yunnan Observatories, Chinese Academy of Sciences, 396 Yangfangwang, Guandu District, Kunming, 650216, People’s Republic of China; \textcolor{blue}{xiongdingrong@ynao.ac.cn}}
\affiliation{Center for Astronomical Mega-Science, Chinese Academy of Sciences, 20A Datun Road, Chaoyang District, Beijing, 100012, People’s Republic of China}
\affiliation{Key Laboratory for the Structure and Evolution of Celestial Objects, Chinese Academy of Sciences, 396 Yangfangwang, Guandu District, Kunming, 650216, People’s Republic of China}

\author{Hai-Qin Wang}
\affiliation{Department of Physics, Anhui Normal University, Wuhu 241002, China; \textcolor{blue}{pengfangkun@163.com}}

\author{Lu-Ming Sun}
\affiliation{Department of Physics, Anhui Normal University, Wuhu 241002, China; \textcolor{blue}{pengfangkun@163.com}}

\author[0000-0001-7171-5132]{Fang-Kun Peng}
\altaffiliation{Corresponding authors}
\affiliation{Department of Physics, Anhui Normal University, Wuhu 241002, China; \textcolor{blue}{pengfangkun@163.com}}

\author{Jirong Mao}
\affiliation{Yunnan Observatories, Chinese Academy of Sciences, 396 Yangfangwang, Guandu District, Kunming, 650216, People’s Republic of China; \textcolor{blue}{xiongdingrong@ynao.ac.cn}}
\affiliation{Center for Astronomical Mega-Science, Chinese Academy of Sciences, 20A Datun Road, Chaoyang District, Beijing, 100012, People’s Republic of China}
\affiliation{Key Laboratory for the Structure and Evolution of Celestial Objects, Chinese Academy of Sciences, 396 Yangfangwang, Guandu District, Kunming, 650216, People’s Republic of China}

\begin{abstract}
Recently, the Large High Altitude Air Shower Observatory (LHAASO) collaboration presented the first catalog of $\gamma$-ray sources using 508 days of LHAASO data, from March 2021 to September 2022. This catalog contains four blazars and a possible liner-type AGN counterpart. In this work, we establish averaged multi-wavelength SEDs by combining data from the \textsl{Fermi}-Large Area Telescope, \textsl{Swift}, ZTF, and WISE covering the same period as the LHAASO detection. In general, these five AGNs are found in low states at all wavelengths. To study the multi-wavelength properties of these AGNs, several jet emission models, including the one-zone leptonic model, the one-zone leptonic and hadronuclear ($pp$) model, the one-zone proton-synchrotron model, and the spine-layer model are applied to reproduce their averaged SEDs, respectively. We find that the one-zone leptonic model can reproduce most of the SEDs, except for the high-energy tail of the LHAASO spectra of Mrk 421 and Mrk 501. To improve the fitting, emission from $pp$ interactions is favoured in the framework of a one-zone model. The spine-layer model, which can be treated as a multi-zone scenario, can also provide good spectral fits. The influence of different extragalactic background light models on fitting LHAASO energy spectrum is also discussed.

\end{abstract}
\keywords{Gamma-ray sources (633); High energy astrophysics (739); Relativistic jets (1390)}

\section{Introduction} \label{sec:intro}
Very high energy (VHE, $0.1\sim 100\,$TeV) $\gamma$-rays are one of the most important messengers for the investigation of the most extreme phenomena in the Universe. More than 90 extragalactic sources have been detected in the VHE band, the majority of which are jetted active galactic nuclei \citep[AGNs;][]{2021Univ....7..421D}. These VHE AGNs include blazars with powerful jets pointing toward the observer, and radio galaxies, which are considered as the misaligned counterparts of blazars \citep{1978PhyS...17..265B, 1995PASP..107..803U}. In addition, other subclasses with GeV detection \citep[e.g., narrow-line Seyfert 1 galaxies;][]{2023MNRAS.523..404L}, or jets are potential VHE emitters as well.

Recently, the Large High Altitude Air Shower Observatory (LHAASO) collaboration presented the first LHAASO catalog of VHE $\gamma$-ray sources, in which four blazars and a possible AGN counterpart are reported \citep{2023arXiv230517030C}. The four blazars, i.e. Mrk 421, Mrk 501, 1ES J1727+502 and 1ES 2344+514, are classified as the high-synchrotron-peaked \citep[HSP;][]{2010ApJ...710.1271A} type
and are known extragalactic VHE emitters that have been extensively studied and included in the TeVCat\footnote{\url{http://tevcat2.uchicago.edu}} website. The possible AGN counterpart is the Liner-type AGN, namely NGC 4278. Although the \textsl{Fermi} telescope does not detect GeV photons from NGC 4278\footnote{The space science data center (http://tools.ssdc.asi.it/SED/) shows that the GeV emission of NGC 4278 was detected by EGRET and AGILE, but no reference is provided.}, the parsec-scale jet discoved by radio observation is still a possible site for the acceleration of relativistic particles \citep{2004AJ....127..119L, 2005ApJ...622..178G}. The new LHAASO observations shed light on the VHE radiation mechanism of AGNs.

The physical origin of the VHE emission of jetted AGNs is complex and under debate. Due to the lack of strong external photon fields for HSP blazars, the most commonly used interpretation is the synchrotron self-Compton process \citep[SSC;][]{1996ApJ...461..657B, 1996A&AS..120C.537M, 2011ApJ...727..129A}. Since the Klein-Nishina effect softens the SSC spectrum, the SSC model predicts a soft VHE spectra naturally. However, the intrinsic hard VHE spectra of some AGNs imply a different physical interpretation. Several models are proposed, such as the spine-layer model \citep{2005A&A...432..401G, 2020ApJS..247...16A}, the proton-synchrotron model \citep{2000NewA....5..377A, 2001APh....15..121M, 2003APh....18..593M, 2015MNRAS.448..910C, 2023PhRvD.107j3019X}, and the ultra-high-energy cosmic-ray propagation model \citep{2011APh....35..135E, 2012ApJ...757..183P, 2020ApJ...889..149D, 2022A&A...658L...6D}. The reported minute-scale variability at VHE band also implies a multi-zone origin of the multi-wavelength emission of the jet. \citep[e.g.,][]{2008MNRAS.384L..19B, 2023MNRAS.526.5054L}. On the other hand, many associations between high-energy neutrinos and AGNs have been reported \citep{2018Sci...361.1378I,2018Sci...361..147I,2021ApJ...912...54R,2020PhRvL.124e1103A,2020A&A...640L...4G,2022MNRAS.511.4697P,2023MNRAS.519.1396S}, which suggest that hadronic interactions, including photohadronic ($p\gamma$) and hadronuclear ($pp$) interactions, in the jet could not be simply ignored. Multi-wavelength modeling finds that the emission of secondary electrons/positrons could contribute in the VHE band \citep{2019MNRAS.483L..12C, 2019NatAs...3...88G, 2019PhRvD..99f3008L, 2019ApJ...886...23X, 2021RAA....21..305W}. Our preliminary work suggests that $pp$ interactions in the jet can generate detectable VHE emission and successfully predicts that Mrk 421, Mrk 501 and 1ES 2344+514 would be detected by LHAASO \citep{2022PhRvD.106j3021X}. 

In this work, to understand the radiation mechanism of these LHAASO-detected AGNs comprehensively, we build averaged multi-wavelength spectral energy distributions (SEDs) by combining observations of WISE in the infrared band, Zwicky Transient Facility (ZTF) in the optical band, \textsl{Swift} in the X-ray band and the ultraviolet band, and \textsl{Fermi}-Large Area Telescope (LAT) data in the $\gamma$-ray band. Several models are applied to fit SEDs, especially the VHE spectra. This paper is organized as follows: In Sect.~\ref{data}, we describe multi-wavelength observations and data reduction. The model description and fitting results are presented in Sect.~\ref{model}.
Finally, we end with discussions and conclusions in Sect.~\ref{DC}. The cosmological parameters $H_{0}=70\ \rm km\ s^{-1}Mpc^{-1}$, $\Omega_{0}=0.3$, and $\Omega_{\Lambda}$= 0.7 are adopted.

\section{Observations and data reduction} \label{data}
In this section, we present the multi-wavelength observations of Mrk 421, Mrk 501, 1ES 1727+502, 1ES 2344+514, and NGC 4278 from 2021 March 5 to 2022 September 30 during the operation of the Water Cherenkov Detector Array (WCDA) of LHAASO and describe the process of data reduction. Detailed information on these five LHAASO AGNs is given in Table~\ref{tabel:agn}.

\begin{table*}
\renewcommand\arraystretch{1.2}
\setlength\tabcolsep{5pt}
\caption{The Sample. Columns from left to right: (1) the source name, (2) right ascension (R.A.), (3) declination (Decl.), (4) the redshift of the source, (5) the SMBH mass in units of the solar mass, $M_{\odot}$, (6) Test statistic (\textsl{Fermi}-LAT), (7) Spectral Index (\textsl{Fermi}-LAT), (8) the type of AGNs. }
\centering
\begin{tabular}{cccccccc}
\hline\hline
Source name	&	R.A. (J2000) 	&	Decl. (J2000)	&	$z$	&	$M_{\rm BH}$	&	TS$\,$(500$\,$d)	&	$\Gamma_{\rm index} \,$(500$\,$d) &	Type	\\
~(1) & (2) & (3) & (4) & (5) & (6) & (7) & (8) \\
\hline
Mrk 421	&	11 04 19	&	+38 11 41	&	0.031	&	$1.35\times10^9$ \citep{2002AA...389..742W}	 & 16652 & 1.83$\pm$0.01	&	HSP blazar \\
Mrk 501	&	16 53 52.2	&	+39 45 37	&	0.034	&	$1.00\times10^9$ \citep{2001AA...367..809K}	 & 4644 & 1.78$\pm$0.02	&	HSP blazar\\
1ES 1727+502 &	17 28 18.6	&	+50 13 10	&	0.055	&	$5.62\times10^{7}$ \citep{2002AA...389..742W}   & 287 & 1.72$\pm$0.01 & HSP blazar \\
1ES 2344+514	&	23 47 04	&	+51 42 49	&	0.044	&	$6.31\times10^8$ \citep{2002AA...389..742W}   & 390 & 1.82$\pm$0.01	&	HSP blazar \\
NGC 4278	&	12 20 06.8	&	+29 16 50.7	&	0.002	&	$3.10\times10^{8}$ \citep{2003MNRAS.340..793W}  & 9 & -	&	Liner \\
\hline
\label{tabel:agn}
\end{tabular}
\end{table*}

\subsection{LHAASO}
LHAASO consists of two VHE emission detector arrays, WCDA sensitive to $\gamma$-rays with energies between $100\,{\rm GeV}$-$30\,{\rm TeV}$ and a $1.3\,{\rm km}^2$ array (KM2A) sensitive to $\gamma$-rays with energies above $10\,{\rm TeV}$ \citep{Ma_2022}. The VHE data were collected from the first LHAASO catalog. All five AGN sources were detected only by WCDA and not by KM2A. Then the parameters of the power-law SED of the VHE spectra obtained by WCDA were given in Table 1 of \cite{2023arXiv230517030C}, and all of them were evaluated without correcting the spectra for extragalactic background light (EBL) absorption. The $95\%$ statistic upper limits of the KM2A component were also given in the same table.

\subsection{Fermi-LAT}

\begin{figure*}[htbp]
	\centering
	\subfloat{\includegraphics[width=0.4\linewidth]{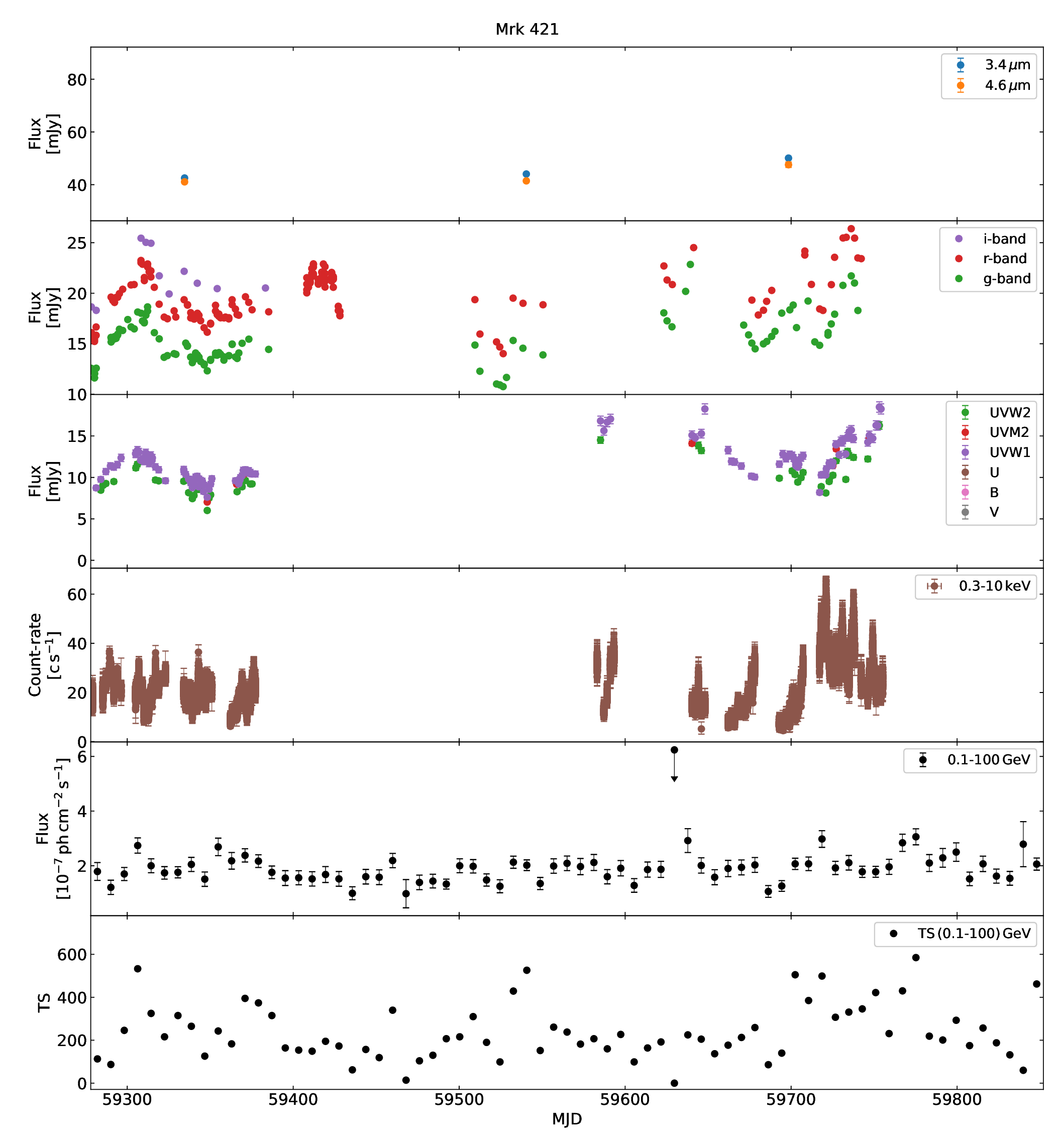}}
	\subfloat{\includegraphics[width=0.4\linewidth]{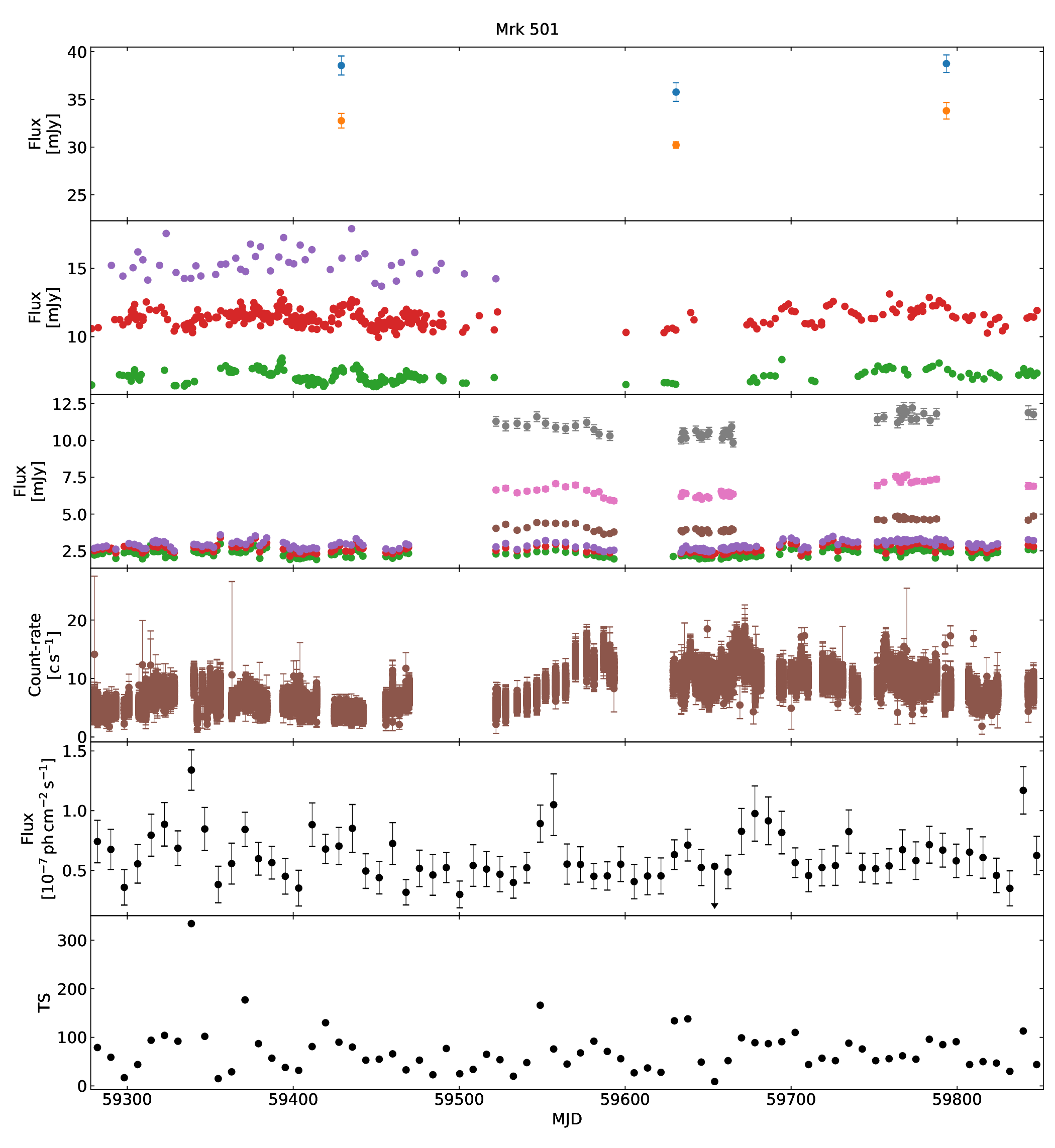}}
	\hfill
	\subfloat{\includegraphics[width=0.4\linewidth]{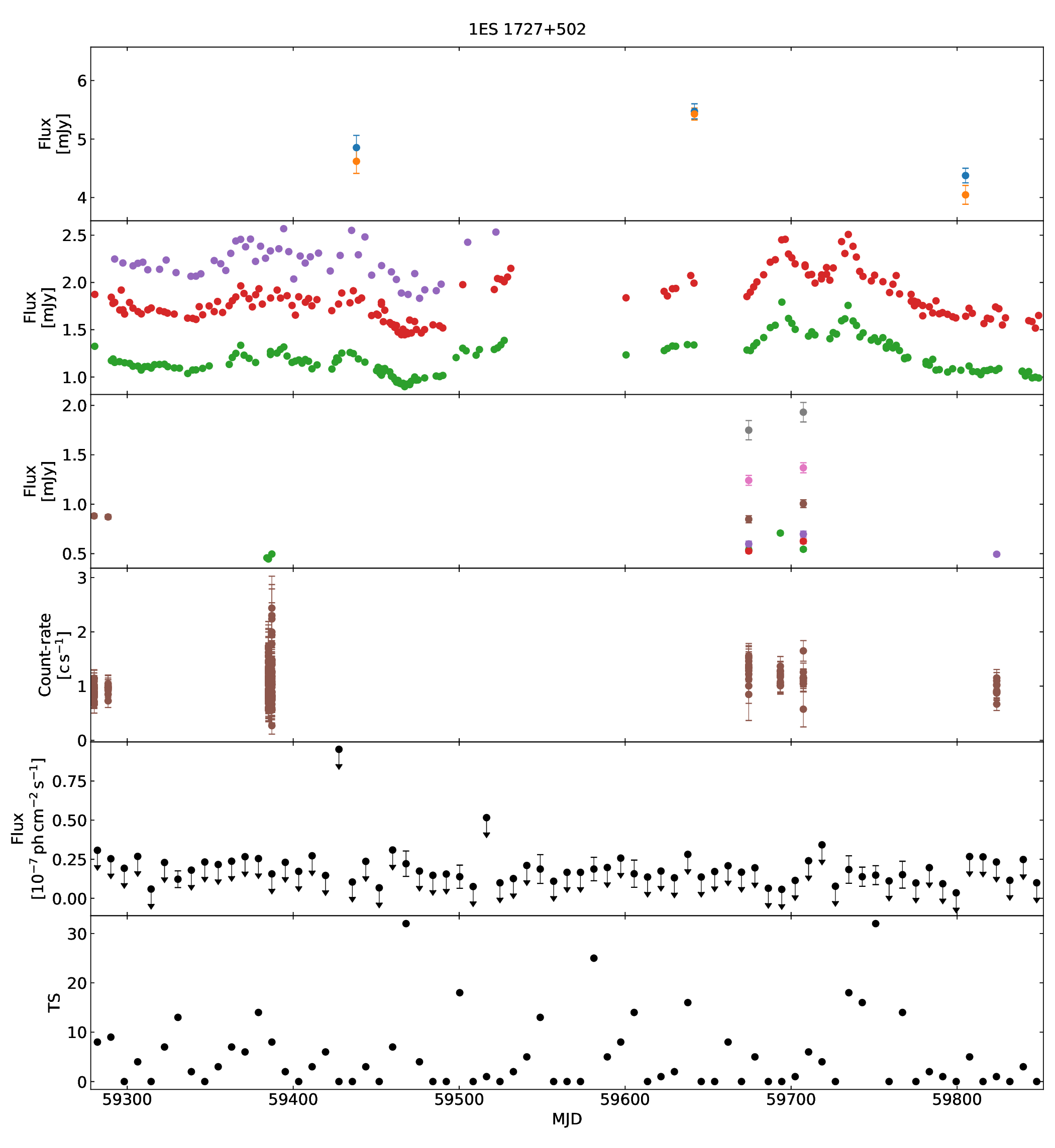}}
	\subfloat{\includegraphics[width=0.4\linewidth]{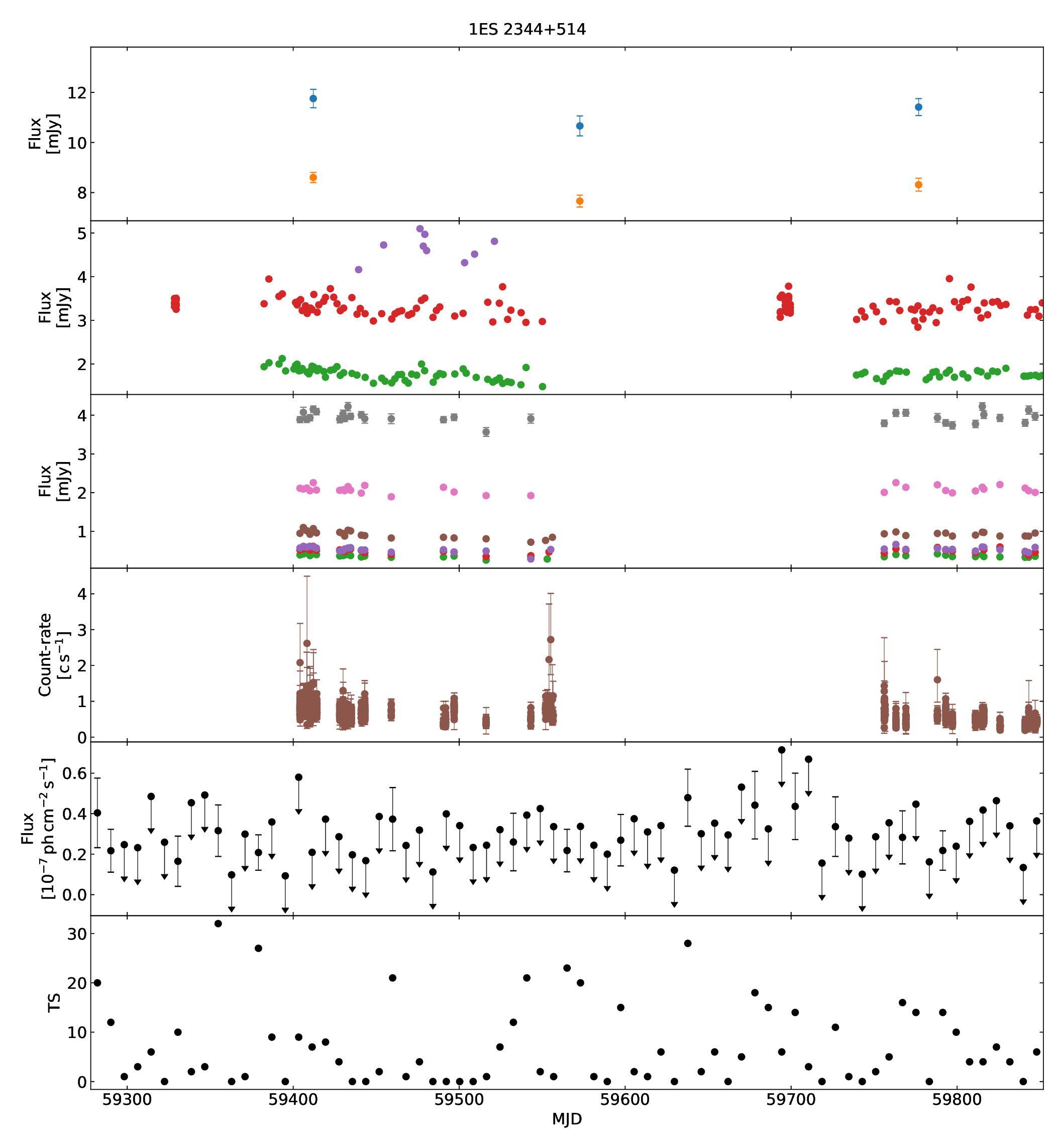}}
	\hfill
	\subfloat{\includegraphics[width=0.4\linewidth]{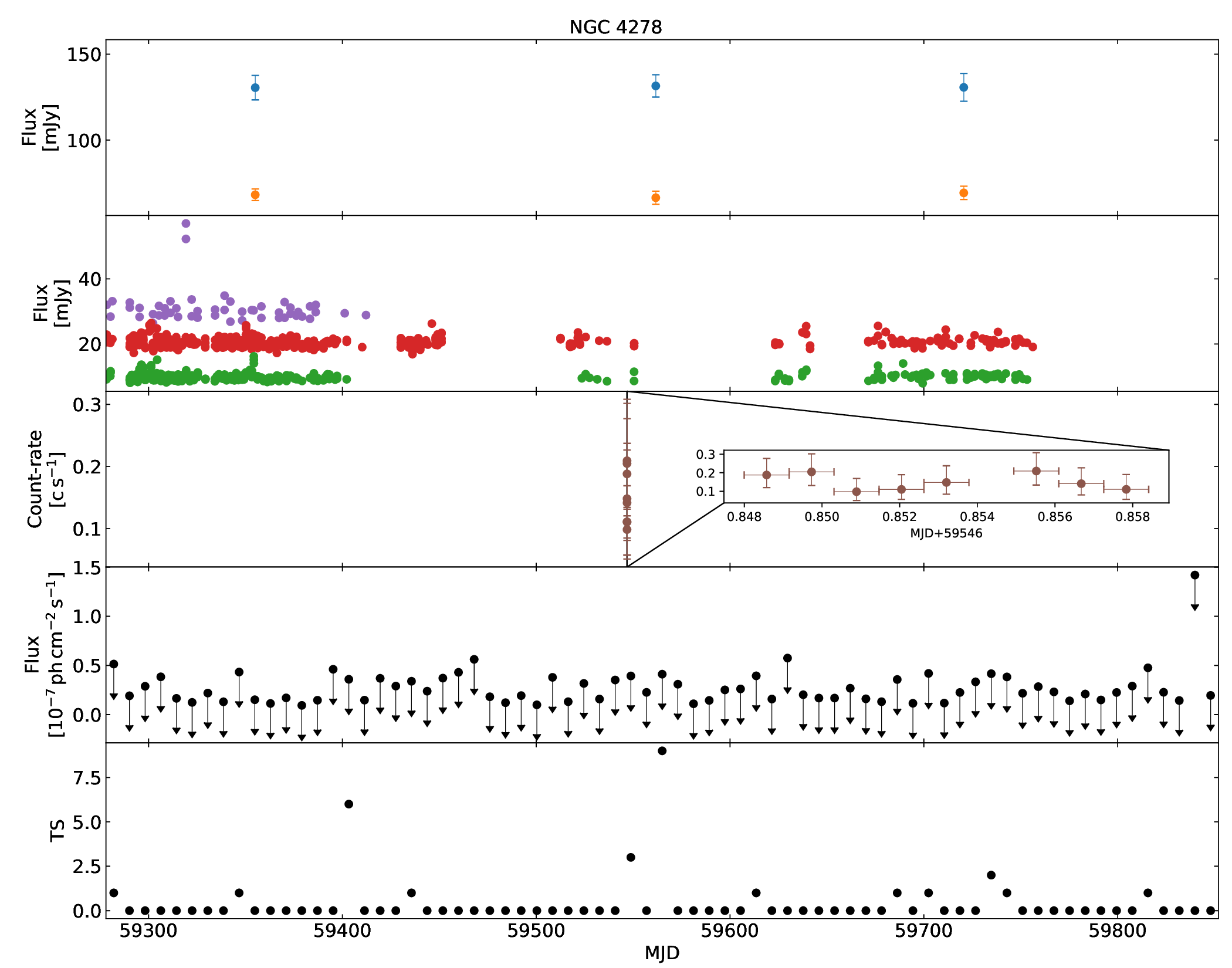}}
	
	\caption{Multiwavelength light curves (LCs) of LHAASO sources. Panels from top to bottom in these six figures: LCs of WISE, ZTF, \textsl{Swift}-UVOT, \textsl{Swift}-XRT, \textsl{Fermi}-LAT, and TS of GeV detection. There is no \textsl{Swift}-UVOT figure for NGC 4278. The meaning of symbols is given in the legend of Mrk 421. Note that the y-axis of some panels does not start from zero.
		\label{fig:lc}}
\end{figure*}

The LAT on board the \textsl{Fermi} mission is a pair-conversion instrument that is sensitive to GeV emission \citep{2009ApJ...697.1071A}. Data are analyzed with the \texttt{fermitools} version 2.2.0. A binned maximum likelihood analysis is performed on a region of interest (ROI) with a radius of $10^{\circ}$ centered at the right ascension (RA) and declination (Decl) of each source. The recommended event selections for data analysis are ``\texttt{FRONT+BACK}'' ({\texttt{evtype} $=3$}) and \texttt{evclass} $=128$. We apply a maximum zenith angle cut of $z_{\rm zmax} = 90^{\circ}$ to reduce the effect of the Earth albedo background. The standard \texttt{gtmktime} filter selection with an expression of (\texttt{DATA\_QUAL} $> 0\ \&\& \ $ \texttt{LAT\_{CONFIG}} $== 1$) is set. A source model is generated containing the position and spectral definition for all the point sources and diffuse emission from the 4FGL-DR3 catalog \citep{2022ApJS..260...53A} within $15^{\circ}$ of the ROI center. The analysis includes the standard Galactic diffuse emission model ($\rm gll\_iem\_v07.fits$) and the isotropic component (iso\_P8R3\_SOURCE\_V3\_v1.txt), respectively. We bin the data in count maps with a scale of $0.1^{\circ}$ per pixel and set ten logarithmically-spaced bins per decade in energy. An energy dispersion correction is made when event energies extending down to 100 MeV are taken into consideration. 
The spectral parameters of weak sources located within $10^{\circ}$ of the center of the ROI are fixed during the maximum likelihood fitting. In a few cases, we fix or delete some sources to obtain a convergent fit. We divide this SED into six equal logarithmic energy bins in the 0.1-100\,GeV, and an additional bin in the 100-800\,GeV for these LHAASO sources. We built GeV lightcurves using about 8-day intervals between 0.1-100\,GeV photons, shown in Fig.$\,$\ref{fig:lc}. For the data points with poorly measured fluxes (where the likelihood Test Statistic ${\rm TS} < 10$ or the nominal uncertainty of the flux is larger than half the flux
itself), upper limits at the 95\% confidence level are given. The $\rm TS$ and spectral index can be found in Table \ref{tabel:agn}.

\subsection{Swift-XRT}
\begin{figure*}[htbp]
  \centering
  \subfloat{\includegraphics[height=0.45\linewidth, angle=-90]{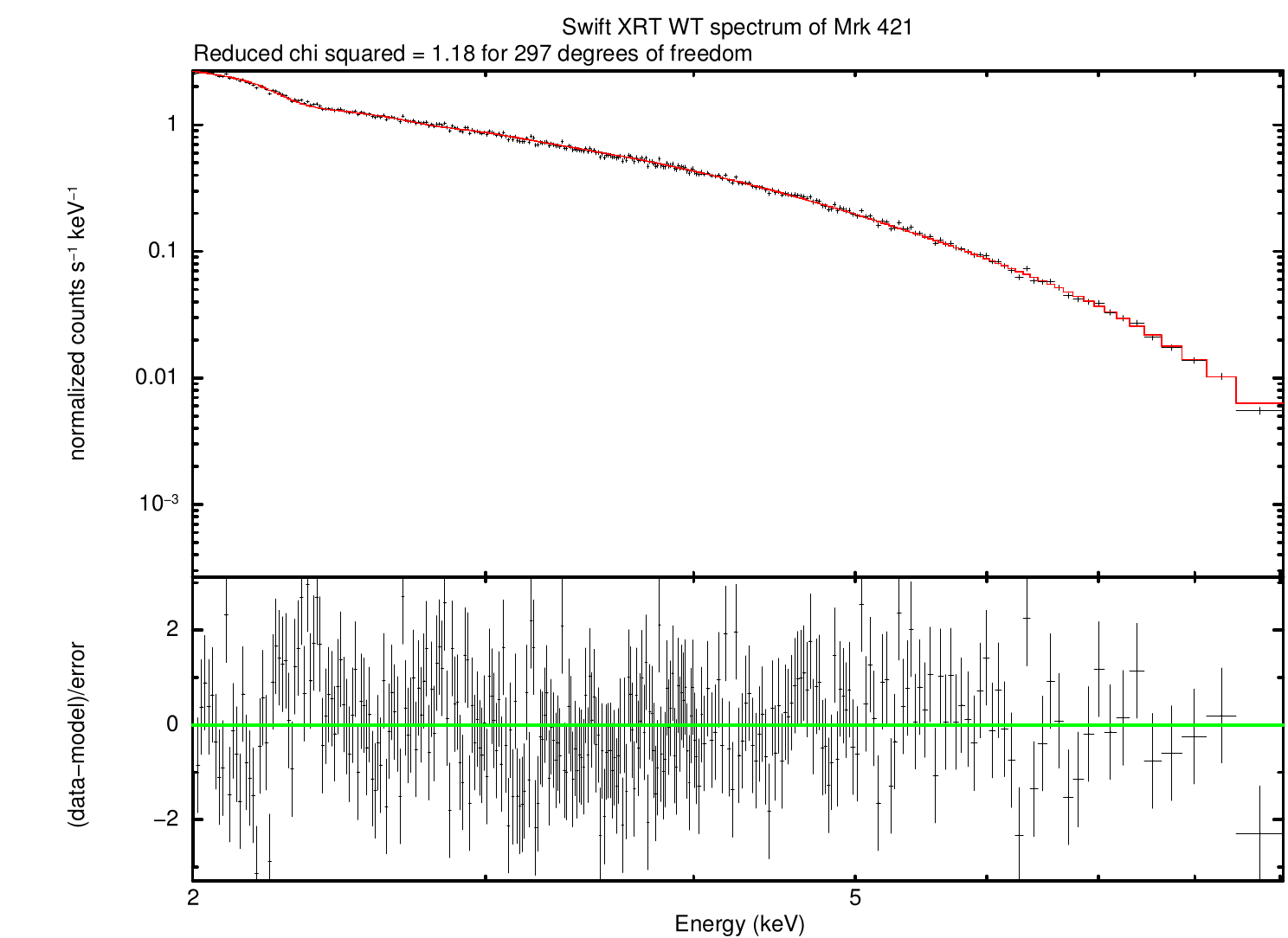}}
  \subfloat{\includegraphics[height=0.45\linewidth, angle=-90]{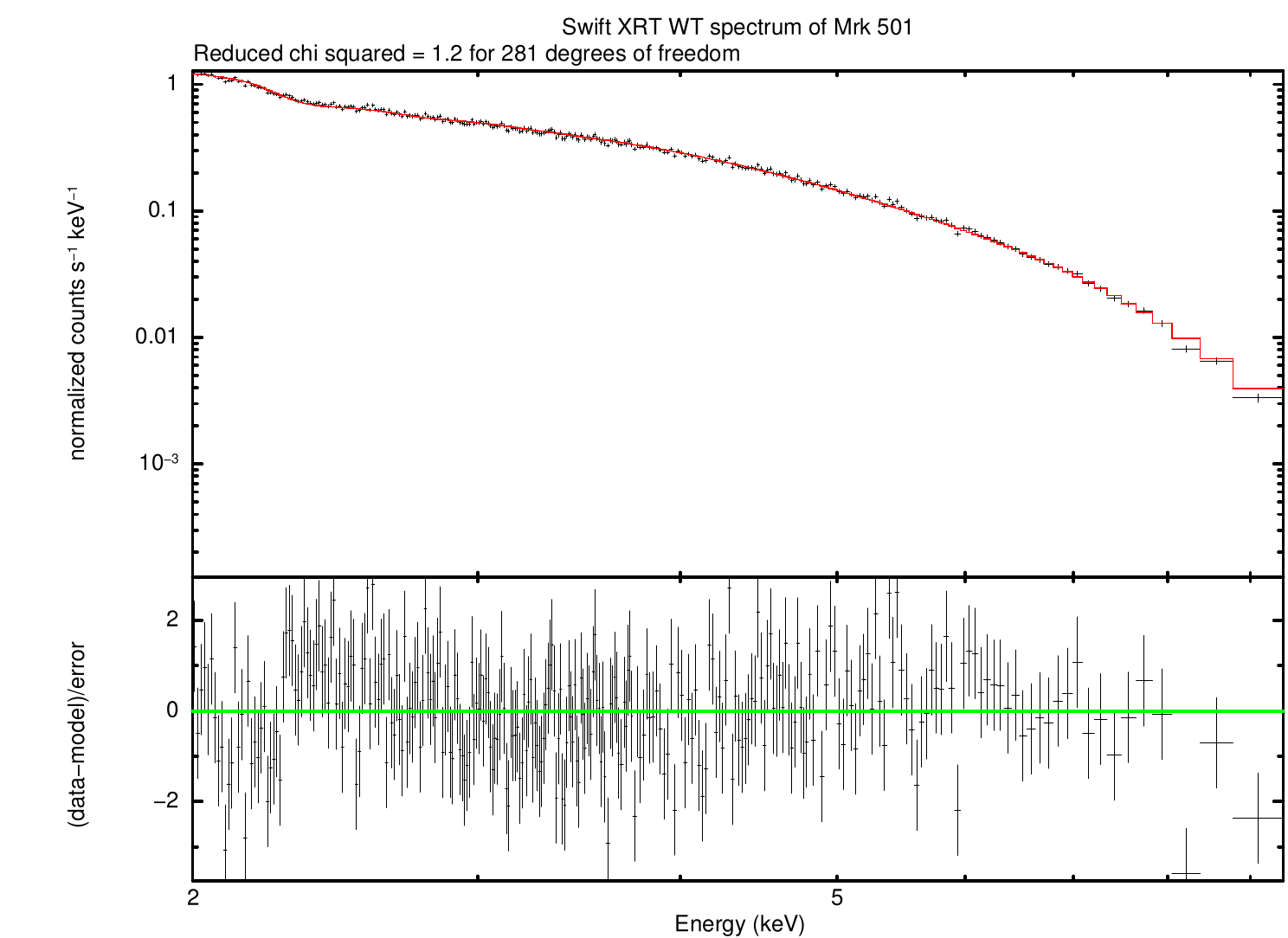}}
  \hfill
  \subfloat{\includegraphics[height=0.45\linewidth, angle=-90]{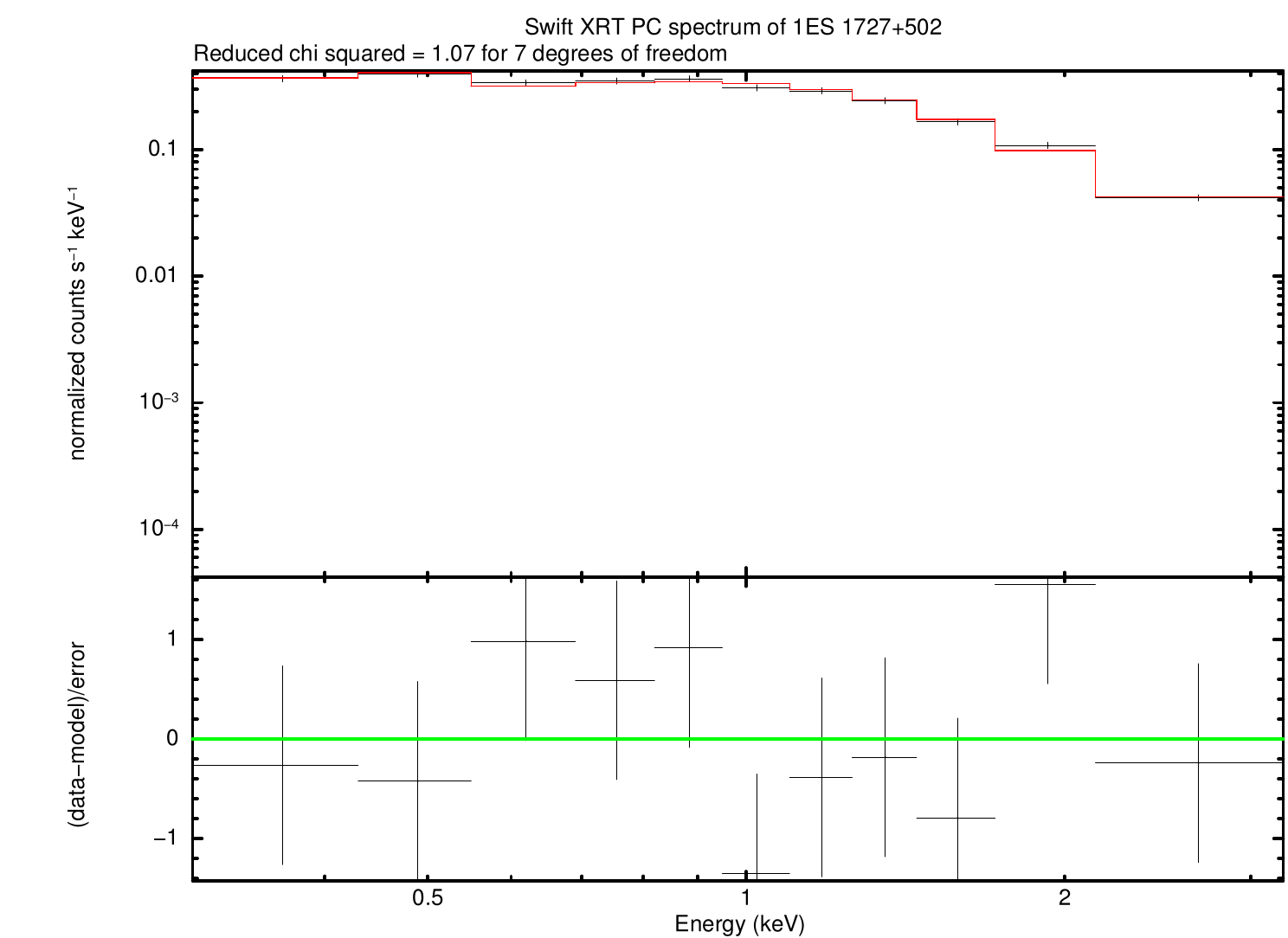}}
  \subfloat{\includegraphics[height=0.45\linewidth, angle=-90]{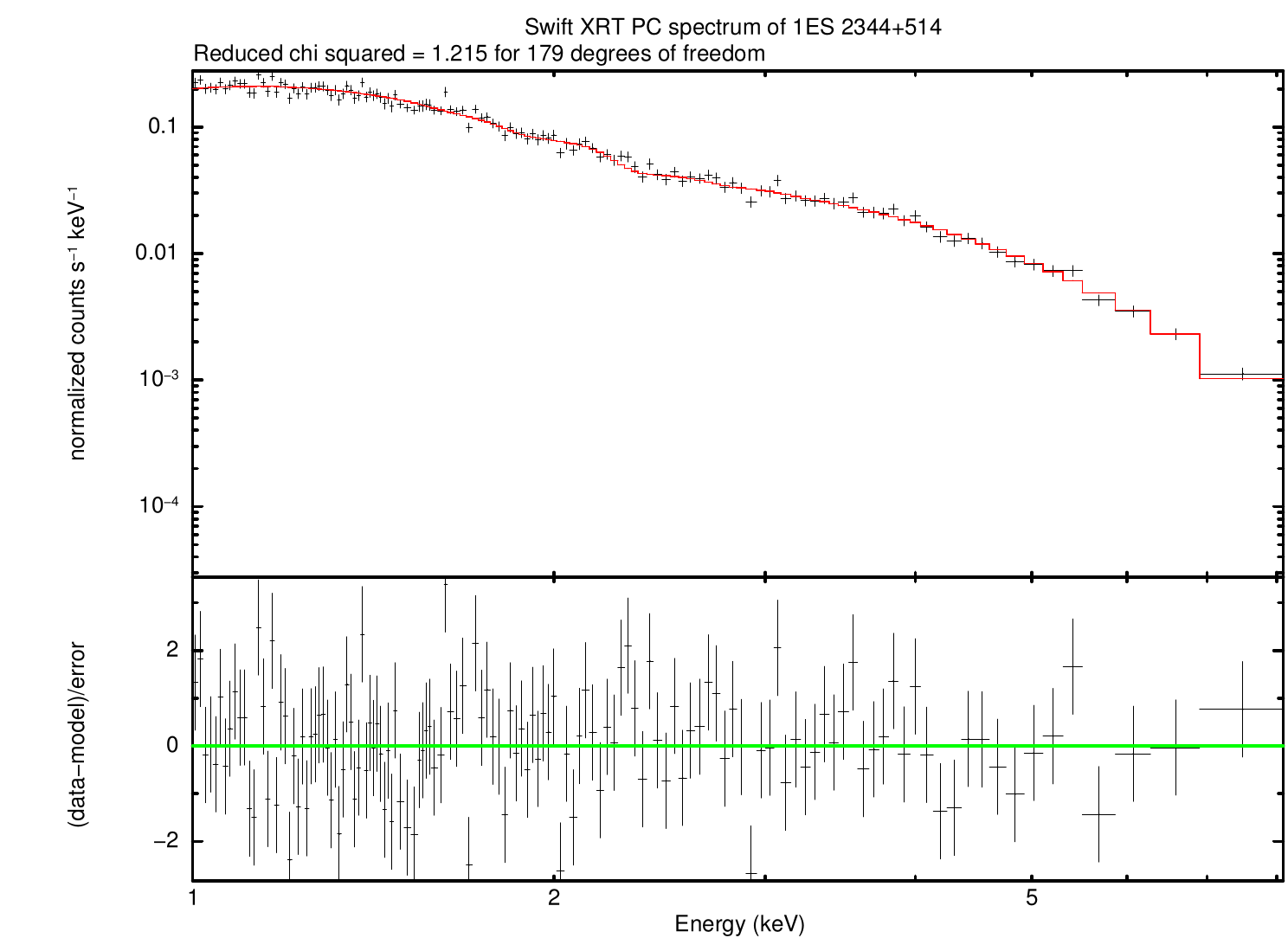}}
  \hfill
  \subfloat{\includegraphics[height=0.45\linewidth, angle=-90]{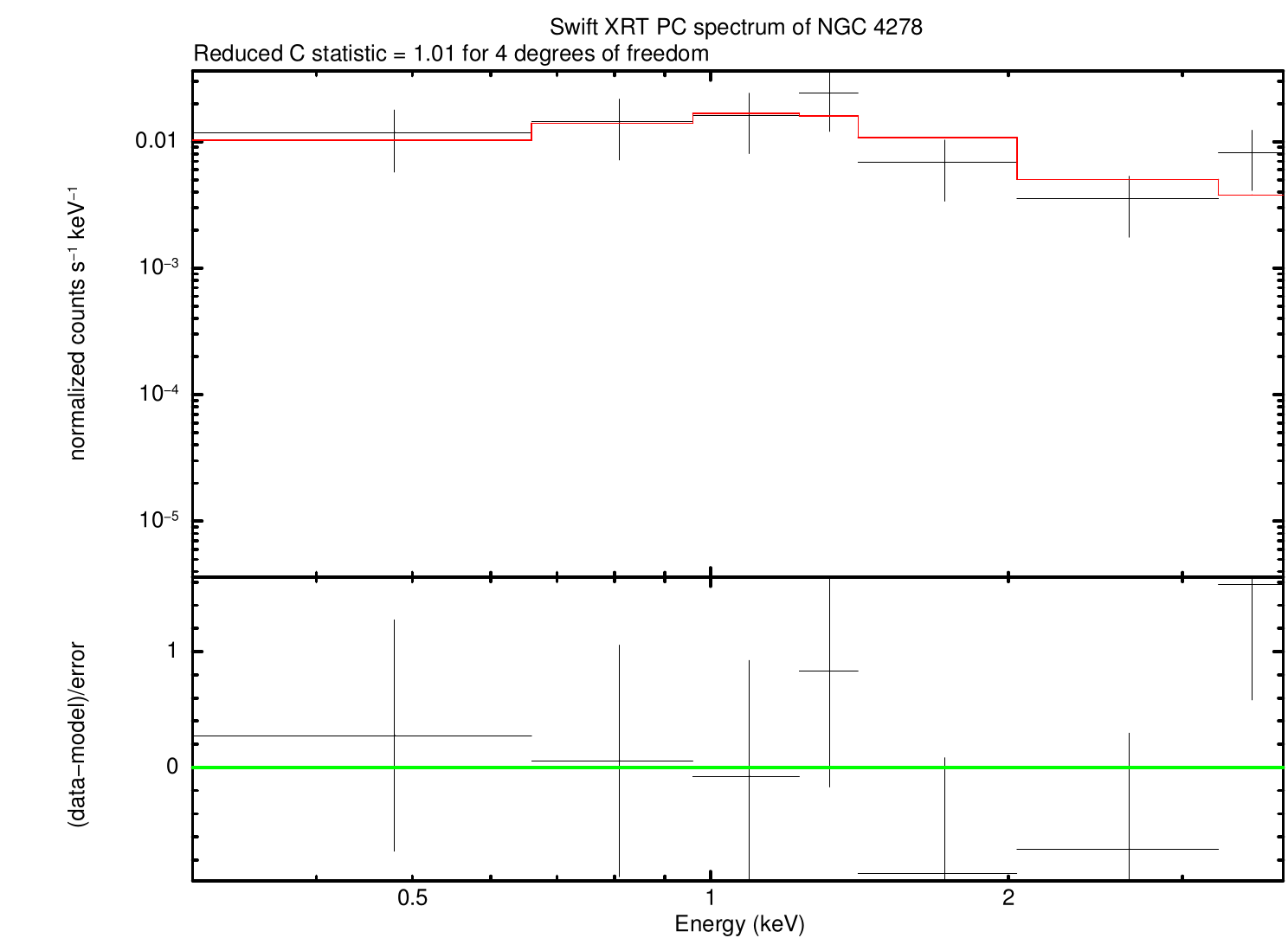}}
  
  \vspace{20pt}
  \caption{The fitting results of the \textsl{Swift}-XRT spectra.\label{fig:XRT}}
\end{figure*}

We make use of the \textsl{Swift}-XRT data products generator\footnote{\url{https://www.swift.ac.uk/user_objects/}}  (\texttt{xrt\_prods}) to obtain 0.3-10\,keV X-ray light curves and spectra. Version 1.10 of the \texttt{xrt\_prods} module is released as part of \texttt{swift tools} v3.0. This facility allows the creation of publication-ready X-ray light curves and spectra. Processing is performed using \texttt{HEASOFT} v6.29. Instrumental artifacts such as pile up and the bad columns on the CCD are corrected as suggested by \cite{2007A&A...469..379E, 2009MNRAS.397.1177E}\footnote{\url{https://www.swift.ac.uk/user_objects/docs.php}}. These spectra and X-ray light curves are produced by specifying the same covering times as the optical band data. Other settings have adopted default values from \texttt{xrt\_prods}. Those obtained spectra are not the single observed spectrum but the average spectra over the entire considered time window whose timescale for the time binning is from MJD=59278 to MJD=59852, which are observed by photon-counting (PC) mode and windowed timing (WT) mode. The WT mode spectra are taken for the Mrk 421 and Mrk 501 because there are only a few short exposure observations in PC mode and have longer exposure observations on the same day in WT mode. The PC mode spectra are chosen for 1ES 1727+502, 1ES 2344+514 and NGC 4278 due to the reasons similar to the above or no observations in WT mode. After downloading those average spectra, we chose \texttt{XSPEC} (version 12.9) to fit them, and fit the spectra of the two modes separately. The specific fitting process is as follows. In order to obtain smaller flux errors, we apply the \texttt{grppha} command to rebin channels, setting a minimum number of groups greater than 29. The group min of NGC 4278 is equal to 4 due to insufficient photon counts.  The power law (\texttt{po}) model is often considered for fitting of X-ray spectrum. It is good to reference the logarithmic parabolic (\texttt{logpar}) model \citep{2004A&A...413..489M}. Thus, the models of both \texttt{TBabs*TBabs*cflux*po} and \texttt{TBabs*TBabs*cflux*logpar} are considered for fitting these spectra. The first \texttt{TBabs} stands for the Galactic absorption $N_H$. It is taken from the \texttt{HEASARC} tool\footnote{\url{https://heasarc.gsfc.nasa.gov/cgi-bin/Tools/w3nh/w3nh.pl}} \citep{2016A&A...594A.116H}, and the value is frozen during fitting. The reduced chi-squared or C-statistic values are used to measure the goodness of the fit. The fitting statistic values are required to be less than 1.3. For different models, such as \texttt{po} and \texttt{logpar}, we choose those with statistical values closer to 1. However, when the statistical values of different models are close, even if we choose a model with a statistical value closer to 1, it does not mean that other models are completely excluded. For both Mrk 421 and Mrk 501, the \texttt{logpar} with $E_{\rm min}=2\,{\rm keV}$ model is selected because of statistical values closer to 1 compared to that from \texttt{po} model. The fitting results are shown in Table \ref{tab:XRT} and Fig.\ref{fig:XRT}. For NGC 4278, model of \texttt{TBabs*TBabs*cflux*po} with index of 1.157$\pm$0.327 is chosen to fit the spectrum while the model of \texttt{TBabs*TBabs*cflux*logpar} can not be completely excluded. Also the model of \texttt{TBabs*TBabs*cflux*(po+bbody)} does not significantly improve the fitting compared to the results from \texttt{po} or \texttt{logpar}. 
After the fitting is completed, we use the \texttt{eeufspec} command to convert them into unfolded spectra whose flux value can be transformed into $\nu F_{\nu}$ of SED. The spectra absorption is corrected by multiplying the ratio of non-absorbed and absorbed model values. For Mrk 421 and Mrk 501, no absorption correction was made because the absorption is weak when $E$ is greater than 2 keV.

\begin{table}[]
\caption{\label{tab:XRT} The fitting results for \textsl{Swift-XRT}.}
\begin{tabular}{llll}
\hline\hline
Name         & Mode & Model  & Fit Statistic\footnote{Chi-Squared value/d.o.f  for the first four sources and C-statistic value/d.o.f for NGC 4278.} \\ \hline
Mrk 421      & WT   & \texttt{po}     & 402/298        \\
             &      & \texttt{logpar} & 350/297        \\ \hline
Mrk 501      & WT   & \texttt{po}     & 349/281        \\
             &      & \texttt{logpar} & 338/281        \\ \hline
1ES 1727+502 & PC   & \texttt{po}     & 7.58/7         \\
             &      & \texttt{logpar} & 7.52/7         \\ \hline
1ES 2344+514 & PC   & \texttt{po}     & 172/131        \\
             &      & \texttt{logpar} & 156/129        \\ \hline
NGC 4278     & PC   & \texttt{po}     & 4.05/4         \\
             &      & \texttt{logpar} & 3.26/3         \\ 
             &      & \texttt{po+bb}  & 1.34/2         \\ \hline
\end{tabular}
\end{table}

\subsection{Swift-UVOT}
The ultraviolet and optical data can be obtained by the \textsl{Swift} ultraviolet and optical Telescope (UVOT) which is equipped with broadband ultraviolet (UVW1, UVM2, and UVW2) and optical (V, B, and U) filters \citep{2005SSRv..120...95R}. Based on the HEASARC Archive Search Web\footnote{\url{https://heasarc.gsfc.nasa.gov/cgi-bin/W3Browse/swift.pl}}, \textsl{Swift}-UVOT images from the 5 sources between MJD=59278 and MJD=59852 were retrieved. NGC 4278 only has one UVOT observation (obsid: 03109562002), but it is not located in the detection window. Therefore, no UVOT images are available for this source. There is only data in the ultraviolet band for Mrk 421. The latest version of \texttt{HEASoft} 6.32.1 and calibration files \texttt{CALDB} version 20211108 are used during data processing. According to UVOT analysis threads\footnote{\url{https://www.swift.ac.uk/analysis/uvot/}}, we check whether level 2 images (sw[obsid]u<filter>\underline sk.img) are correctly aligned to the world coordinate system. The small scale sensitivity check is performed by default by the software. The \texttt{uvotimsum} command is then used to sum extensions within an image, and the \texttt{uvotsource} command is used to perform aperture photometry with a circular source region of 5 arcsec radii and a circular (annular) background region of 15 to 40 arcsec (inner) radii. The output results include the magnitude of the AB system, corrected count rate, and the flux density in mJy, etc. The corrected count rate is converted into magnitude of the AB system using the new AB zeropoints \citep{2011AIPC.1358..373B}, and magnitude of the AB system into flux density by considering zero-point flux density. The flux density is corrected for Galactic extinction. The specific process is as follows: a reddening coefficient of E(B-V) is obtained from \cite{2011ApJ...737..103S} with $R_V$ = 3.1. Then the extinction value $A_{V}$ is calculated using the dust extinction laws of \cite{1999PASP..111...63F} are chosen. Based on this extinction curve, we obtain the extinction value for the Swift ultraviolet and optical bands, and then multiply our flux by 10$^{\left(0.4 \, A_{\rm Swift\,band}\right)}$ to perform the extinction correction.

\subsection{ZTF}
The optical magnitudes in g, r and i bands are collected from the 17th ZTF public data release\footnote{\url{https://irsa.ipac.caltech.edu/cgi-bin/Gator/nph-scan?submit=Select&projshort=ZTF}} \citep{2019PASP..131a8003M}. If the parameter \texttt{catflags} for a ZTF image has a value less than 32768 (i.e., does not contain bit 15), the photometry at that epoch is probably usable \citep{2019PASP..131a8003M}. Thus, in order to obtain good observation data, we require \texttt{catflags score} $=0$ for other sources apart from Mrk 421 that there are not data with \texttt{catflags score} $=0$. It should be noted that these data with \texttt{catflags score} $=4096$ are chosen for Mrk 421. We convert the g, r, i magnitudes into fluxes following \cite{2020ApJS..247...49X}. In addition, the Galactic extinctions in the g, r and i bands are corrected, and the NASA/IPAC Extragalactic Database \footnote{\url{https://ned.ipac.caltech.edu/}; The NASA/IPAC Extragalactic Database (NED) is funded by the National Aeronautics and Space Administration and operated by the California Institute of Technology.} provide extinction values \citep[also see][]{2011ApJ...737..103S}. We construct the SED using the average magnitudes and average errors during the selected period.

\subsection{WISE}
The WISE \citep{2010AJ....140.1868W} telescope has been operating a repetitive all-sky survey since 2010, except for a gap between 2011 and 2013. The WISE telescope visits each location every half a year and takes $>10$ exposures during one day. Although initially four filters were used, most of the time only two filters, named W1 and W2, are used at the moment. The central wavelengths of the two filters are $3.4 \ \mu $m and $4.6 \ \mu$m. We collected the magnitudes of the five sources by point spread functions (PSF) fitting from the NASA/IPAC InfRared Science Archive (IRSA)\footnote{\url{https://irsa.ipac.caltech.edu/applications/Gator/}}.
Following \citet{2021ApJS..252...32J}, we selected magnitudes with good image quality (\texttt{qi\_fact} $>0$) and unaffected by charged particle hits (\texttt{saa\_sep} $>0$), scattered moon light (\texttt{moon\_masked} $<1$) or artifacts (\texttt{cc\_flags} $=0$), and then binned the magnitudes every half a year since we did not detect any intraday variabilities. The magnitudes are in the Vega system, then we can convert WISE Vega magnitudes to flux density units with $F_{\nu}=F_{\nu0}\times10^{-0.4\,m}$, where the zero magnitude flux density ($F_{\nu0}$) for the W1 and W2 bands in 309.5 Jy and 171.8 Jy, respectively, with $m$ being the calibrated WISE magnitude.

\section{SED modelling}\label{model}

\begin{table*}
\resizebox{\linewidth}{!}{
\begin{threeparttable}[b]
\caption{\label{tab:par}The fitting parameters. The minimum electron Lorent factor $\gamma_{\rm e,min}$ is set to $1\times10^{2}$ because it is insensitive in fitting. The '-' sign indicates that the parameters do not exist in the One-zone SSC or SSC+$pp$ scenario.}
\begin{tabular}{cccccccccccccll}
\hline\hline
\multicolumn{11}{c}{SSC model and SSC+$pp$ model  \tnote{a}}                                                                                                                                                                                                                                             &                        &                        &                              &                                    \\
Source name                  & $\theta$ & $\Gamma$ & $L_{\rm e}^{\rm inj}\left({\rm erg}\,{\rm s}^{-1}\right)$ & $\gamma_{\rm e,b}$ & $\gamma_{\rm e,max}$ & $p_{\rm e,1}$ & $p_{\rm e,2}$ & $B\left({\rm G}\right)$ & $R\left({\rm cm}\right)$         & $L_{\rm p}^{\rm inj}$/$L_{\rm Edd}$ \tnote{b} & $\chi^2/{\rm d.o.f}$ \tnote{c}               & $\chi^2_{\rm WCDA}/{\rm d.o.f}$ \tnote{d}         & \multicolumn{1}{c}{$\chi^2_{pp}/{\rm d.o.f}$} & \multicolumn{1}{c}{$\chi^2_{pp,{\rm WCDA}}/{\rm d.o.f}$} \\ \hline
Mrk 421                      & 1.8      & 23       & 1.00E+44                                                  & 1.70E+05           & 1.00E+07             & 2.20          & 4.20          & 0.06                    & 1.40E+16                         & 2.78E-01                            & 3.27                   & 39.30                  & \multicolumn{1}{c}{2.25}     & \multicolumn{1}{c}{24.88}          \\
Mrk 501                      & 1.8      & 23       & 4.40E+43                                                  & 3.50E+05           & 1.00E+07             & 2.23          & 4.50          & 0.09                    & 5.00E+15                         & 4.17E-01                            & 6.18                   & 70.08                  & \multicolumn{1}{c}{3.20}     & \multicolumn{1}{c}{26.21}          \\
1ES 1727+502                 & 1.8      & 23       & 4.20E+43                                                  & 3.00E+04           & 1.00E+07             & 2.03          & 3.00          & 0.05                    & 6.50E+15                         & 6.00E+01                            & 20.07                  & 19.86                  & \multicolumn{1}{c}{20.02}    & \multicolumn{1}{c}{20.17}          \\
1ES 2344+514(SSC)            & 1.8      & 23       & 6.00E+43                                                  & 6.00E+05           & 1.00E+07             & 2.50          & 3.50          & 0.08                    & 1.90E+15                         & -                                   & 6.97                   & 19.43                  & \multicolumn{1}{c}{-}        & \multicolumn{1}{c}{-}              \\
1ES 2344+514(SSC+$pp$)       & 1.8      & 23       & 8.00E+43                                                  & 8.00E+05           & 1.00E+07             & 2.60          & 4.20          & 0.08                    & 2.50E+15                         & 5.56E-03                            & -                      & -                      & \multicolumn{1}{c}{6.50}     & \multicolumn{1}{c}{21.44}          \\
NGC 4278$^{\rm a}$(SSC)      & 1.8      & 5        & 8.00E+40                                                  & 9.00E+05           & 1.00E+07             & 1.00          & 3.00          & 0.04                    & 8.50E+13                         & -                                   & 61.58                  & 18.93                  & \multicolumn{1}{c}{-}        & \multicolumn{1}{c}{-}              \\
NGC 4278$^{\rm a}$(SSC+$pp$) & 1.8      & 5        & 2.15E+43                                                  & 4.00E+03           & 1.00E+07             & 1.50          & 4.90          & 0.01                    & 1.00E+15                         & 1.22E-04                            & -                      & -                      & \multicolumn{1}{c}{50.14}    & \multicolumn{1}{c}{8.56}           \\
NGC 4278$^{\rm b}$(SSC)      & 30       & 3        & 1.30E+42                                                  & 3.00E+06           & 5.00E+07             & 1.00          & 2.30          & 0.20                    & 1.50E+14                         & -                                   & 60.55                  & 18.80                  & \multicolumn{1}{c}{-}        & \multicolumn{1}{c}{-}              \\
NGC 4278$^{\rm b}$(SSC+$pp$) & 30       & 3        & 3.90E+43                                                  & 5.00E+03           & 1.00E+07             & 1.50          & 4.90          & 0.05                    & 3.00E+15                         & 2.78E-01                            & -                      & -                      & \multicolumn{1}{c}{47.76}    & \multicolumn{1}{c}{5.28}           \\ \hline\hline
\multicolumn{11}{c}{SSC+proton-synchrotron   model (two zone) \tnote{e}}                                                                                                                                                                                                                                &                        &                        &                              &                                    \\
Source name                  & $\theta$ & $\Gamma$ &                                                           &                    &                      &               &               & $B\left({\rm G}\right)$ & $R\left({\rm cm}\right)$         & $L_{\rm p}^{\rm inj}$/$L_{\rm Edd}$ & $\chi^2/{\rm d.o.f}$               & $\chi^2_{\rm WCDA}/{\rm d.o.f}$     &                              &                                    \\ \hline
1ES 2344+514                 & 1.8      & 12       &                                                           &                    &                      &               &               & 9.00                    & 1.00E+17                         & 2.00E-07 \tnote{f}                            & 6.45                   & 23.72                &                              &                                    \\ \hline\hline
\multicolumn{11}{c}{spine-layer model   (EC) \tnote{g}}                                                                                                                                                                                                                                                 &                        &                        &                              &                                    \\
Source name                  & $\theta$ & $\Gamma$ & $L_{\rm e}^{\rm inj}\left({\rm erg}\,{\rm s}^{-1}\right)$ & $\gamma_{\rm e,b}$ & $\gamma_{\rm e,max}$ & $p_{\rm e,1}$ & $p_{\rm e,2}$ & $B\left({\rm G}\right)$ & $R_{\rm c}\left({\rm cm}\right)$ & $L\left({\rm cm}\right)$            & $\chi^2/{\rm d.o.f}$               & $\chi^2_{\rm WCDA}/{\rm d.o.f}$     &                              &                                    \\ \hline
Mrk 421 (spine)              & 1.8      & 21       & 4.20E+43                                                  & 2.10E+05           & 1.00E+07             & 1.90          & 4.30          & 0.03                    & 9.00E+16                         & 1.00E+17                            & \multirow{2}{*}{3.33}  & \multirow{2}{*}{59.03} &                              &                                    \\
Mrk 421 (layer)              & 1.8      & 4        & 7.30E+41                                                  & 2.50E+04           & 1.00E+07             & 1.90          & 4.30          & 0.03                    & 1.08E+17                         & 5.00E+17                            &                        &                        &                              &                                    \\
Mrk 501   (spine)            & 1.8      & 18       & 4.40E+44                                                  & 5.00E+05           & 1.00E+07             & 2.10          & 3.90          & 0.02                    & 9.00E+16                         & 1.00E+17                            & \multirow{2}{*}{4.75}  & \multirow{2}{*}{63.12} &                              &                                    \\
Mrk 501 (layer)              & 1.8      & 3        & 5.00E+41                                                  & 3.00E+04           & 1.00E+07             & 2.10          & 3.90          & 0.02                    & 1.08E+17                         & 5.00E+17                            &                        &                        &                              &                                    \\
1ES 1727+502   (spine)       & 1.8      & 21       & 2.50E+43                                                  & 6.00E+04           & 1.00E+07             & 1.90          & 3.10          & 0.02                    & 5.00E+16                         & 1.00E+17                            & \multirow{2}{*}{21.79} & \multirow{2}{*}{26.96} &                              &                                    \\
1ES 1727+502 (layer)         & 1.8      & 4        & 4.00E+41                                                  & 3.00E+04           & 1.00E+07             & 1.90          & 3.10          & 0.02                    & 6.00E+16                         & 5.00E+17                            &                        &                        &                              &                                    \\
1ES 2344+514   (spine)       & 1.8      & 23       & 1.00E+43                                                  & 2.00E+05           & 1.00E+07             & 2.00          & 3.50          & 0.02                    & 7.00E+16                         & 1.00E+17                            & \multirow{2}{*}{7.16}  & \multirow{2}{*}{26.78} &                              &                                    \\
1ES 2344+514 (layer)         & 1.8      & 4        & 1.50E+42                                                  & 1.20E+04           & 1.00E+07             & 2.00          & 3.50          & 0.02                    & 8.40E+16                         & 5.00E+17                            &                        &                        &                              &                                    \\
NGC 4278$^{\rm a}$ (spine)   & 1.8      & 18       & 3.50E+38                                                  & 8.00E+05           & 1.00E+07             & 1.00          & 3.40          & 0.15                    & 5.20E+15                         & 1.00E+17                            & \multirow{2}{*}{60.17} & \multirow{2}{*}{1.98}  &                              &                                    \\
NGC 4278$^{\rm a}$ (layer)   & 1.8      & 2        & 9.00E+39                                                  & 1.00E+04           & 1.00E+07             & 1.00          & 3.40          & 0.15                    & 6.24E+15                         & 5.00E+17                            &                        &                        &                              &                                    \\
NGC 4278$^{\rm b}$ (spine)   & 30       & 25       & 2.00E+41                                                  & 8.00E+02           & 1.00E+07             & 1.00          & 3.20          & 0.30                    & 7.00E+16                         & 1.00E+17                            & \multirow{2}{*}{66.78} & \multirow{2}{*}{26.43} &                              &                                    \\
NGC 4278$^{\rm b}$ (layer)   & 30       & 2        & 4.60E+40                                                  & 5.00E+06           & 5.00E+07             & 1.00          & 3.20          & 0.30                    & 8.40E+16                         & 5.00E+17                            &                        &                        &                              &                                    \\ \hline\hline
\multicolumn{11}{c}{spine-layer model   (two zone) \tnote{h}}                                                                                                                                                                                                                                           &                        &                        &                              &                                    \\
Source name                  & $\theta$ & $\Gamma$ & $L_{\rm e}^{\rm inj}\left({\rm erg}\,{\rm s}^{-1}\right)$ & $\gamma_{\rm e,b}$ & $\gamma_{\rm e,max}$ & $p_{\rm e,1}$ & $p_{\rm e,2}$ & $B\left({\rm G}\right)$ & $R_{\rm c}\left({\rm cm}\right)$ & $L\left({\rm cm}\right)$            & $\chi^2/{\rm d.o.f}$               & $\chi^2_{\rm WCDA}/{\rm d.o.f}$    &                              &                                    \\ \hline
Mrk 421 (spine)              & 1.8      & 18       & 3.50E+43                                                  & 1.20E+05           & 1.00E+07             & 1.90          & 4.20          & 0.11                    & 1.80E+16                         & 1.00E+17                            & \multirow{2}{*}{5.79}  & \multirow{2}{*}{43.82} &                              &                                    \\
Mrk 421 (layer)              & 1.8      & 6        & 2.00E+42                                                  & 1.00E+07           & 5.00E+07             & 1.90          & 4.20          & 0.04                    & 2.16E+16                         & 5.00E+17                            &                        &                        &                              &                                    \\
Mrk 501   (spine)            & 1.8      & 18       & 1.60E+43                                                  & 2.70E+05           & 1.00E+07             & 2.10          & 3.60          & 0.16                    & 1.00E+16                         & 1.00E+17                            & \multirow{2}{*}{3.57}  & \multirow{2}{*}{51.00} &                              &                                    \\
Mrk 501 (layer)              & 1.8      & 8        & 3.00E+42                                                  & 1.00E+07           & 5.00E+07             & 2.10          & 3.60          & 0.06                    & 1.20E+16                         & 5.00E+17                            &                        &                        &                              &                                    \\
1ES 1727+502   (spine)       & 1.8      & 15       & 2.40E+43                                                  & 3.00E+04           & 1.00E+07             & 2.00          & 3.00          & 0.09                    & 8.50E+15                         & 1.00E+17                            & \multirow{2}{*}{24.66} & \multirow{2}{*}{28.37} &                              &                                    \\
1ES 1727+502 (layer)         & 1.8      & 7        & 1.00E+42                                                  & 1.00E+07           & 5.00E+07             & 2.00          & 3.00          & 0.02                    & 1.02E+16                         & 5.00E+17                            &                        &                        &                              &                                    \\
1ES 2344+514   (spine)       & 1.8      & 23       & 2.30E+43                                                  & 1.50E+05           & 1.00E+07             & 2.30          & 3.50          & 0.08                    & 6.00E+15                         & 1.00E+17                            & \multirow{2}{*}{6.50}  & \multirow{2}{*}{28.28} &                              &                                    \\
1ES 2344+514 (layer)         & 1.8      & 8        & 2.80E+43                                                  & 1.00E+07           & 5.00E+07             & 2.30          & 3.50          & 0.03                    & 7.20E+15                         & 5.00E+17                            &                        &                        &                              &                                    \\
NGC 4278$^{\rm a}$ (spine)   & 1.8      & 10       & 3.00E+39                                                  & 4.00E+05           & 1.00E+07             & 1.00          & 4.20          & 0.04                    & 3.40E+14                         & 1.00E+17                            & \multirow{2}{*}{62.78} & \multirow{2}{*}{1.86}  &                              &                                    \\
NGC 4278$^{\rm a}$ (layer)   & 1.8      & 2        & 2.80E+41                                                  & 1.00E+07           & 5.00E+07             & 1.00          & 4.20          & 0.04                    & 4.08E+14                         & 5.00E+17                            &                        &                        &                              &                                    \\
NGC 4278$^{\rm b}$ (spine)   & 30       & 18       & 7.00E+41                                                  & 7.00E+06           & 1.00E+07             & 1.00          & 3.50          & 0.06                    & 5.00E+14                         & 1.00E+17                            & \multirow{2}{*}{69.04} & \multirow{2}{*}{27.31} &                              &                                    \\
NGC 4278$^{\rm b}$ (layer)   & 30       & 2        & 3.00E+41                                                  & 1.00E+07           & 5.00E+07             & 1.00          & 3.50          & 0.06                    & 6.00E+14                         & 5.00E+17                            &                        &                        &                              &                                   \\ \hline
\end{tabular}
   \begin{tablenotes}
     \item[a] The SSC+$pp$ model is fitted based on the SSC model for cases of Mrk 421, Mrk 501 and 1ES 1727+502, and its leptonic parameters are identical to those of the SSC model, barring an extra proton injection luminosity. In the other three cases, the SSC+$pp$ model has different leptonic parameters than the SSC model. For more detailed discussion, please refer to Sect. \ref{sec:pp}.
     \item[b] In the SSC+$pp$ model, we assume that the power of the cold protons in the jet is $0.5\,L_{\rm Edd}$. So the power of the jet should be at least $0.5\,L_{\rm Edd}+L_{\rm p}^{\rm inj}$.
     \item[c] The corresponding chi-square value per degrees of freedom for each object is calculated by $\chi^2/{\rm d.o.f} = \frac{1}{m-n}\sum_{i=1}^{m}(\frac{\hat {y}_i-y_i}{\sigma_i})^2$, where $m$ is the number of the observational data points, $n$ is the number of free parameters in the fitting model, $\hat {y}_i$ are the expected values from the model, $y_i$ are the observed data and $\sigma_i$ is the standard deviation for each data point. For the WCDA spectrum with only the power-law bow-tie, we divide it into an average of 30 bins in the logarithmic energy space to calculate the chi-square value.
     \item[d] The subscript 'WCDA' indicates that we use only WCDA data to calculate the chi-square value.
     \item[e] We consider two radiation zones here. The initial zone maintains the same leptonic parameters as the SSC+$pp$ model, while the second zone primarily considers proton-synchrotron radiation. The parameters for the latter zone are shown here. For more detailed discussion, please refer to Sect. \ref{sec:psyn}.
     \item[f] Although the proton injection luminosity is much smaller than the Eddington luminosity, the magnetic power is comparable to the Eddington luminosity in this case.
     \item[g] We present two strategies for modeling with the spine-layer model. Here we consider using the EC process to fit the high-energy hump independently and show its parameters here. A more detailed discussion can be found in Sect. \ref{sec:sl}.
     \item[h] Here are the parameters for another fitting strategy of the spine-layer model. The SEDs are reproduced by the superposition of emissions from two components.
   \end{tablenotes}
  \end{threeparttable}}
\end{table*}

These five LHAASO AGNs do not show significant flare\footnote{The criterion here for a significant flare is that the peak flux is more than three times the average flux in a given band. It is worth noting that although the X-ray light curve of Mrk 421 around MJD 59721 does not meet the criteria for significant flares, it shows a very peculiar behaviour. At the peak of the X-ray flares, the optical flux was at its lowest and then it began to brighten.} during the same observation period of LHAASO in all bands, as shown in Fig.$\,$\ref{fig:lc}. Therefore, we use the averaged flux of each band to construct SEDs. As mentioned above, the first LHAASO catalog report five AGNs including four HSP blazars and one Liner-type AGN. In the case of blazars, the multi-wavelength emission, apart from the obvious thermal peaks in the infrared and optical bands, is undoubtedly from the jet. For the Liner-type AGN NGC 4278, the jet/radiatively inefficient accretion flow (RIAF) and the thin accretion disk may alternately dominate as the origin of the radiation, depending on the strength of its X-ray emission \citep{2010A&A...517A..33Y}. In our data analysis, the discovered hard X-ray power-law spectrum favours the jet origin. In order to better understand the radiation mechanisms of these LHAASO detected AGN, we consider four popular jet models to reproduce the SEDs: the one-zone SSC model, the one-zone SSC+$pp$ model, the one-zone proton-synchrotron model and the spine-layer model.

The synchrotron radiation, the inverse Compton (IC) radiation and the $pp$ interactions radiation are calculated using the \texttt{naima} Python package \citep{naima}. We also consider the absorption of $\gamma$-ray photons due to the soft photons in the radiation zone \citep{2022PhRvD.106j3021X} and the extragalactic background light \cite[EBL;][]{2011MNRAS.410.2556D} during propagation in intergalactic space. In addition, the energy of the absorbed $\gamma$-ray photons in the radiation region is converted to lower energies through the cascade process. The calculation of the cascade spectrum is applied as proposed in \cite{2013ApJ...768...54B}.

In the infrared and optical bands of SEDs, most sources exhibit a clear hump, which differs significantly from the trend in the other bands. This is normally suggested as the emission from the host galaxy. We assume that the host galaxy is a 13 Gyr old elliptical galaxy for all of the AGNs \citep{2014MNRAS.442..629R} and use the SWIRE template\footnote{\url{http://www.iasf-milano.inaf.it/~polletta/templates/swire_templates.html}} \citep{2007ApJ...663...81P} to generate the spectrum of host galaxy in the fitting. The host galaxy contribution is based on the results of \cite{2007ApJ...663...81P} for Mrk 501 and 1ES 2344+514. The flux of the host galaxy in the R-band of $\sim1\,{\rm mJy}$ is used for 1ES 1727+502 \citep{2007A&A...475..199N}. For NGC 4278, the host galaxy contribution can be clearly distinguished from the continuous radiation components, so it can be obtained by fitting. The spectrum of infrared, optical, and UV data from Mrk 421 in Fig.$\,$\ref{fig:ssc} has a power-law shape, indicating a weak contribution from the host galaxy.

\subsection{One-zone SSC model}\label{sec:ssc}

\begin{figure*}[htbp]
  \centering
  \subfloat{\includegraphics[width=0.45\linewidth]{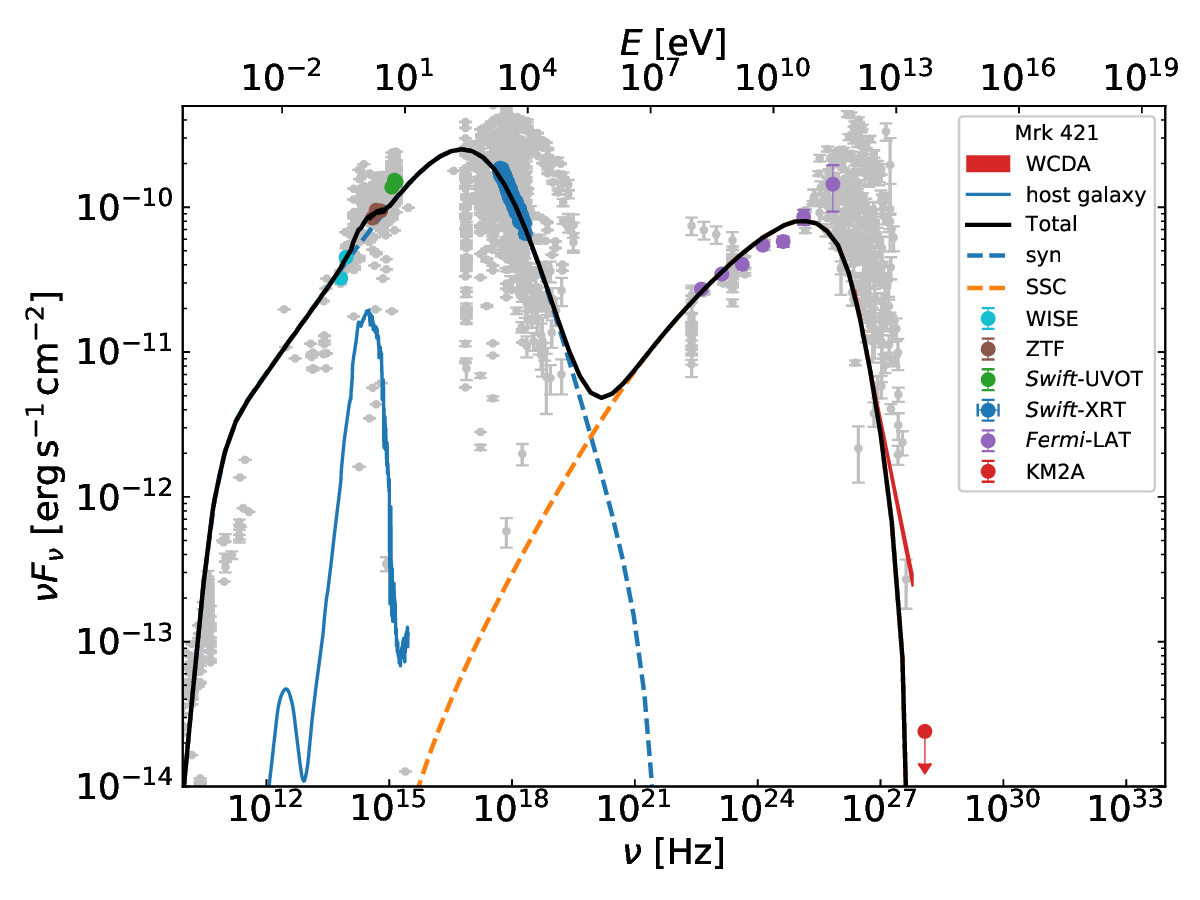}}
  \subfloat{\includegraphics[width=0.45\linewidth]{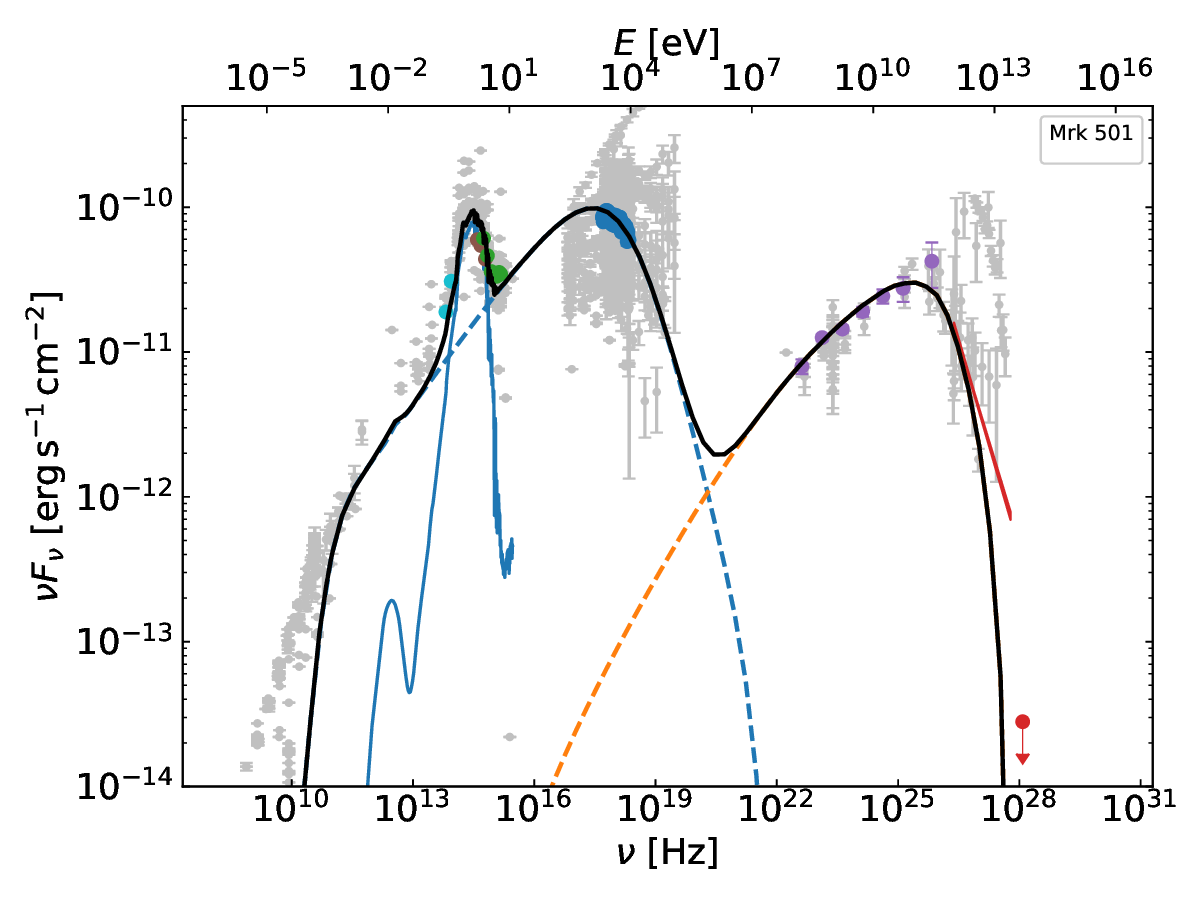}}
  \hfill
  \subfloat{\includegraphics[width=0.45\linewidth]{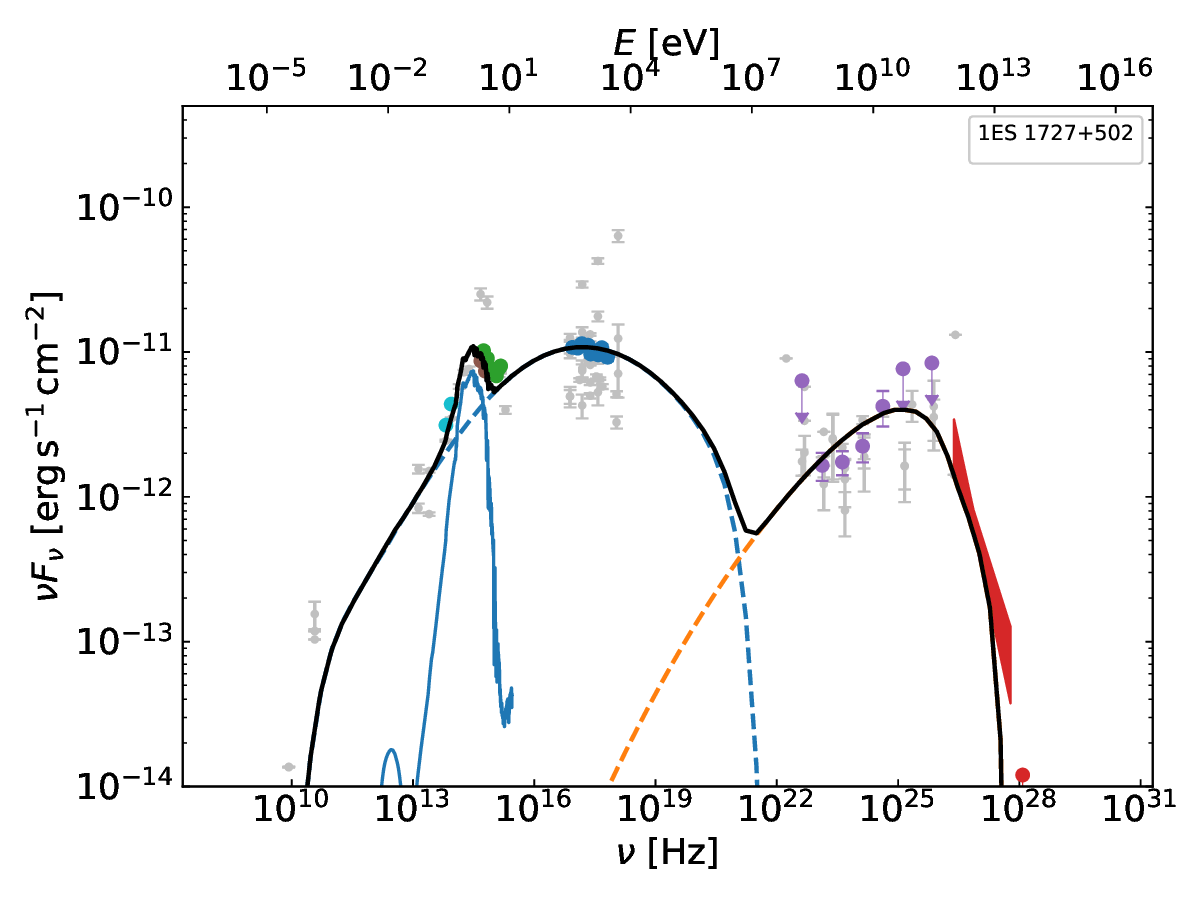}}
  \subfloat{\includegraphics[width=0.45\linewidth]{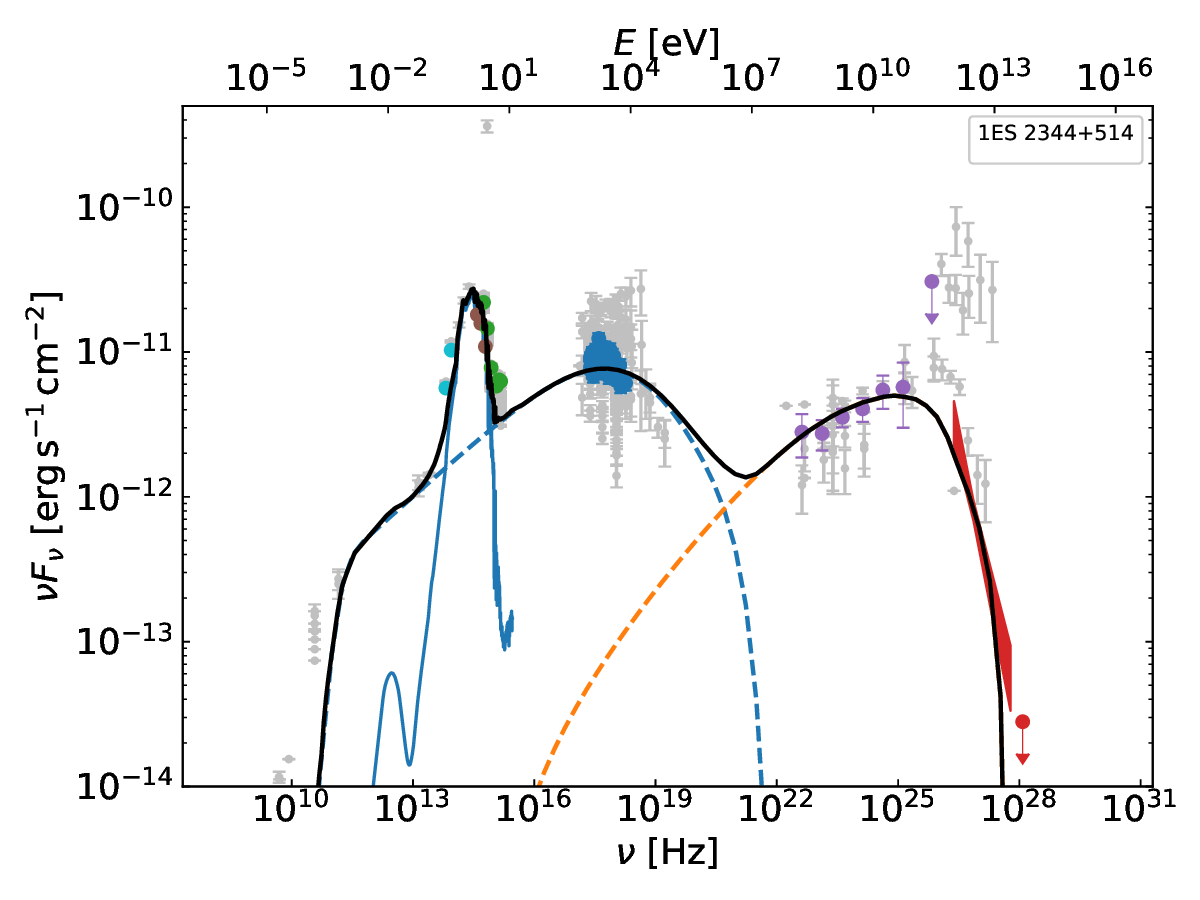}}
  \hfill
  \subfloat{\includegraphics[width=0.45\linewidth]{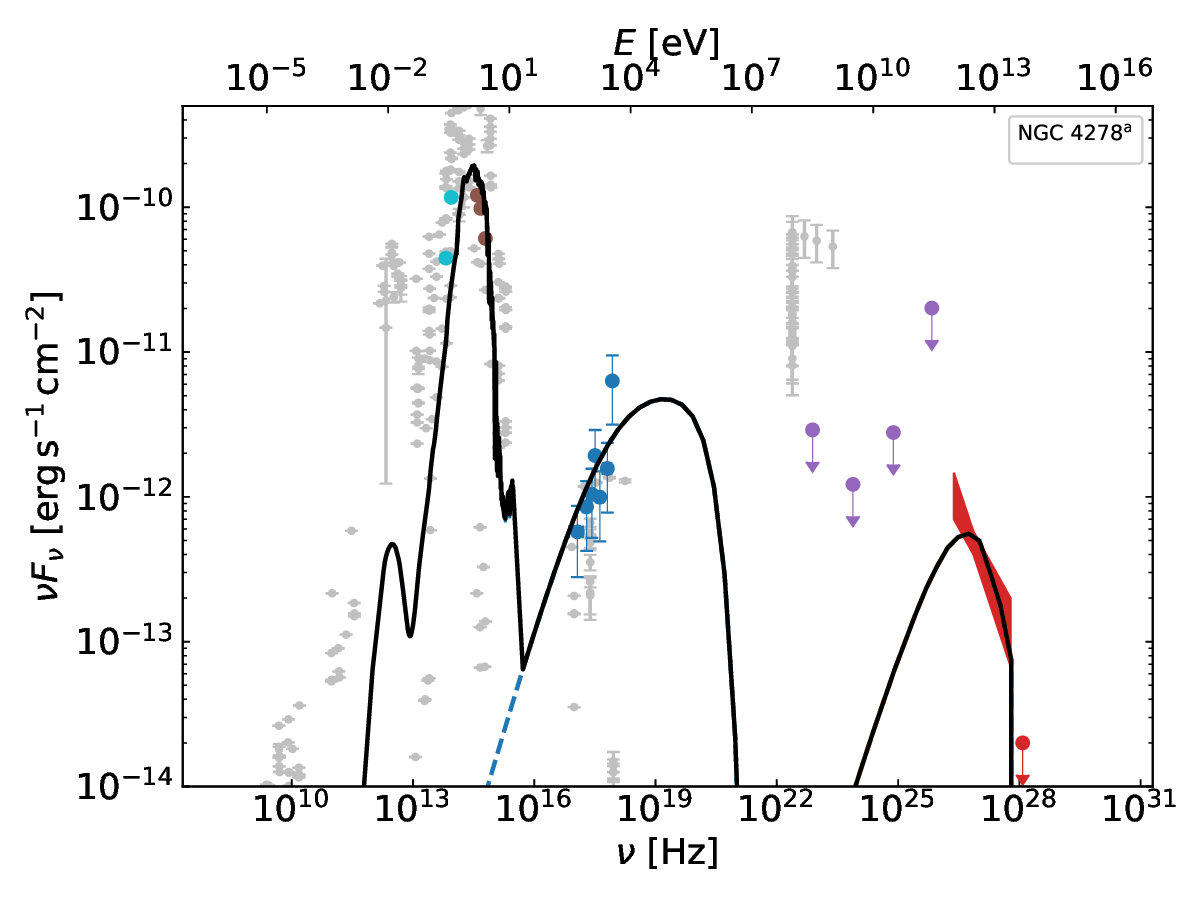}}
  \subfloat{\includegraphics[width=0.45\linewidth]{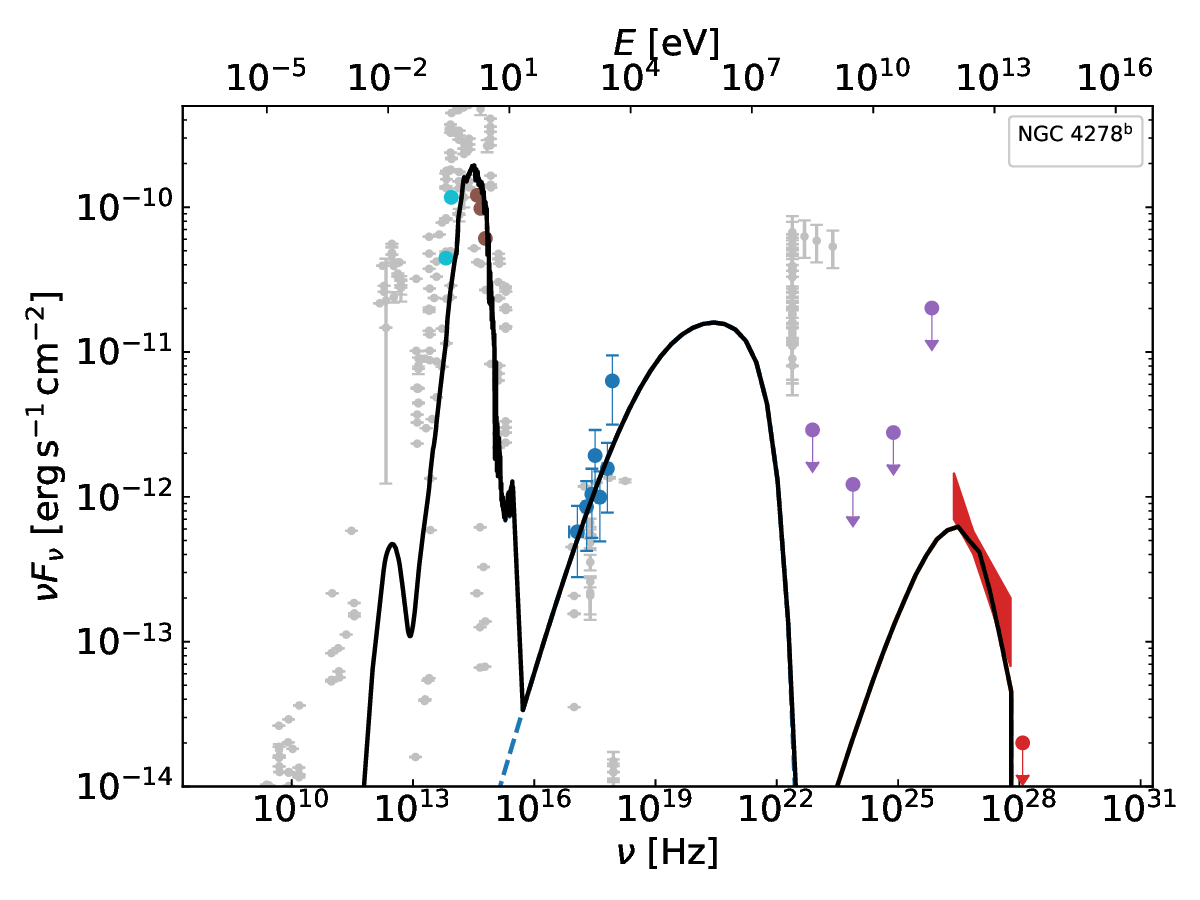}}
  \caption{One-zone SSC modeling. The meanings of line styles are given in the legend of Mrk 421. The light blue data points are infrared data from WISE, the dark blue data points are optical data from ZTF, the orange and green data points are X-ray data from \textsl{Swift}-XRT's PC mode and WT mode, respectively, and the purple data points are $\gamma$-ray data from \textsl{Fermi}-LAT, the red strap shows the observation of WCDA, and the red upper limit point is from KM2A. The gray data points are historical data from the ASI/SSDC SED Builder Tool (\url{https://tools.ssdc.asi.it/}) of the Italian Space Agency \citep{2011arXiv1103.0749S}.\label{fig:ssc}}
\end{figure*}

The one-zone SSC model is the simplest and most commonly used model in the study of jet emission. In this paper, we assume a broken power-law injection electron density distribution. By taking into account the radiative cooling and the escape of the electrons, the steady-state electron density distribution can be calculated with \citep{2019ApJ...886...23X}
\begin{equation}\label{eq:EED}
N_{\rm e}(\gamma_{\rm e})=\frac{3L_{\rm e}^{\rm inj}n_{\rm e}^{\rm inj}(\gamma_{\rm e})}{4\pi R^3m_{\rm e}c^2 \int{\gamma_{\rm e} n_{\rm e}^{\rm inj}(\gamma_{\rm e}){\rm d}{\gamma_{\rm e}}}}{\rm min}\{t_{\rm cool}(\gamma_{\rm e}),t_{\rm esc}\},
\end{equation}
where $n_{\rm e}^{\rm inj}(\gamma_{\rm e})\propto\left\{\begin{array}{ll}\gamma_{\rm e}^{-s_{\rm e,1}}, & \gamma_{\rm e,min}\leq \gamma_{\rm e}\leq  \gamma_{\rm e,b}~\\\gamma_{\rm e,b}^{s_{\rm e,2}-s_{\rm e,1}}\gamma_{\rm e}^{-s_{\rm e,2}}, & \gamma_{\rm e,b}<\gamma_{\rm e}\leq\gamma_{\rm e,max}\end{array}\right.$ is the injection electron density distribution, $\gamma_{\rm e}$ is the electron Lorentz factors, $\gamma_{\rm e,min}$ and $\gamma_{\rm e,max}$ are the minimum and maximum electron Lorentz factors of the distribution, $\gamma_{\rm e,b}$ is the break electron Lorentz factor, $s_{\rm e,1}$ and $s_{\rm e,2}$ are the low-energy and the high-energy indexes of the broken power-law spectrum, $L_{\rm e}^{\rm inj}$ is the electron injection luminosity, $R$ is the radius of radiation zone, $m_{\rm e}$ is the rest mass of the electron, $c$ is the speed of light, $t_{\rm esc}=R/c$ is the escape timescale, $t_{\rm cool}=3m_{\rm e}c/(4\sigma_{\rm T}\gamma_{\rm e}(u_{\rm B}+f_{\rm KN}u_{\rm ph}))$ is the electron cooling timescale, $\sigma_{\rm T}$ is the Thomson scattering cross section, $f_{\rm KN}$ is the factor accounting for Klein-Nishina (KN) effects \citep{2005MNRAS.363..954M}, $u_{\rm B}=B^{2}/\left(8\pi\right)$ is the energy density of the magnetic field, $B$ is the magnetic field strength, and $u_{\rm ph}$ is the energy density of the soft photons\footnote{We use the iterative approach to calculate $u_{\rm ph,\,syn}$ in the SSC process.}. 
The observed emission will be Doppler boosted by a factor $\delta^4$, where $\delta=[\Gamma(1-\beta_{\Gamma} \rm cos\theta)]^{-1}$ is the Doppler factor, $\Gamma$ is the the bulk Lorentz factor, $\beta_{\Gamma}c$ is the velocity of the jet, $\theta$ is the viewing angle of the jet.

Current observational data do not provide good constraints on all parameters in the modeling. Therefore, we fix some of the less sensitive parameters in fitting to reduce the number of free parameters. All the fitting parameters can be found in Table \ref{tab:par}. The relativistic jet of the blazars is close to the line of sight of the observer, so we set the viewing angles for Mrk 421, Mrk 501, 1ES 1727+502, and 1ES 2344+514 to $1.8^{\circ}$ uniformly. \cite{2005ApJ...622..178G} suggest that the viewing angle of NGC 4278 is uncertain, and their study reports that it could have a small viewing angle $\left(2^{\circ}<\theta<4^{\circ}\right)$, alternatively a large viewing angle. We therefore divide it into two cases, one with the same viewing angle ($\theta=1.8^{\circ}$, hence NGC 4278$^{\rm a}$) as the blazar sources and the other with a larger viewing angle ($\theta=30^{\circ}$, hence NGC 4278$^{\rm b}$).

The fitting results of the one-zone SSC model are shown in Fig.$\,$\ref{fig:ssc}. In the case of four blazars, it can be found that the low-energy component of the SED can be reproduced by the superposition of host galaxy emission and electron synchrotron radiation, except that the model slightly underestimates the UV data. This may be due to the use of non-simultaneous data. For example, the absence of \textsl{Swift}-UVOT observations from MJD 59500 to MJD 59550 for Mrk 421, in which period its flux for the optical band is at its lowest state during the entire observation period, and may overestimate its flux in the UV band. The high-energy component can be fitted very well by the SSC model. However, the high-energy tail of the VHE spectrum of LHAASO is poorly interpreted for Mrk 421 and Mrk 501\footnote{It should be noted that the VHE spectra of TeV AGNs are sometimes fitted as log-parabolas, whereas the first LHAASO catalog only reports power-law SEDs.}. This is caused by the KN effects, which steepens the spectrum naturally. When $\gamma_{\rm e} E_{0}/\left(m_{\rm e}c^2\right)>1$, the IC scattering occurs from the Thomson regime into the KN regime, where $E_{0}$ is the energy of the soft photon in the comoving frame. Then we can obtain the critical electron Lorentz factor
\begin{equation}\label{eq:gamma_KN}
    \gamma_{\rm KN}=\frac{m_{\rm e}c^2}{E_{0}},
\end{equation}
and the corresponding critical energy of the IC radiation can be estimated by $E_{\rm KN}\approx \gamma_{\rm KN}^{2}E_{0}=m_{\rm e}^{2}c^4/E_{0}$. In the observer frame,
\begin{equation}\label{eq:KN}
    E_{\rm KN}^{\rm obs}\approx \frac{\delta^{2}m_{\rm e}^{2}c^4}{\left(1+z\right)^{2}}\frac{1}{E_{0}^{\rm obs}}.
\end{equation}
The soft photon energy can be approximately replaced by the peak energy of the low-energy hump. For Mrk 421 and Mrk 501, we can obtain $E_{0}^{\rm obs}\sim 1\,{\rm keV}$. Then substituting $E_{0}^{\rm obs}$ and $\delta$ into Eq. (\ref{eq:KN}), we get the critical energy $E_{\rm KN}^{\rm obs}\approx 0.2\,{\rm TeV}$. This means that the IC radiation spectrum is steeper above $\sim0.2~\rm TeV$ because of KN effects, as shown in Fig.$\,$\ref{fig:ssc}. Therefore, the high-energy tail of LHAASO spectra cannot be fitted with the one-zone SSC model, unless very extreme parameters are considered.

In the case of NGC 4278, due to the lack of GeV $\gamma$-ray data, the fitting parameters have a larger space to choose from. Nevertheless, in order to explain the spectra in both the X-ray and VHE bands simultaneously, extreme parameters are required. For example, the model requires a very hard low-energy slope ($s_{\rm e,1}=1$) or a very large minimum electron Lorentz factor (approaching $\gamma_{\rm e,b}$) to explain the very hard X-ray spectra. If we consider that its X-ray radiation is produced by the electron-synchrotron process, a large $\gamma_{\rm e,b}$ (around $10^{6}$) is required. In addition, the critical energy $E_{\rm KN}^{\rm obs}$ is about $0.7\,{\rm GeV}$ for the case of NGC 4278$^{\rm a}$ and $7\,{\rm MeV}$ for the case of NGC 4278$^{\rm b}$, which requires the low-energy slope $s_{\rm e,1}$ to be close to 1 and the high-energy slope $s_{\rm e,2}\leq3$ in order to counteract the impact of the KN limit and fit the VHE spectra. For NGC 4278$^{\rm b}$, it is also necessary to set $\gamma_{\rm e,max}$ to $5\times10^{7}$.

\subsection{One-zone SSC+\texorpdfstring{$pp$}. model}\label{sec:pp}

\begin{figure*}[htbp]
  \centering
  \subfloat{\includegraphics[width=0.45\linewidth]{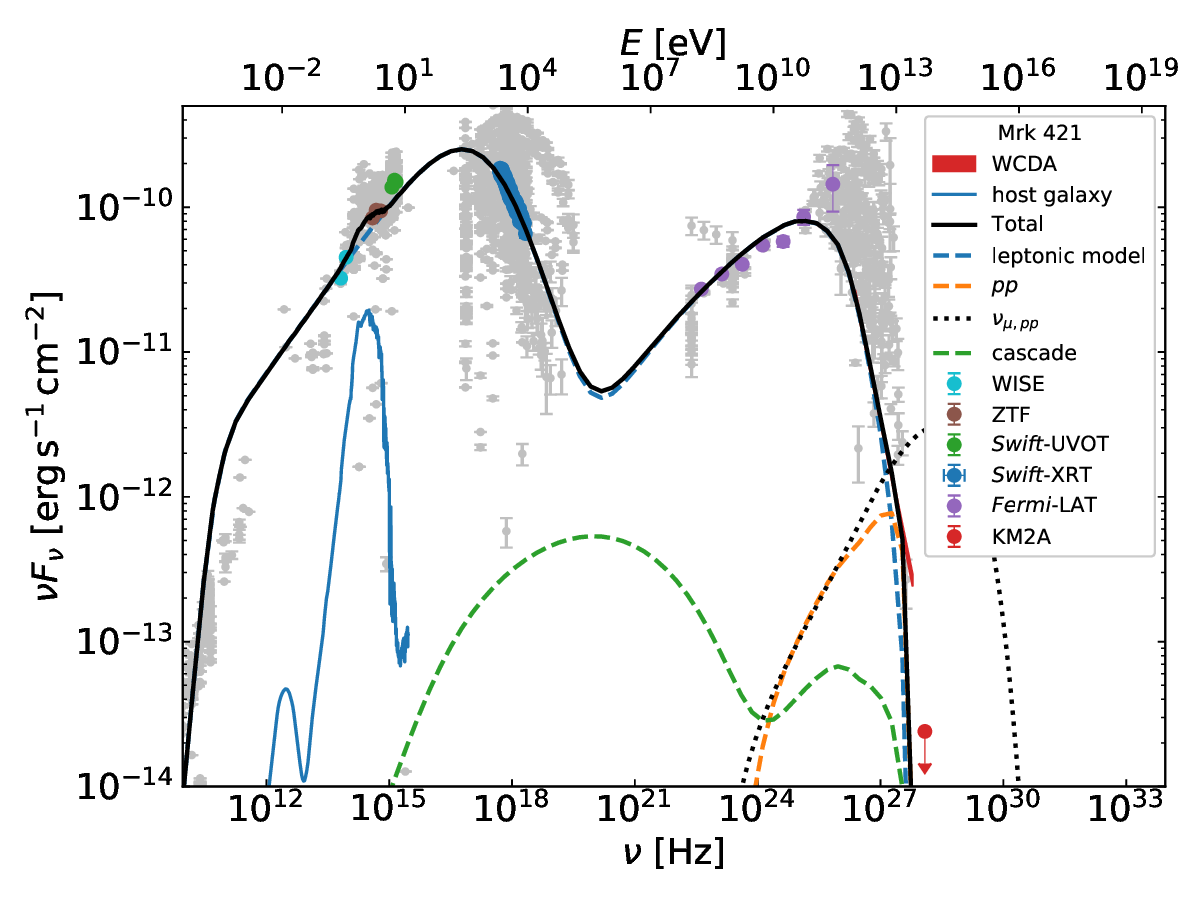}}
  \subfloat{\includegraphics[width=0.45\linewidth]{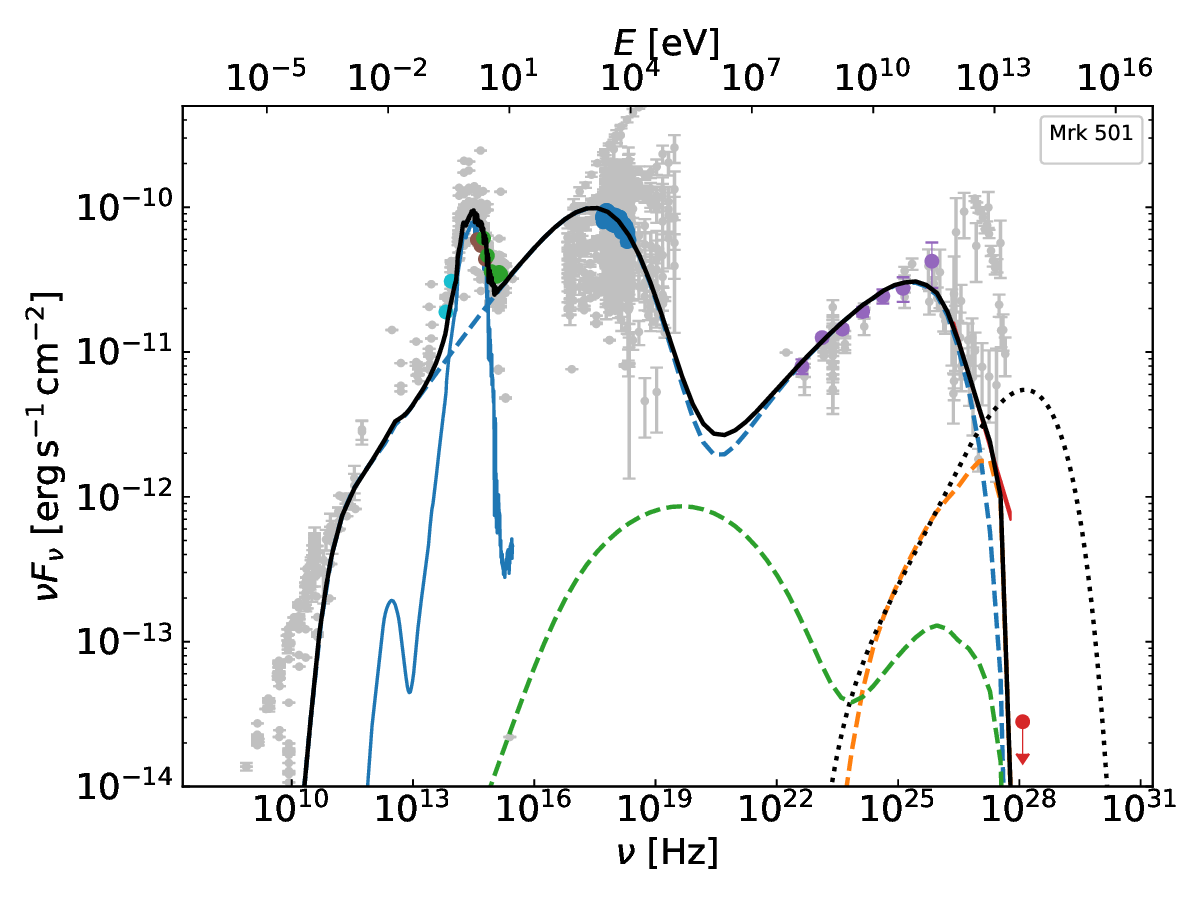}}
  \hfill
  \subfloat{\includegraphics[width=0.45\linewidth]{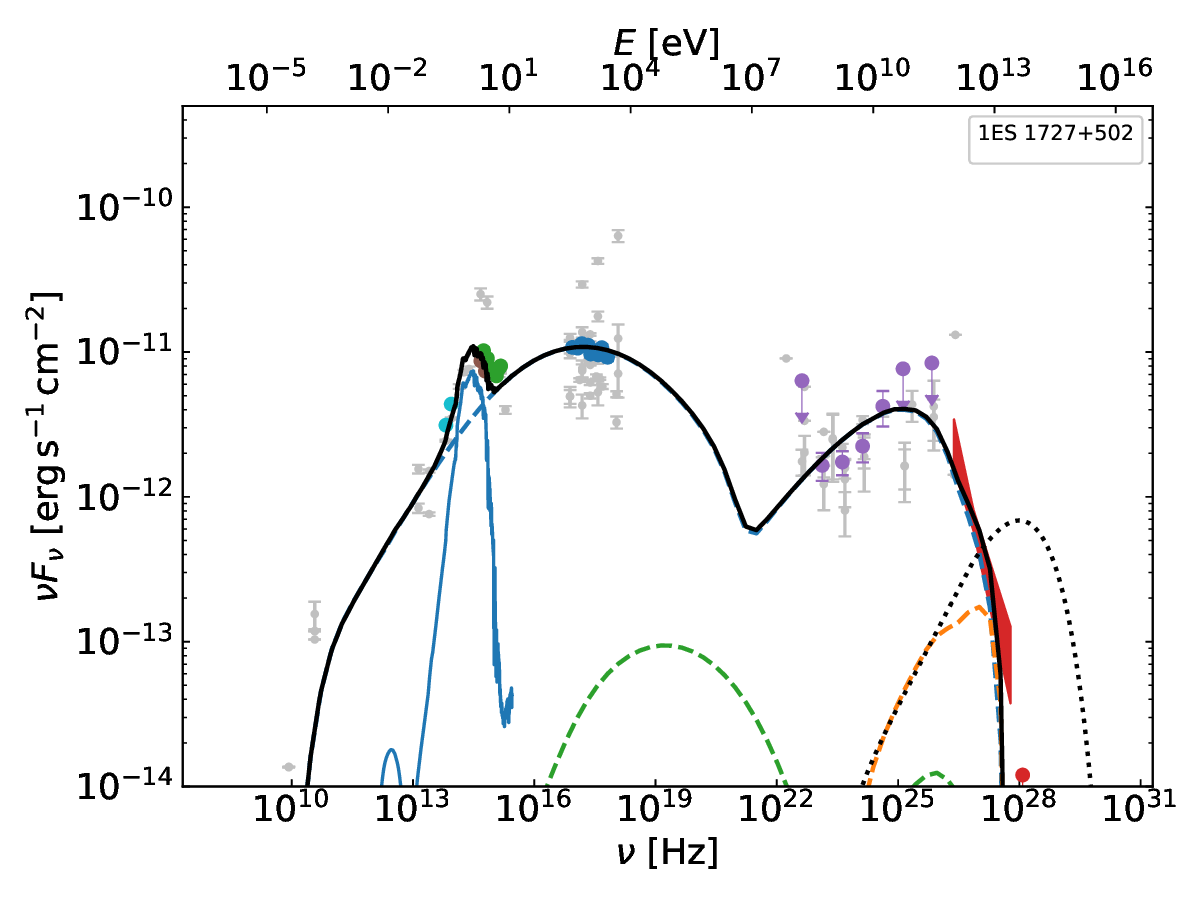}}
  \subfloat{\includegraphics[width=0.45\linewidth]{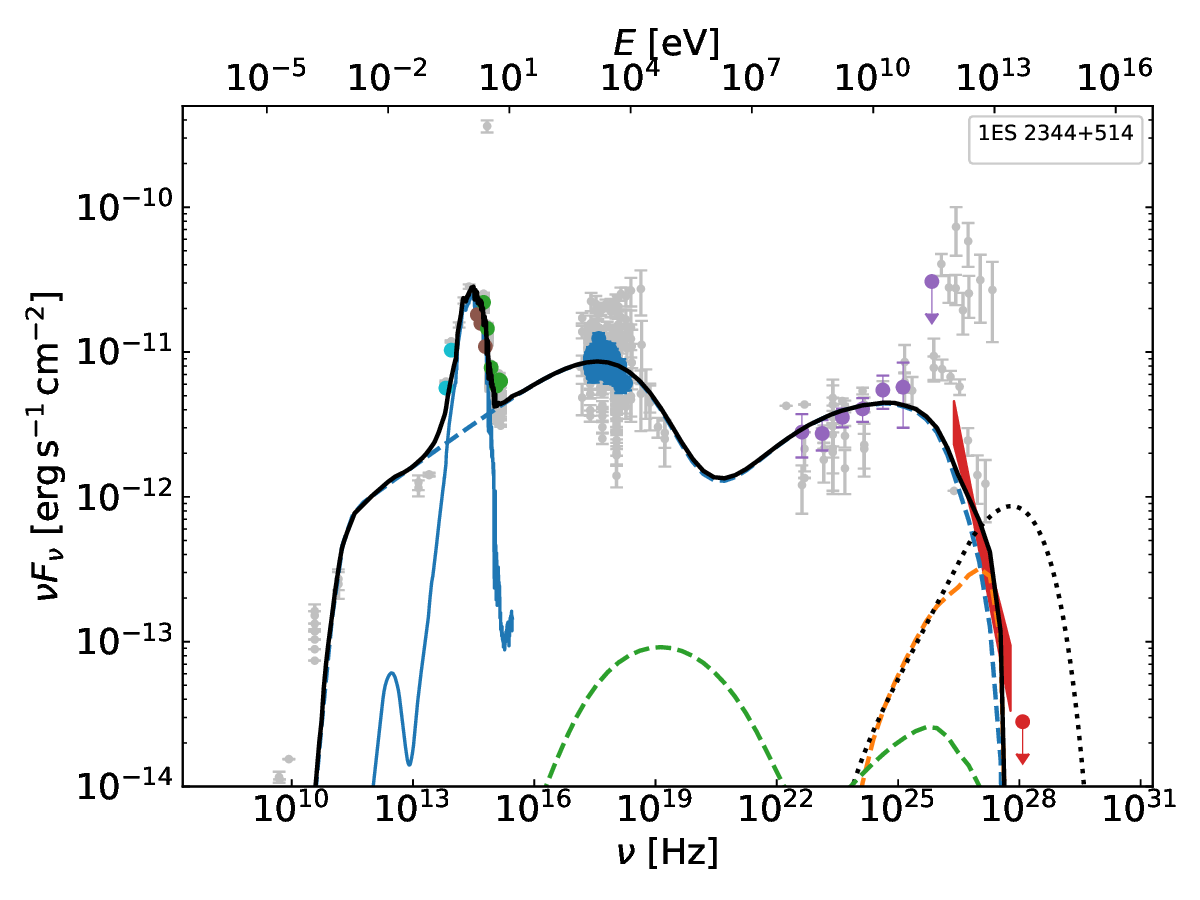}}
  \hfill
  \subfloat{\includegraphics[width=0.45\linewidth]{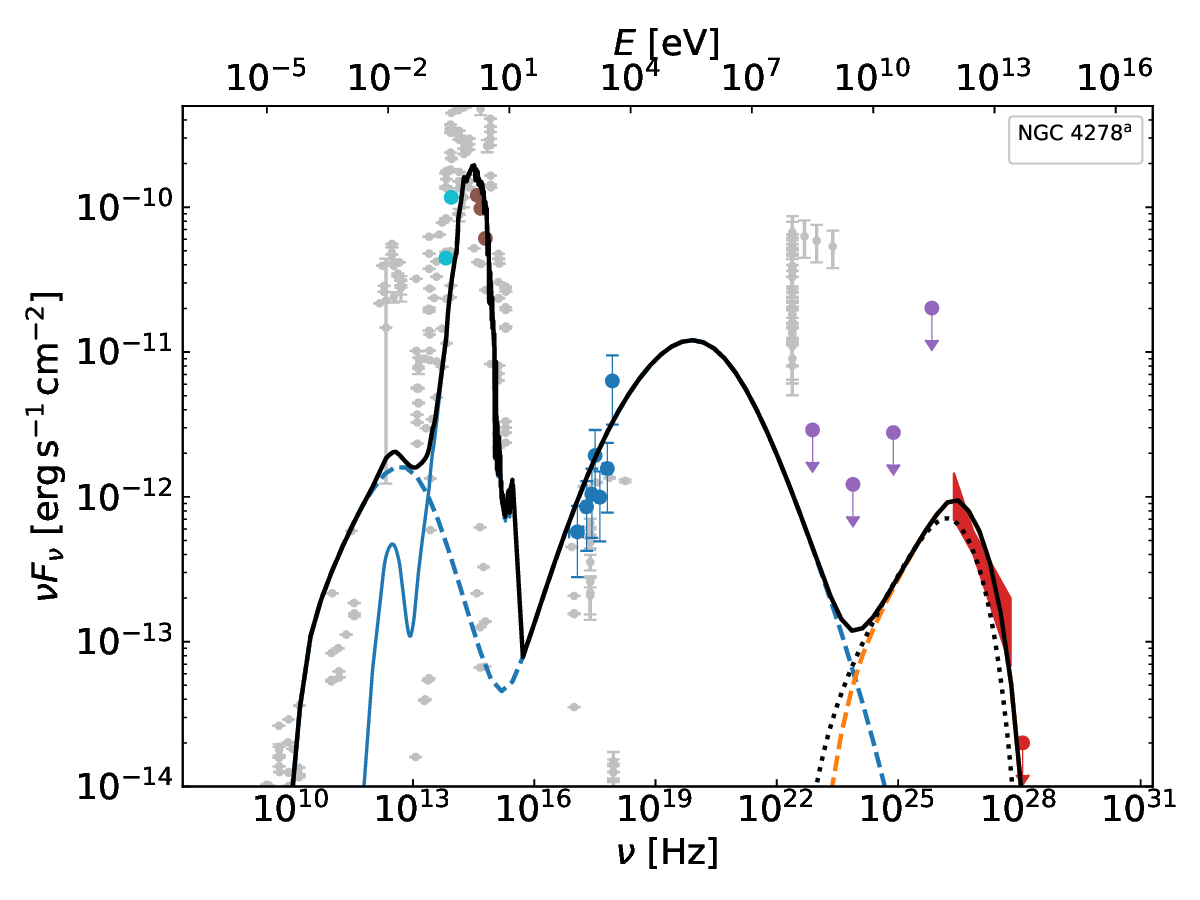}}
  \subfloat{\includegraphics[width=0.45\linewidth]{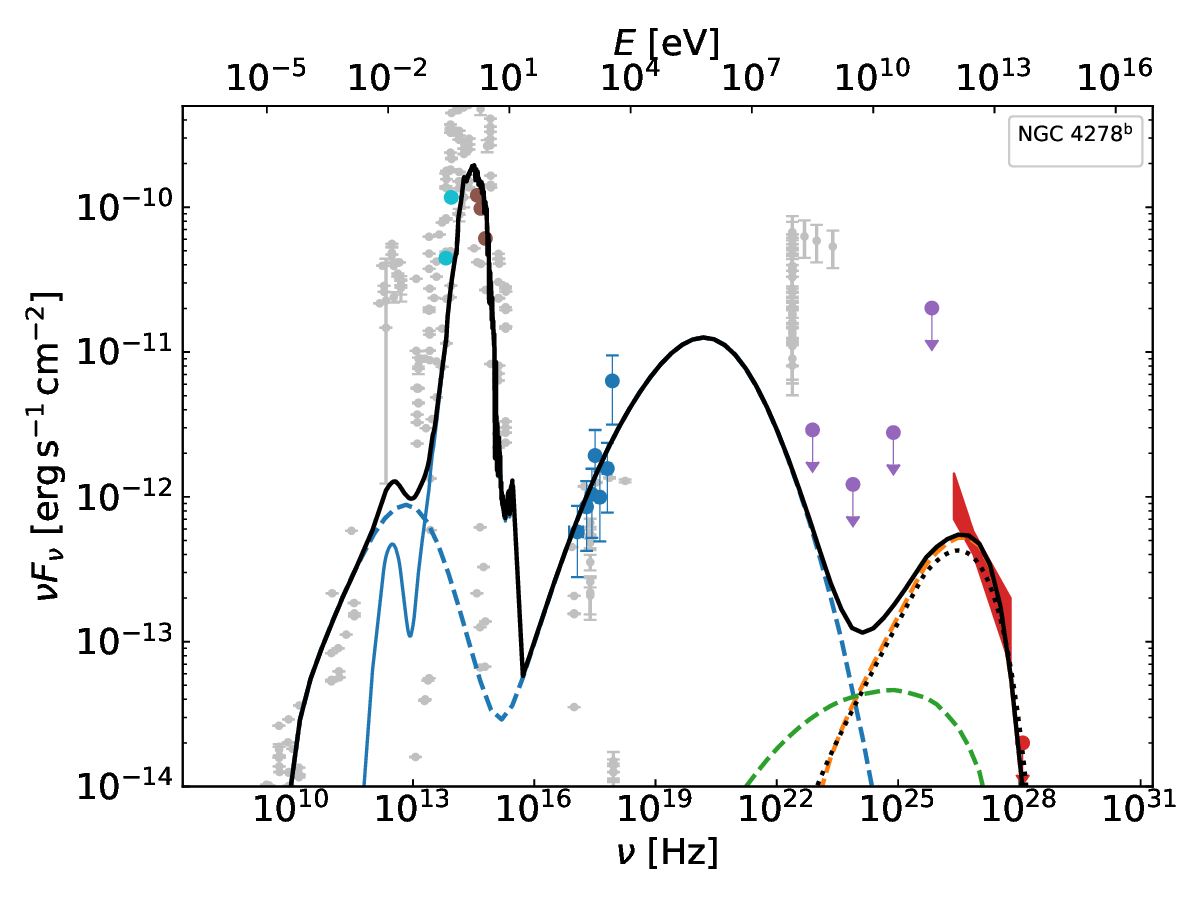}}
  \caption{One-zone SSC+$pp$ modeling. The meanings of symbols and line styles are given in the legend of Mrk 421. \label{fig:pp}}
\end{figure*}

The $pp$ model has a potential to produce the observed TeV spectra of blazars without exceeding the Eddington luminosity, which is difficult to avoid in the $p\gamma$ model \citep{2019ApJ...871...81X}. 
Our recent study \citep{2022PhRvD.106j3021X} shows that the $pp$ interactions in the jet have the potential to generate VHE emission that can be deteced by LHAASO.
Therefore, we incorporate the $pp$ interactions into the one-zone SSC model to reproduce the SEDs. 

In the $pp$ modeling, we assume a power-law injection proton density distribution. By taking into account the radiative cooling and the escape of the protons, then the steady-state proton density distribution can be calculated with \citep{2022PhRvD.106j3021X}
\begin{equation}\label{eq:PED}
N_{\rm p}(\gamma_{\rm p})=\frac{3L_{\rm p}^{\rm inj}n_{\rm p}^{\rm inj}(\gamma_{\rm p})}{4\pi R^3m_{\rm p}c^2 \int{\gamma_{\rm p} n_{\rm p}^{\rm inj}(\gamma_{\rm p}){\rm d}{\gamma_{\rm p}}}}{\rm min}\{t_{\rm cool}(\gamma_{\rm p}),t_{\rm esc}\},
\end{equation}
where $n_{\rm p}^{\rm inj}\left(\gamma_{\rm p}\right)\propto\gamma_{\rm p}^{-s_{\rm p}}$ is the injection proton density distribution, $\gamma_{\rm p}$ is the proton Lorentz factors in the range of $\gamma_{\rm p,min}$ to $\gamma_{\rm p,max}$, $s_{\rm p}$ is the slope of the power-law spectrum, $L_{\rm p}^{\rm inj}$ is the proton injection luminosity, $m_{\rm p}$ is the rest mass of the proton, $t_{\rm cool}(\gamma_{\rm p})$ is the cooling timescale of the proton. More specifically, $t_{\rm cool}(\gamma_{\rm p})$ is dominated by the $pp$ interactions in the SSC+$pp$ scenario and can be approximated by $t_{\rm cool}^{pp}\left(\gamma_{\rm p}\right)=1/\left(K_{pp}\sigma_{pp}n_{\rm H}c\right)$, where $K_{pp}\approx0.5$ is the inelasticity coefficient, $n_{\rm H}$ is the number density of cold protons in the jet, $\sigma_{pp}=\left(34.3+1.88\mathcal{L}+0.25\mathcal{L}^{2}\right)\left[1-\left(\frac{E^{\rm pp}_{\rm th}}{\gamma_{\rm p}m_{\rm p}c^{2}}\right)^{4}\right]^{2}$ is the cross section for inelastic $pp$ interactions \citep{2006PhRvD..74c4018K}, $E^{\rm pp}_{\rm th}=1.22\times10^{-3}\,{\rm TeV}$ is the threshold energy of production of $\pi^{0}$, and $\mathcal{L}=\ln{\left(\frac{\gamma_{\rm p}m_{\rm p}c^{2}}{1\,{\rm TeV}}\right)}$. 

To maximise the efficiency of the $pp$ interaction within a reasonable parameter range, analytical calculations suggest that the power of the cold protons in the jet should be set as half the Eddington luminosity \citep{2022A&A...659A.184L,2022PhRvD.106j3021X}. Then we can get the number density of cold protons $n_{\rm H}=L_{\rm Edd}/\left(2\pi R^{2}\Gamma^{2}m_{\rm p}c^{3}\right)$, where $L_{\rm Edd}=1.26\times10^{38}M_{\rm BH}/M_{\odot}$ is the Eddington luminosity, and $M_{\rm BH}$ is the SMBH mass. We may estimate the maximum proton energy by equating the acceleration timescale with the escape timescale \citep{2019ApJ...886...23X}. Then the maximum proton energy can be calculated by
\begin{equation} \label{eq:pmax}
  \gamma_{\rm p,max}=\frac{eBR}{\alpha m_{\rm p}c^{2}},
\end{equation}
where $e$ is the elementary charge, $\alpha$ is the factor representing the deviation from the highest acceleration rate. We employ $\alpha=1100$, which corresponds to the shock speed measured in the upstream frame of $\sim 0.07c$ in the situation of shock acceleration \citep{2007Ap&SS.309..119R}. We set the minimum proton energy $\gamma_{\rm p,min}=1$ and the slope of the power-law spectrum $s_{\rm p}=1.5$. These two parameters have no effect on the fitting results, only on the required proton injection luminosity. Finally, the only free parameter of the $pp$ model remains the proton injection luminosity $L_{\rm p}^{\rm inj}$, which is shown in Table \ref{tab:par}.

The fitting results can be found in Fig.$\,$\ref{fig:pp}. The radiation produced by the $pp$ interactions fits perfectly the high-energy tail of the LHAASO spectrum of Mrk 421 and Mrk 501, which previously could not be explained by the one-zone SSC model. The dotted curve in Fig.$\,$\ref{fig:pp} shows the neutrino flux produced by the $pp$ interactions, which should be comparable to the photon flux produced at the same time (as shown in two cases of NGC 4278). The sudden drop of photon flux in four cases of blazars is due to the absorption of photon-photon interactions occurring in the jet and in the intergalactic propagation. The fitting parameters in Table \ref{tab:par} show that Mrk 421, Mrk 501, 1ES 2344+514, NGC 4278$^{\rm a}$ and NGC 4278$^{\rm b}$ can be fitted without exceeding the Eddington luminosity, and a quite low proton injection luminosity is required in the cases of 1ES 2344+514 and NGC 4278$^{\rm a}$. The proton injection luminosity used in fitting of 1ES 1727+502 is 60 times of the Eddington luminosity, because of its low SMBH mass ($5.62\times10^{7}~M_{\odot}$). For comparison, the fiducial SMBH mass of the BL Lacs is $10^{8.5-9}~M_{\odot}$ \citep{2013ApJ...764..135S,2022ApJ...925...40X}. 
Moreover, based on the fitting results, it can be found that the cascade process make a negligible contribution to the energy spectrum in the one-zone SSC+$pp$ model.

From the chi-squared test results in Table \ref{tab:par}, it can be seen that for the cases of Mrk 421, Mrk 501, NGC 4278$^{\rm a}$ and NGC 4278$^{\rm b}$ the goodness of fitting has been significantly improved after the introduction of the $pp$ model (especially considering only the chi-squared values of the WCDA data). While for the two cases of 1ES 1727+502 and 1ES 2344+514, the introduction of the $pp$ model in the fitting has no significant advantage.

It is necessary to evaluate the possible neutrino emission when high-energy protons are introduced into the model. Therefore, we calculate the neutrino flux that could be produced in the process of the $pp$ interactions. The time-integrated neutrino sources searches with 10 years of IceCube data collected between 2008 and 2018 reports that the best-fit number of astrophysical neutrino events for Mrk 421 is 2.1, with a local pre-trial p-value of 0.42, and the number of astrophysical neutrino events for Mrk 501 is 10.3, with a p-value of 0.25 \citep{2020PhRvL.124e1103A}. It should be noted that neutrino events from Mrk 421 and Mrk 501 have not yet been detected by IceCube. These best-fitting numbers of astrophysical neutrino events both correspond to very large p-values, which means that this result can only be treated as a rough upper limit \citep[e.g.,][]{2023ApJS..266...37A}. The neutrino event rate can be obtained by $\frac{{\rm d}N_{\nu}}{{\rm d}t}=\int_{E_{\nu,{\rm min}}}^{E_{\nu,{\rm max}}}{\rm d}{E_{\nu}A_{\rm eff}\left(E_{\nu},\delta_{\rm decl}\right)\phi\left(E_{\nu}\right)}$, where $E_{\nu,{\rm min}}$ and $E_{\nu,{\rm max}}$ are the lower and upper bounds of the neutrino energy, respectively, $A_{\rm eff}$ is the IceCube point-source effective area for (anti)muon neutrinos \citep{2019ICRC...36..851C}, $\delta_{\rm decl}$ is the declination, and $\phi\left(E_{\nu}\right)$ is the differential neutrino energy flux. Then the expected neutrino event rates\footnote{To minimise the impact of the model parameters on the results, we consider only the part of the $pp$ interaction emission that is necessary to fit the high-energy tail of the LHAASO spectrum, i.e., the neutrino emission corresponding to photons with energy less than 25\,TeV.} from the sources are 0.15\,events/yr for 1ES 1727+502, 0.21\,events/yr for 1ES 2344+514, 0.11\,events/yr for NGC 4278$^{\rm a}$, 0.10\,events/yr for NGC 4278$^{\rm b}$, 0.47\,events/yr for Mrk 421 and 1.06\,events/yr for Mrk 501, which are slightly exceed the upper limits for Mrk 421 and Mrk 501. Thus,
although the SSC+$pp$ model reproduces the SEDs best and has the smallest $\chi^2/{\rm d.o.f}$ for Mrk 421 and Mrk 501, it remains to be investigated whether $pp$ interactions can reasonably explain the high-energy tail of the LHAASO spectra, after obtaining the simultaneous SEDs or the variability in the VHE band.

\subsection{One-zone proton-synchrotron model}\label{sec:psyn}
\begin{figure*}[htbp]
  \centering
  \subfloat{\includegraphics[width=0.35\linewidth]{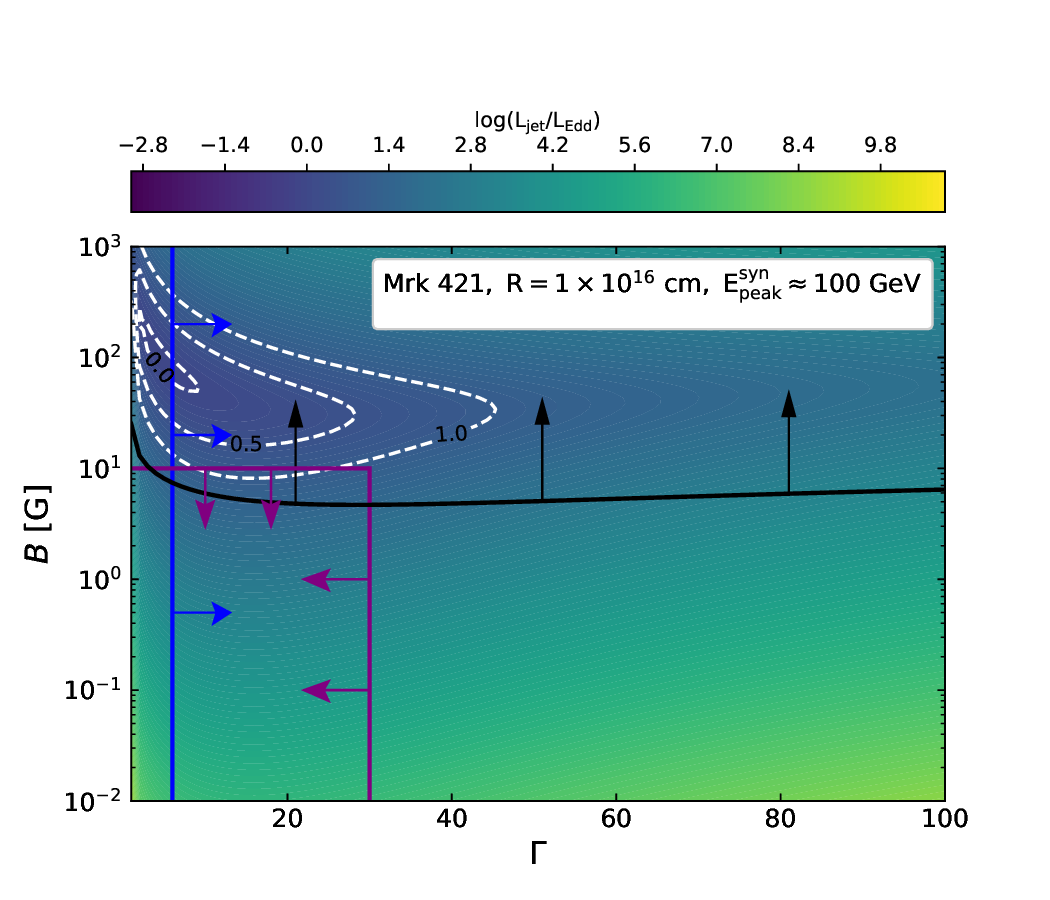}}
  \subfloat{\includegraphics[width=0.35\linewidth]{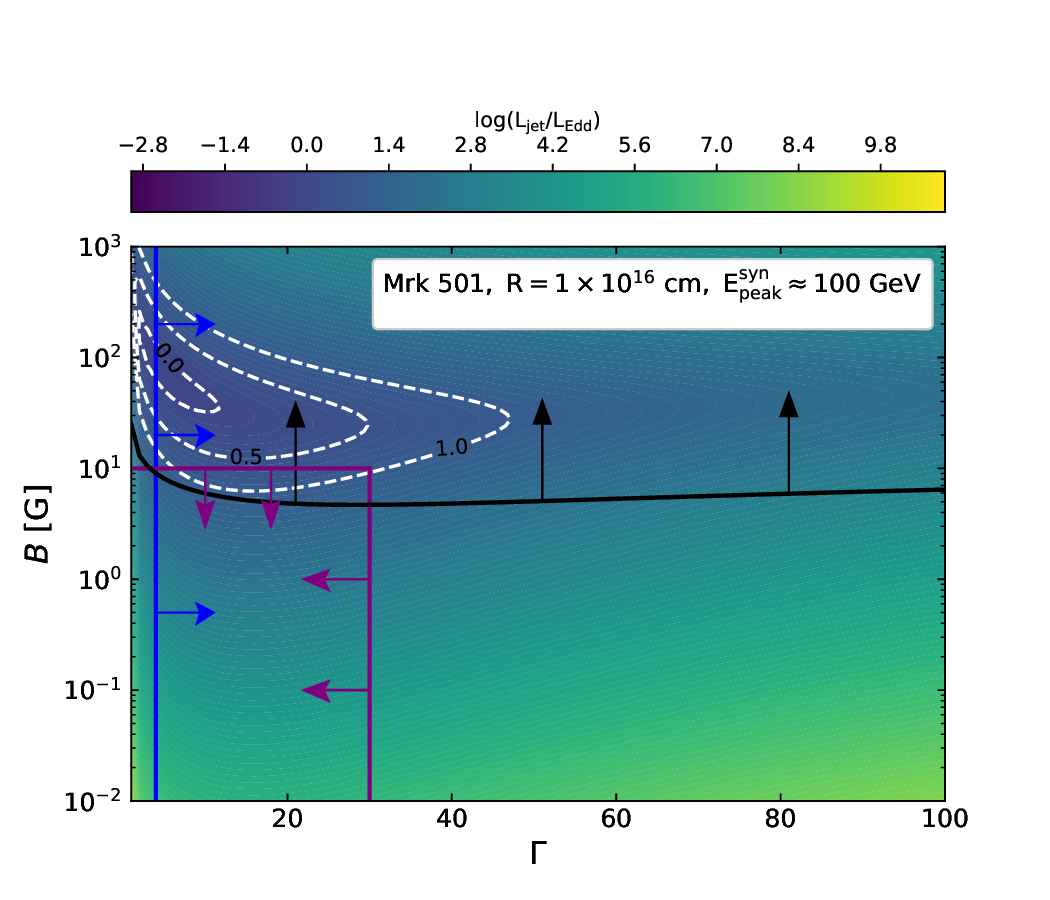}}
  \subfloat{\includegraphics[width=0.35\linewidth]{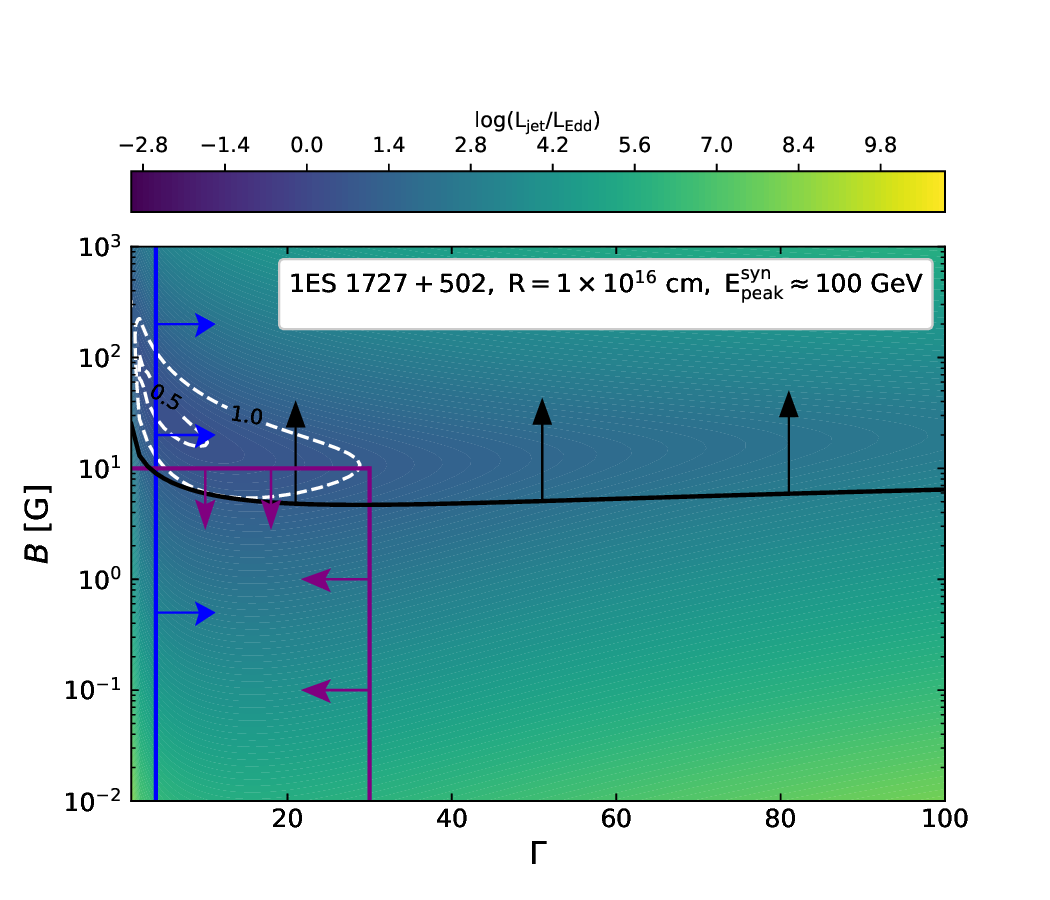}}
  \hfill
  \subfloat{\includegraphics[width=0.35\linewidth]{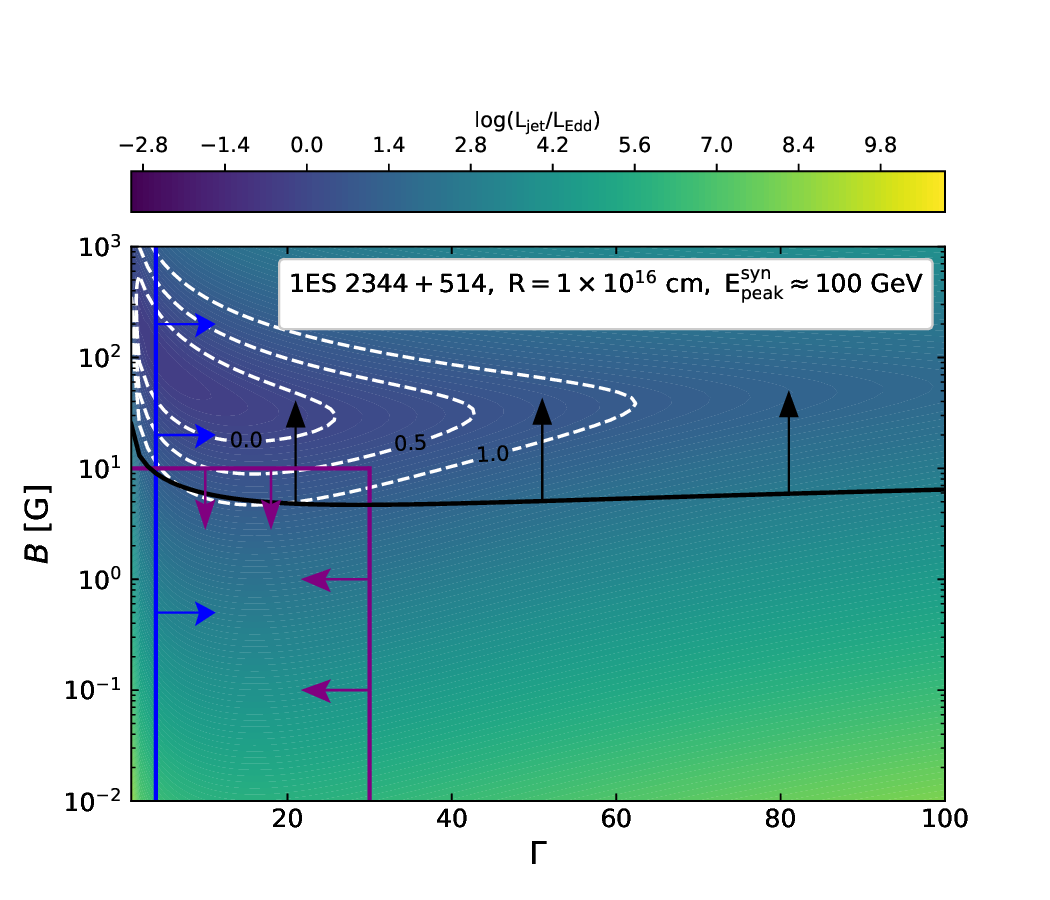}}
  \subfloat{\includegraphics[width=0.35\linewidth]{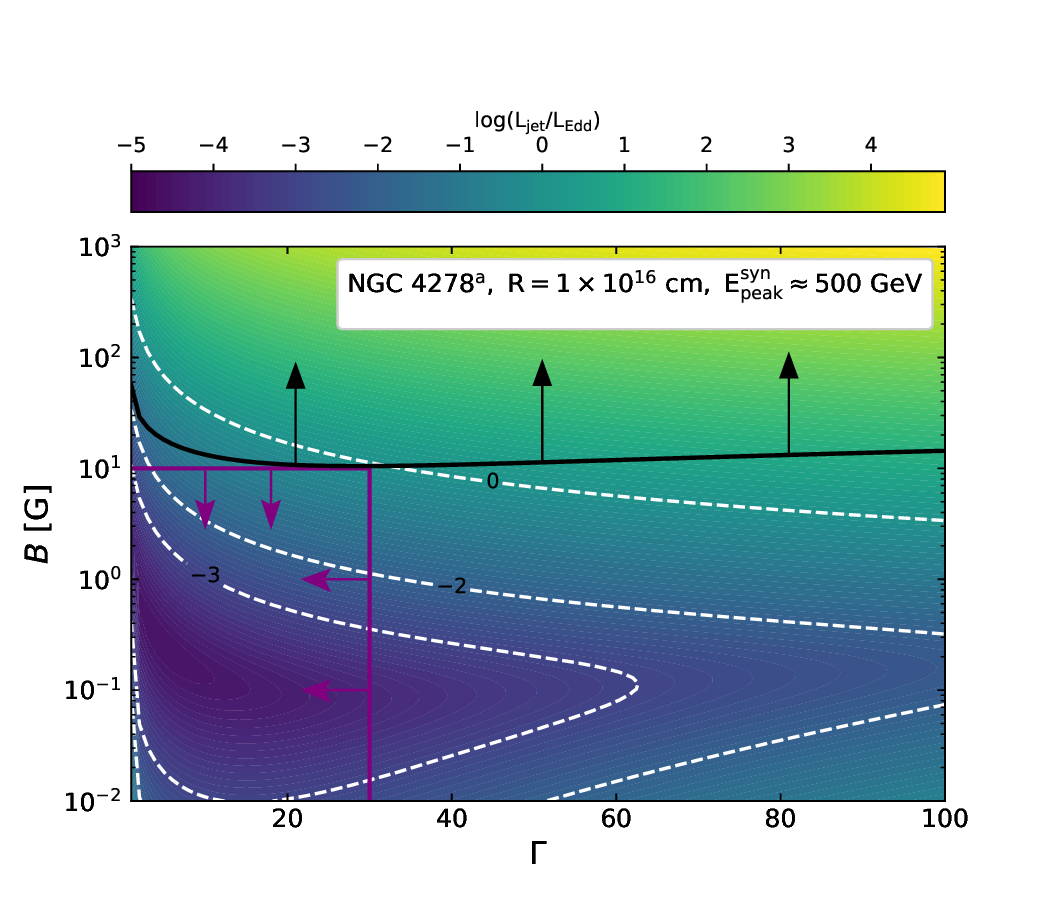}}
  \subfloat{\includegraphics[width=0.35\linewidth]{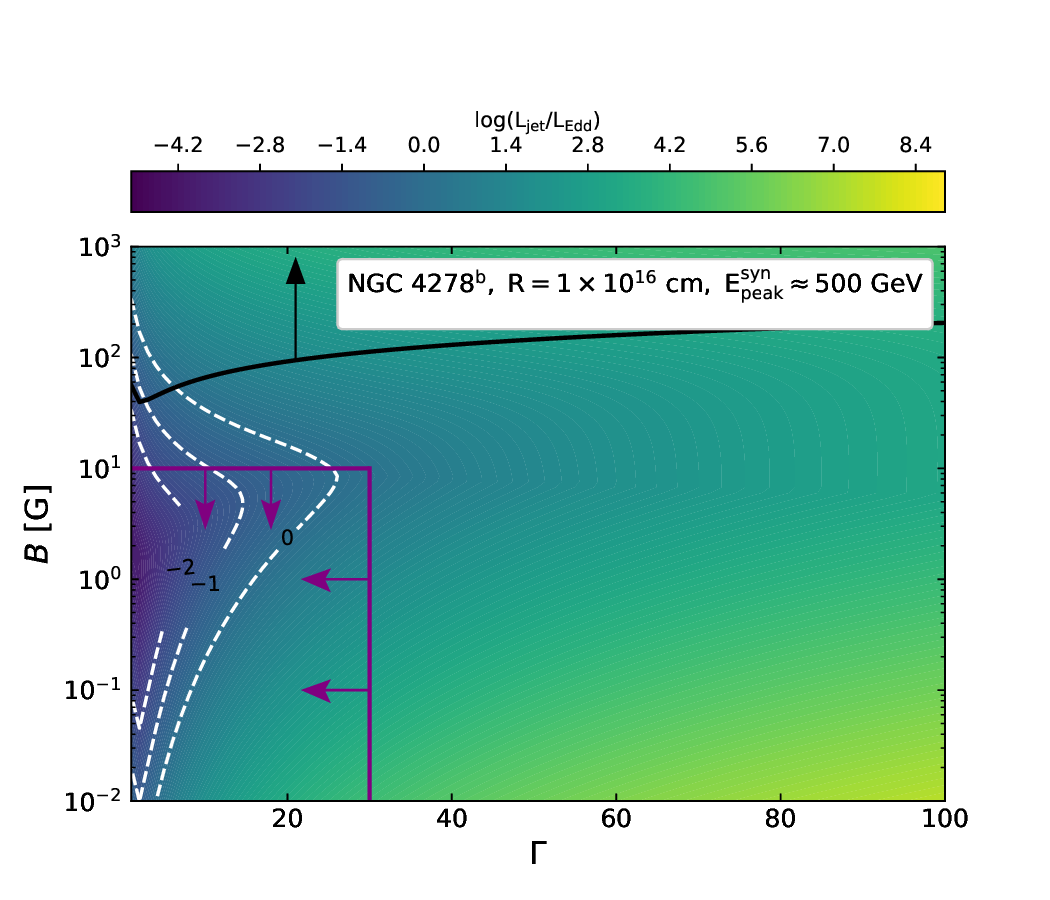}}
  \hfill
  \subfloat{\includegraphics[width=0.35\linewidth]{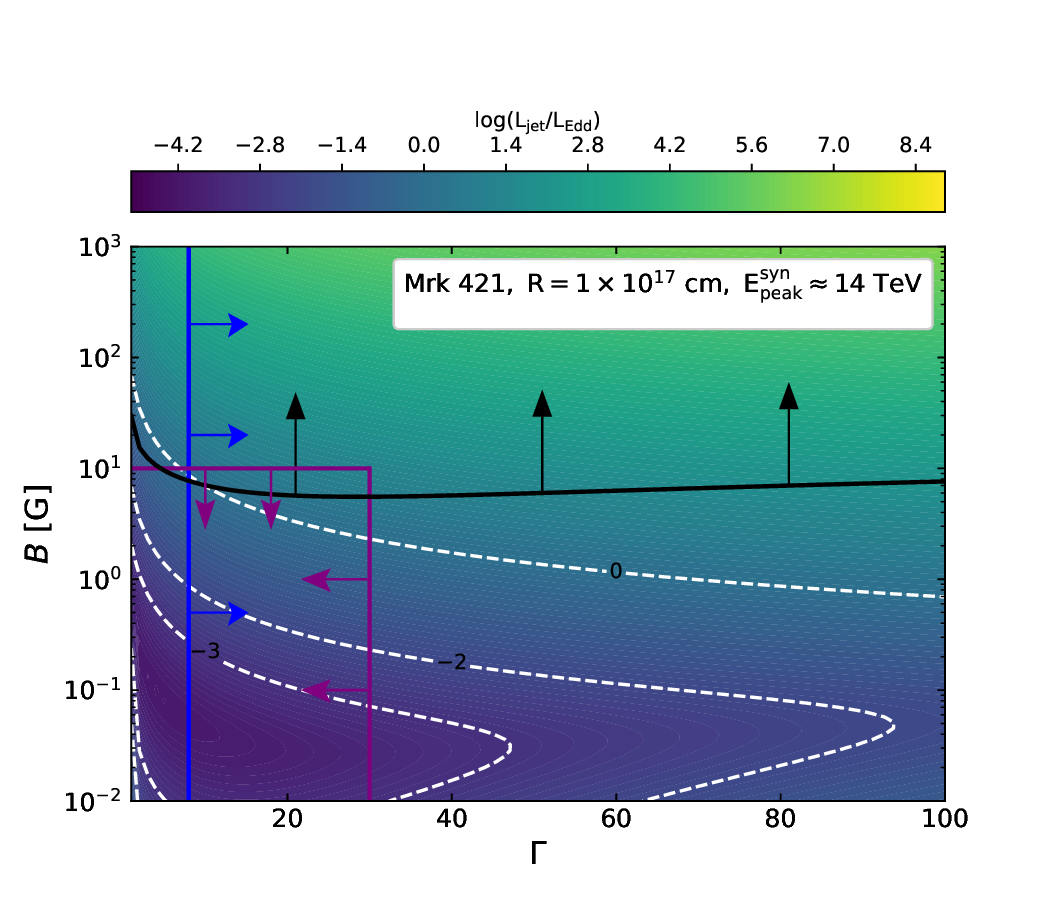}}
  \subfloat{\includegraphics[width=0.35\linewidth]{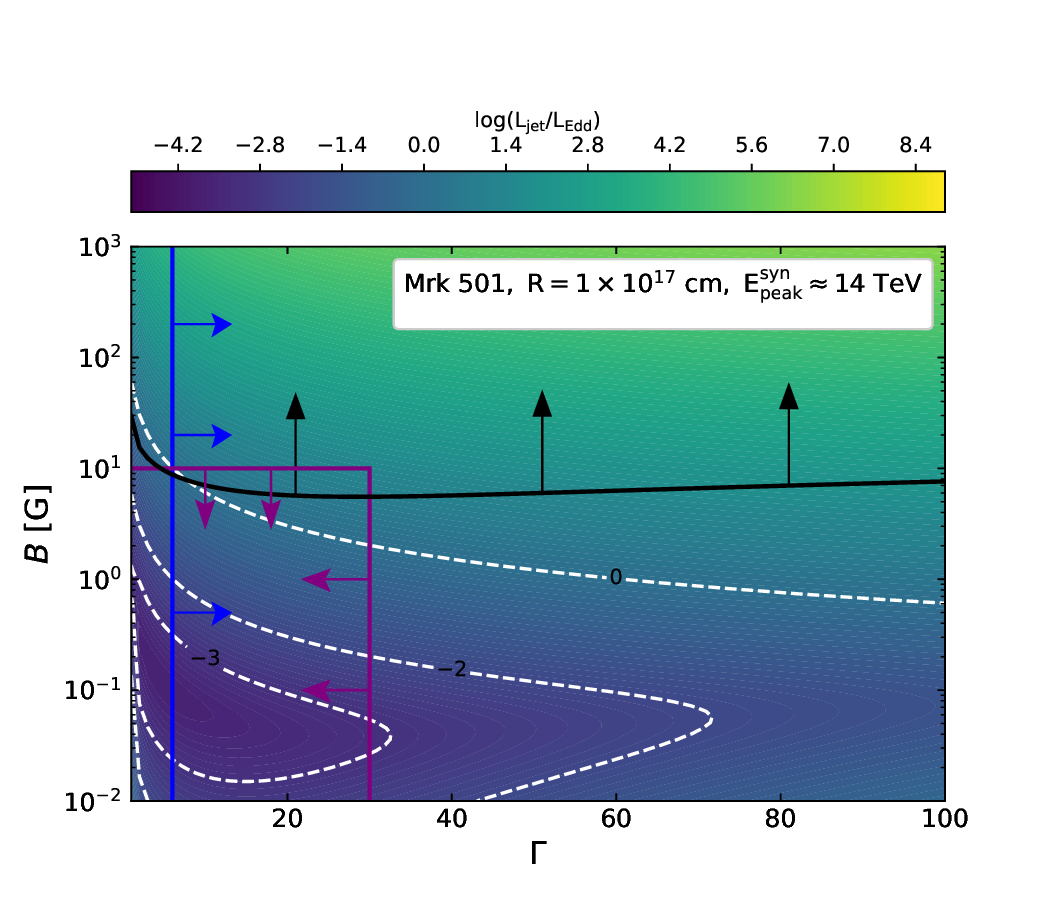}}
  \subfloat{\includegraphics[width=0.35\linewidth]{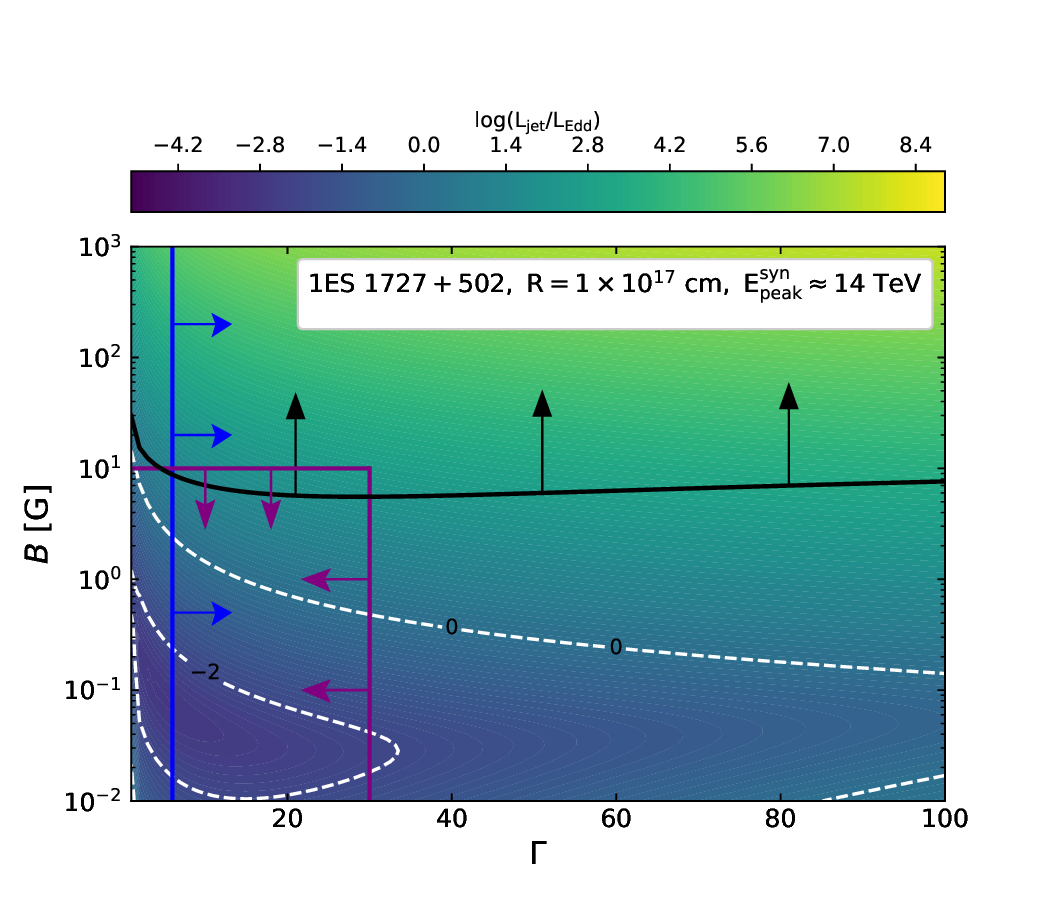}}
  \hfill
  \subfloat{\includegraphics[width=0.35\linewidth]{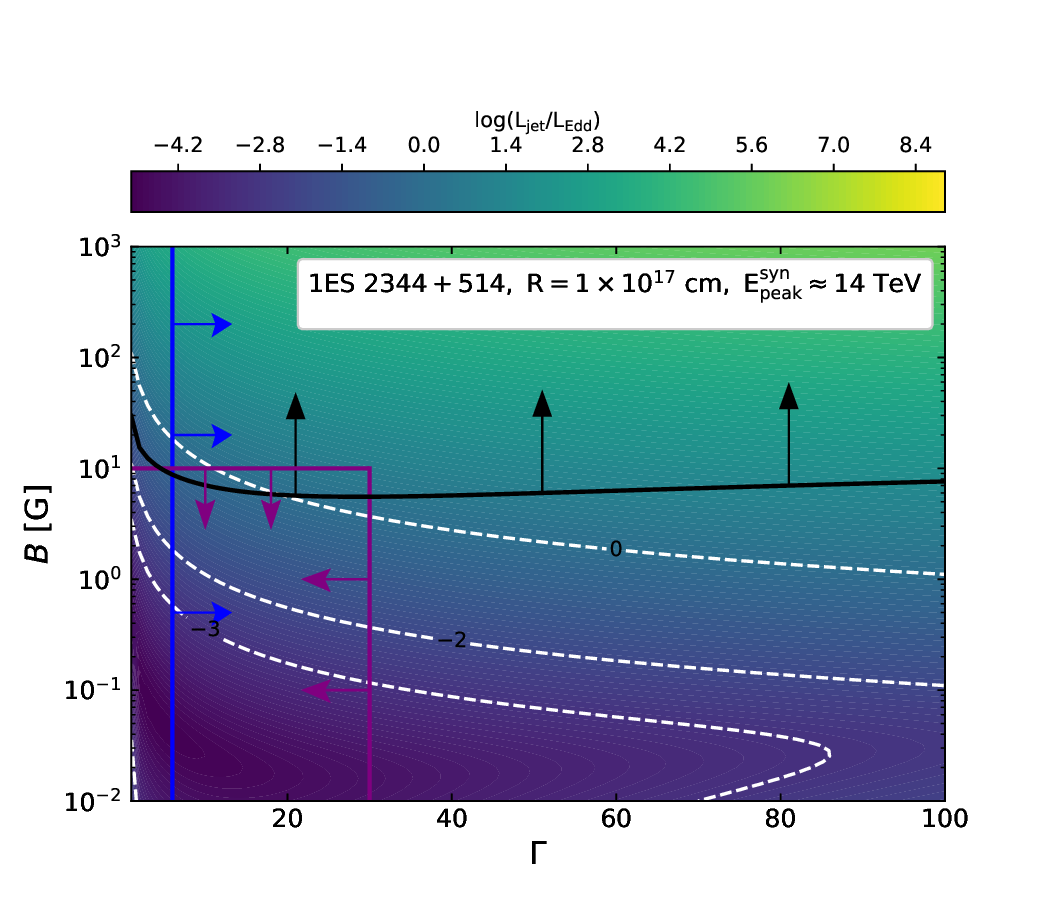}}
  \subfloat{\includegraphics[width=0.35\linewidth]{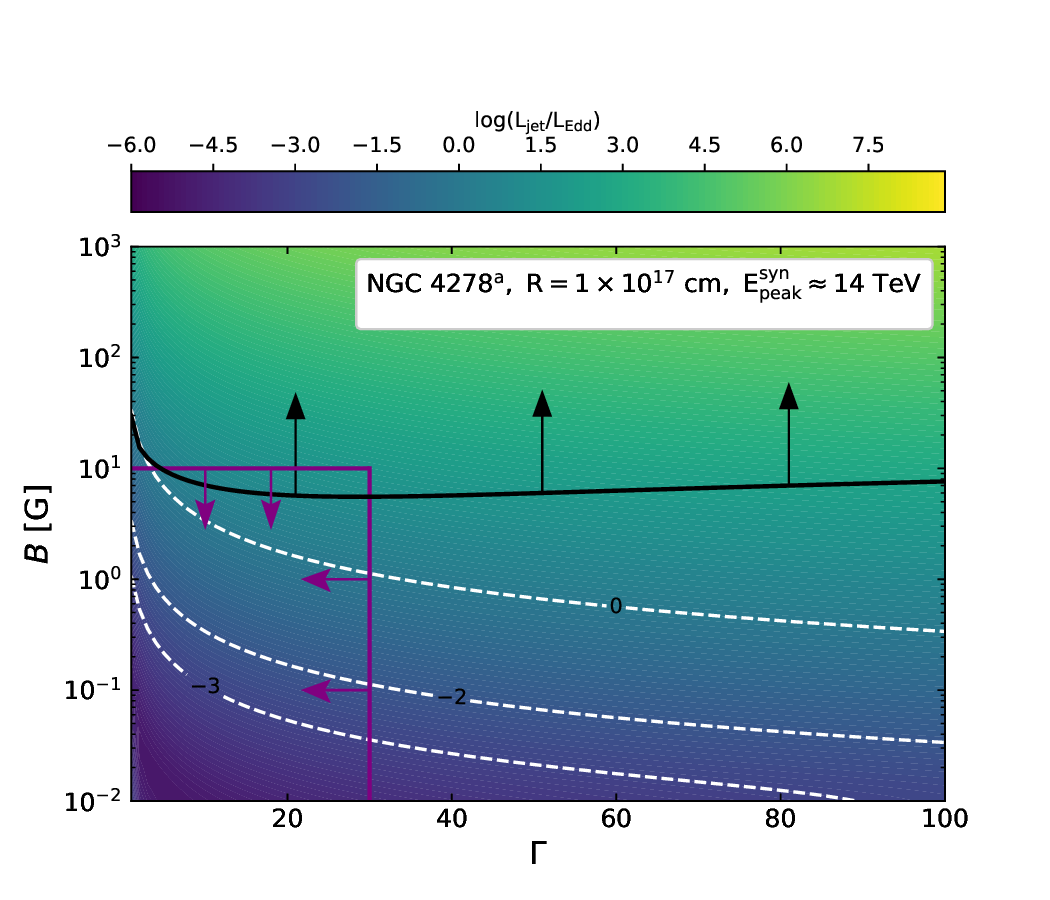}}
  \subfloat{\includegraphics[width=0.35\linewidth]{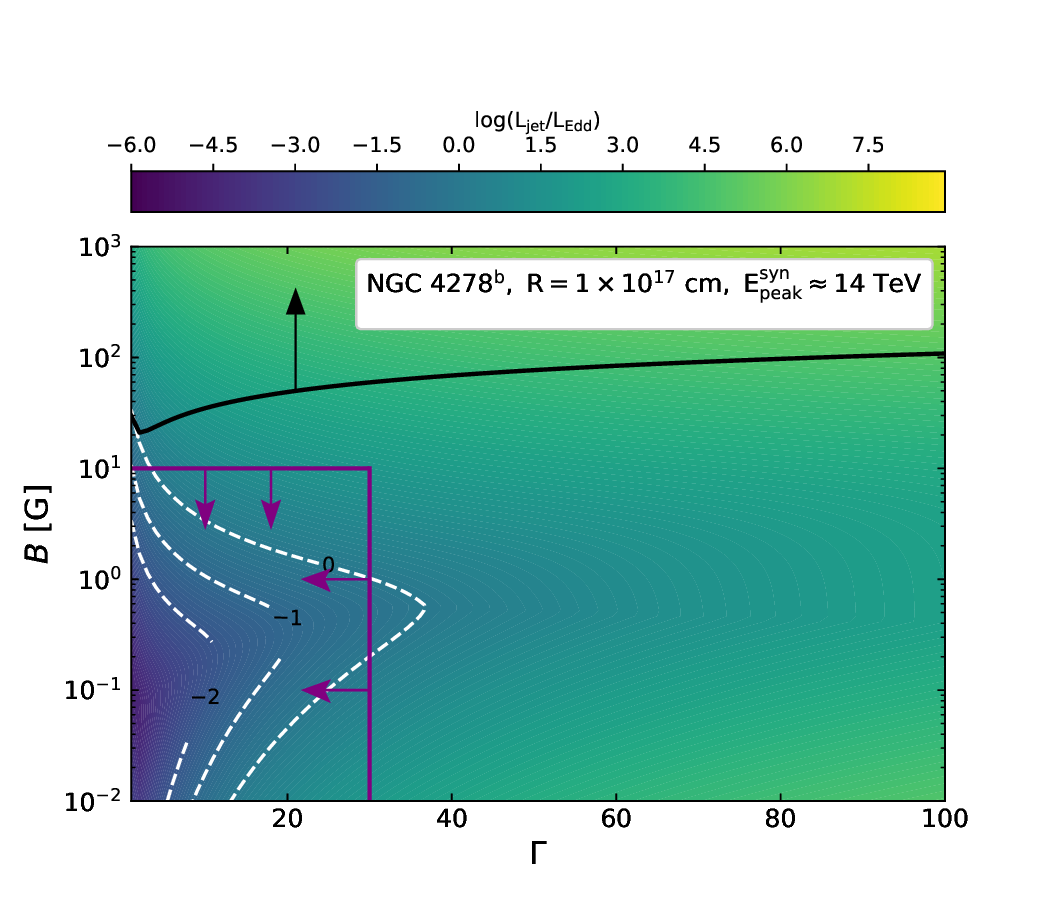}}
  \caption{The ratio of $L_{\rm jet}/L_{\rm Edd}$ in the $\Gamma-B$ diagram for the One-zone proton-synchrotron model. The upper six panels show the results that applying proton-synchrotron emission to explain the entire high-energy component, and the lower six panels show the results that applying proton-synchrotron emission to explain the high-energy tail of LHAASO spectra. The black curves with arrows represent the parameter space that satisfied the Hillas condition. The vertical blue curves with arrows show the lower limit of $\Gamma$ that allows the escape of maximum energy $\gamma$-ray photons. The vertical and horizontal purple curves show the space that $B\lesssim10~\rm G$ and $\Gamma\lesssim30$. The white dashed contours denote specific values of log($L_{\rm jet}/L_{\rm Edd}$) associated with the color bar. \label{fig:psynspace}}
\end{figure*}

The proton-synchrotron emission in the framework of one-zone model is often suggested as a possible interpretation of the high-energy component of HSP AGNs \citep[e.g.,][]{2000NewA....5..377A,2013ApJ...768...54B,2020ApJS..247...16A}, although extreme physical parameters, such as a super-Eddington jet power and a strong magnetic field, are usually introduced \citep[cf.,][]{2015MNRAS.448..910C, 2016ApJ...825L..11P,2023PhRvD.107j3019X}. In this subsection, before fitting SEDs of these five LHASSO AGNs, we search the proton-synchrotron modeling parameter space with an analytical method proposed in our recent work \citep{2023PhRvD.107j3019X}. In this method, the parameter space is limited by three constraints, which are the total jet power (dominated by the injection power of relativistic protons and the power carried in magnetic field) does not exceed the Eddington luminosity, relativistic protons can be accelerated to the required maximum energy, and the emitting region is transparent to VHE photons, respectively. In addition, observation results suggest that the magnetic field in the inner jet of AGNs is typically lower than 10 G \citep{2009MNRAS.400...26O, 2012A&A...545A.113P, 2017A&A...597A..80H, 2022MNRAS.510..815K}, and the bulk Lorentz factor of jet is lower than 30 \citep[e.g.,][]{2009A&A...494..527H}. If a reasonable parameter space that satisfies observational constraints can be found (i.e., $B\lesssim10~\rm G$ and $\Gamma \lesssim30$), we fit their SEDs.

There are two strategies to fit the LHAASO spectra with proton-synchrotron emission. The first one is to use proton-synchrotron emission to account for the entire high-energy component. The second one is to use proton-synchrotron emission to fit the high-energy tail of the LHAASO spectra, while the rest of the high-energy component is still attributed to the leptonic SSC emission. For the first strategy, the index $s_{\rm p}$ of injected proton energy distribution can be obtained by the photon index $\Gamma_{\rm index}$ of \textsl{Fermi}-LAT spectrum, i.e., $s_{\rm p} = 2\Gamma_{\rm index}-1$. Then we derive values of $s_{\rm p}$ of Mrk 421, Mrk 501, 1ES 1727+502, and 1ES 2344+514, which are 2.66, 2.56, 2.44, and 2.64, respectively. For NGC 4278, since only upper limits are given by \textsl{Fermi}-LAT, we default $s_{\rm p} = 2$. Based on the leptonic modeling in Sect.~\ref{sec:ssc}, we set the peak energies $E_{\rm peak}^{\rm syn}$ of proton-synchrotron emission of Mrk 421, Mrk 501, 1ES 1727+502, and 1ES 2344+514 are 100 GeV, and $E_{\rm peak}^{\rm syn}=500~\rm GeV$ for NGC 4278. \cite{2023PhRvD.107j3019X} find that the reasonable parameter space might be found if considering a relative large blob radius ($R\gtrsim 10^{16}~\rm cm$), and the size of parameter space is inversely proportional to $R$. So here we only check if the parameter space can be found when $R=10^{16}~\rm cm$. For the second strategy, we default $E_{\rm peak}^{\rm syn}=14~\rm TeV$ and $s_{\rm p} = 2$ for all five AGNs, since there is no constraint on $s_{\rm p}$. Since the default peak energy is quite large, it is only necessary to check if the parameter space exists in a large emitting region. Here, we set $R=10^{17}~\rm cm$. 

\begin{figure}[htbp]
\includegraphics[width=0.47\textwidth]{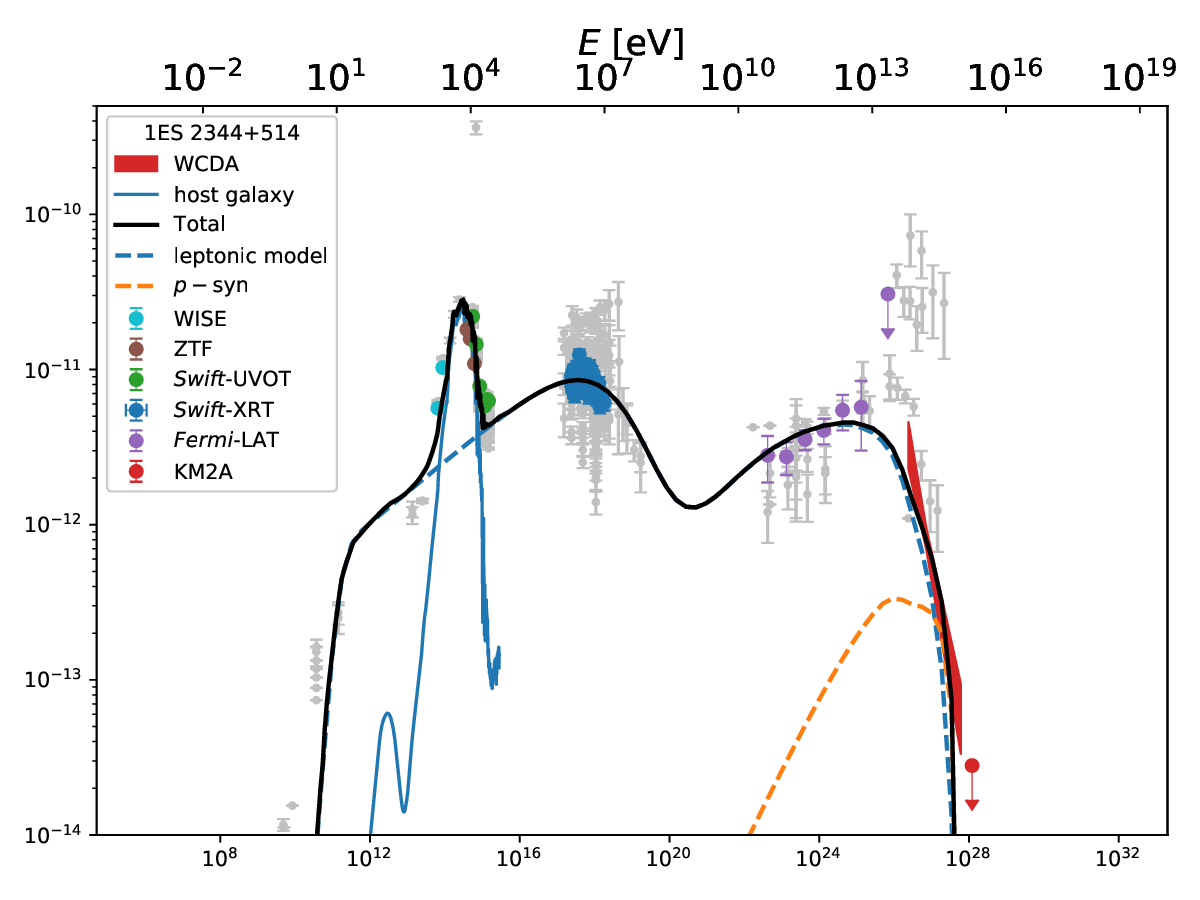}
\caption{Two-zone proton-synchrotron modeling. The meanings of symbols and line styles are given in the legend.
\label{twozone2344}}
\end{figure}

The results of the parameter space scans are shown in Fig.~\ref{fig:psynspace}. In the first strategy (the upper six panels), it can be seen that no valid parameter space is found for all five AGNs (SED fitting with strong magnetic fields is given in Appendix~\ref{Bsyn}). Among them, the super-Eddington jet power is needed for four blazars because soft proton indexes are suggested by the \textsl{Fermi}-LAT spectra. For NGC 4278, there is a large parameter space to get the sub-Eddington jet power, however this space is in conflict with the Hillas condition (black curves with arrows). In the second strategy (the lower six panels), it can be seen that only 1ES 2344+514 can find a reasonable parameter space. However, with $R=10^{17}~\rm cm$, if we set $B=7~\rm G$ and $\Gamma=10$, the energy density of low-energy component $U_{\rm syn}\approx 8.2\times10^{-7}~\rm erg~cm^{-3}$ would be much lower than that of magnetic field $U_{\rm B}\approx 1.9~\rm erg~cm^{-3}$. Therefore, in the framework of one-zone model, it is impossible to fit the GeV data of 1ES 2344+514 with SSC emission when using the proton-synchrotron emission to explain the LHAASO spectrum. A second emitting region has to be introduced. In Fig.~\ref{twozone2344}, we show that the SED of 1ES 2344+514, including the LHAASO spectrum, can be explained as a superposition of leptonic emission from first emitting zone and proton-synchrotron emission from second emitting zone. The leptonic emission from the first emitting zone is the same as that obtained in Sect.~\ref{sec:ssc}. For the second emitting zone, we set $R=10^{17}~\rm cm$, $B=9~\rm G$, and $\Gamma=12$ as indicated by the obtained parameter space.

Overall, the one-zone proton-synchrotron model seems to be difficult to interpret the currently observed LHAASO spectrum within a reasonable parameter space (i.e., $B\lesssim10~\rm G$ and $\Gamma \lesssim30$). If we introduce a second emitting zone, the proton-synchrotron emission is only applicable to 1ES 2344+514. However, the fitting results of the two-zone proton-synchrotron model do not show any advantages over the one-zone SSC model, either in terms of fitting results or chi-squared test. As shown in Appendix \ref{Bsyn}, the chi-square test shows that the one-zone proton-synchrotron model has no advantage in the fitting of the SEDs over the one-zone SSC model after removing the parameter constraints.

\subsection{Spine-layer model}\label{sec:sl}

\begin{figure*}[htbp]
  \centering
  \subfloat{\includegraphics[width=0.45\linewidth]{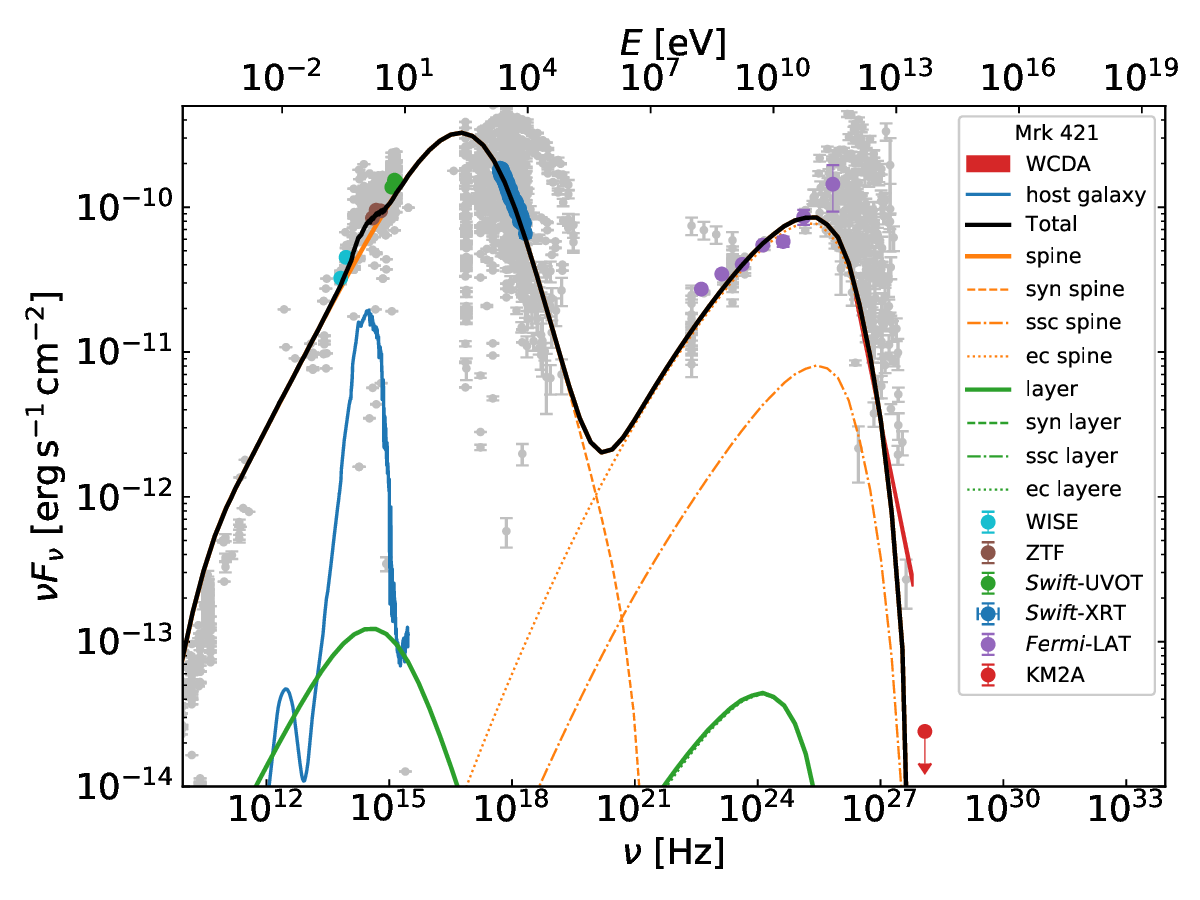}}
  \subfloat{\includegraphics[width=0.45\linewidth]{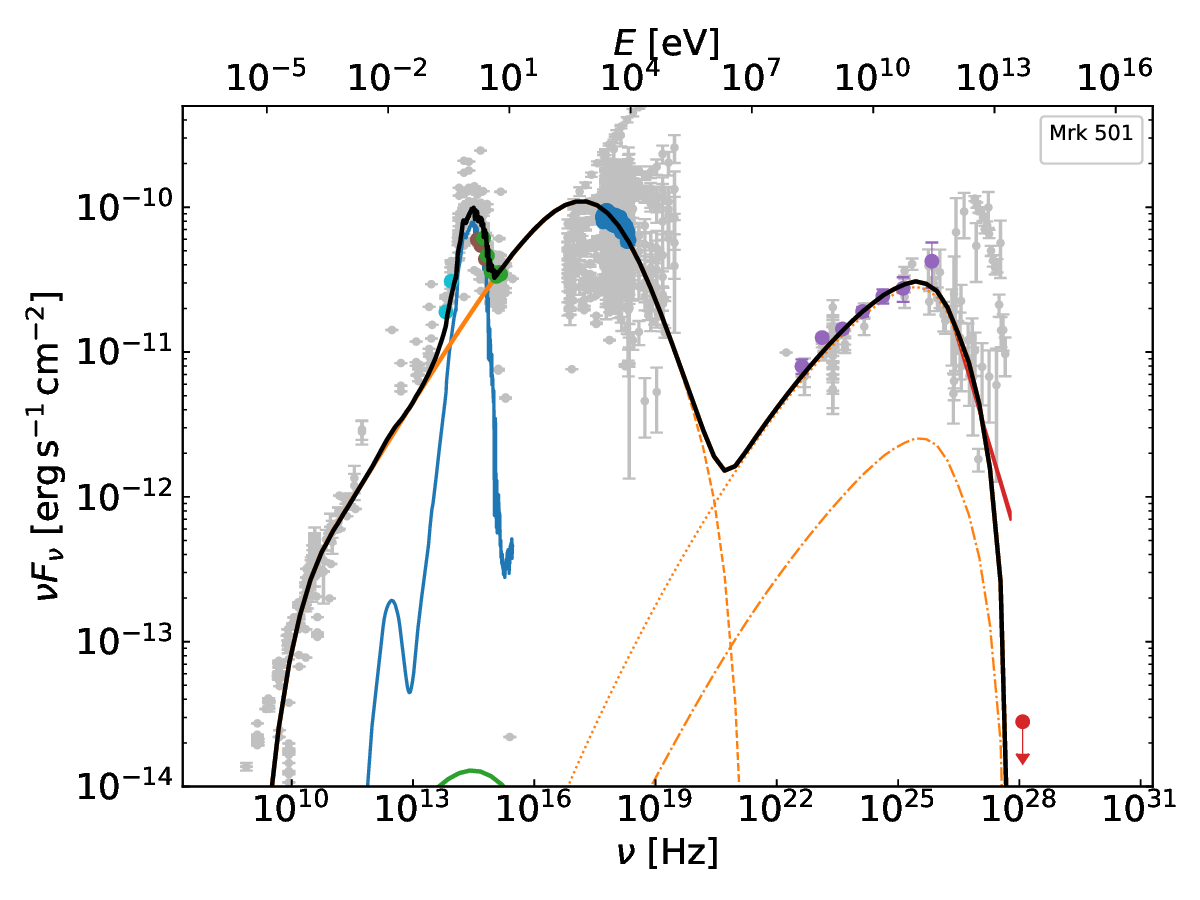}}
  \hfill
  \subfloat{\includegraphics[width=0.45\linewidth]{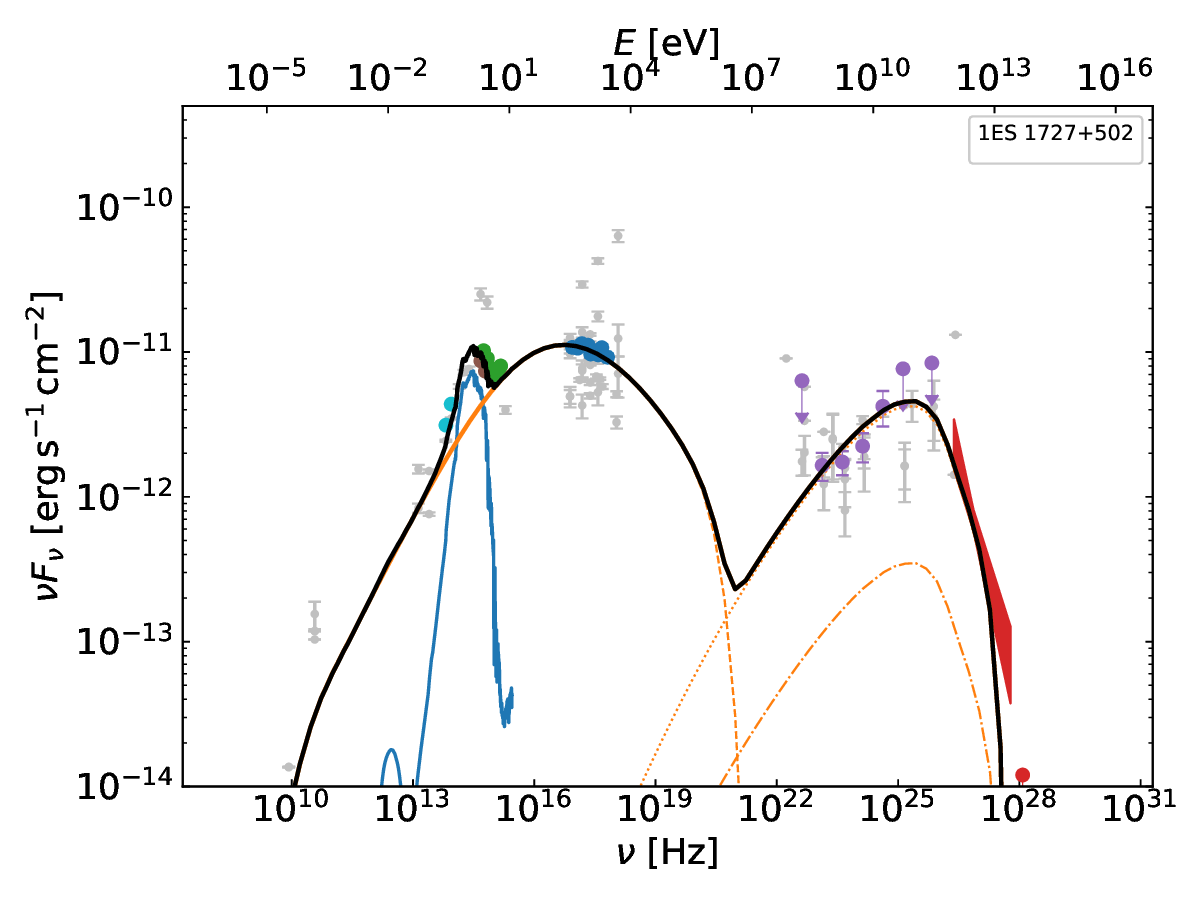}}
  \subfloat{\includegraphics[width=0.45\linewidth]{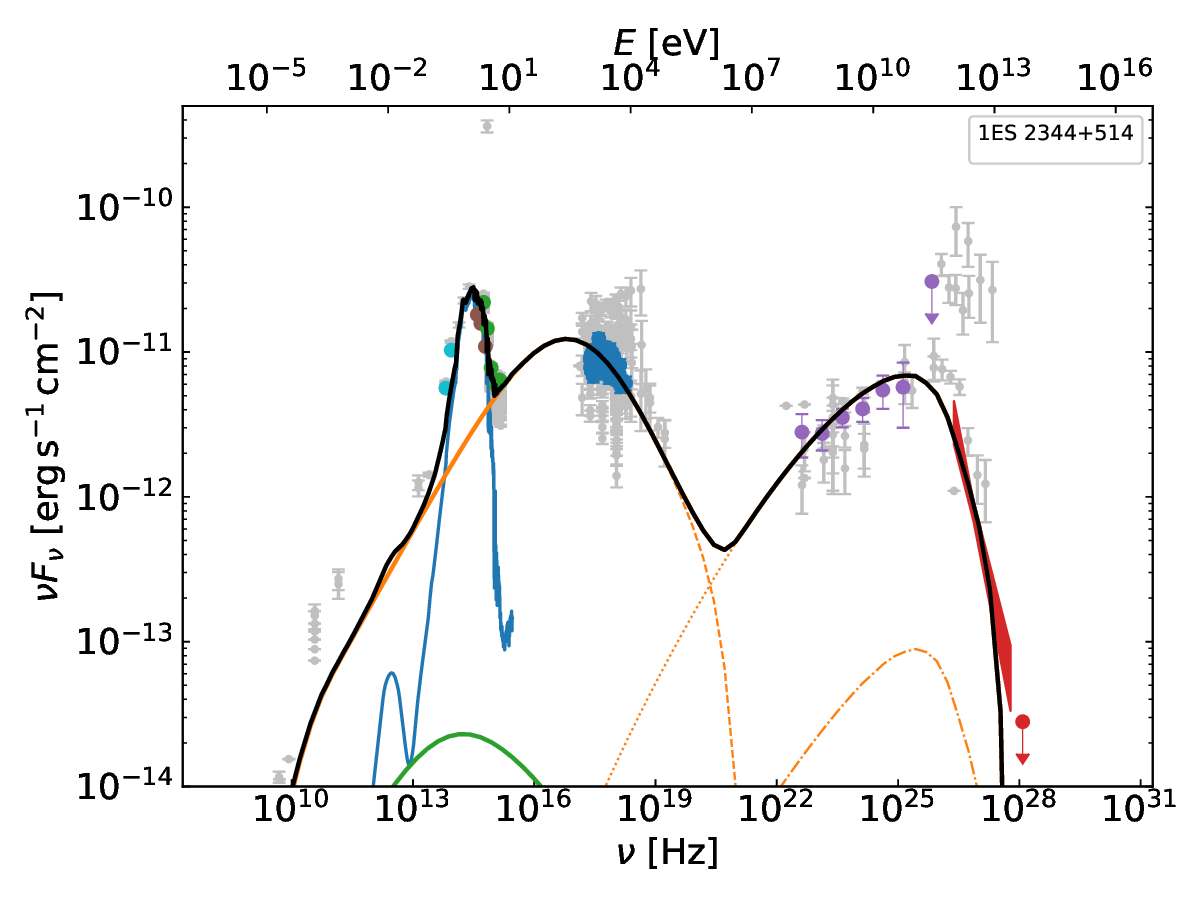}}
  \hfill
  \subfloat{\includegraphics[width=0.45\linewidth]{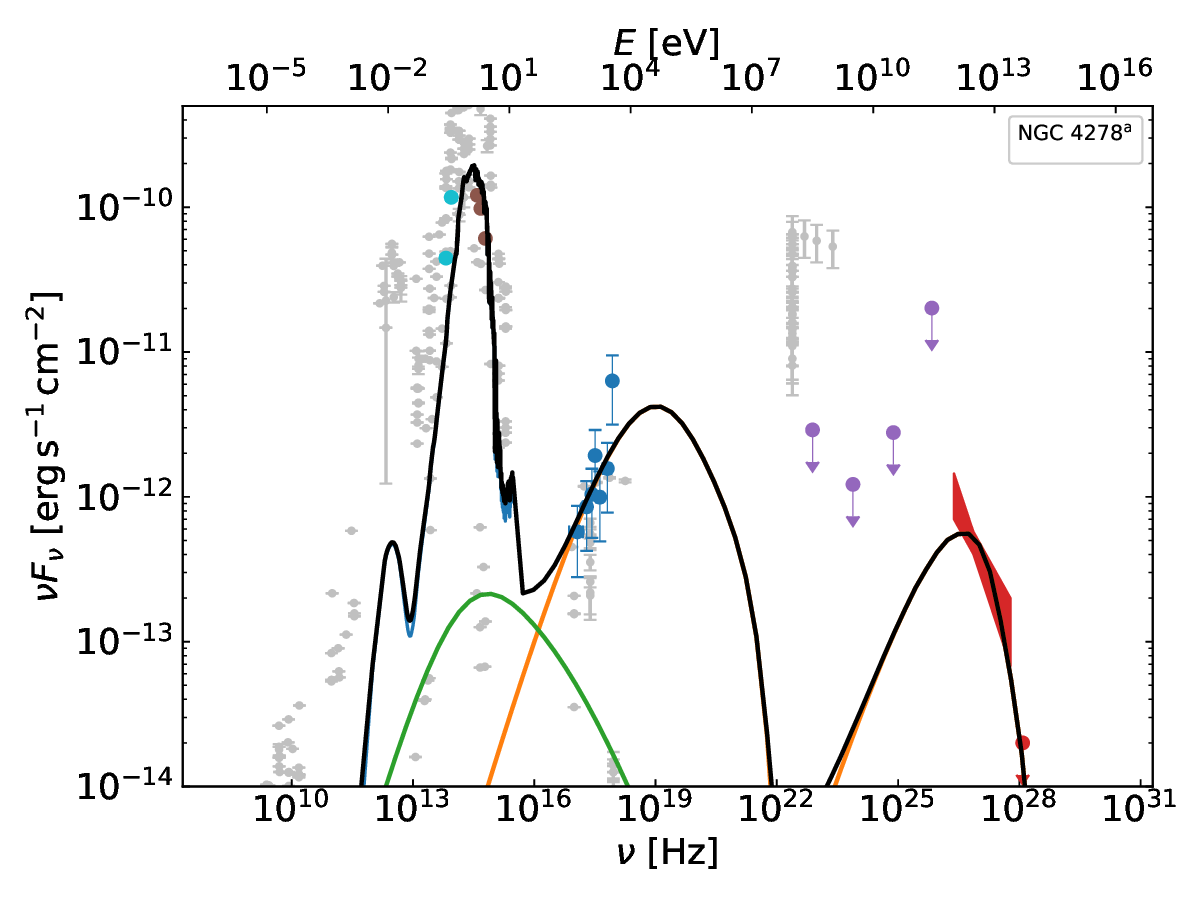}}
  \subfloat{\includegraphics[width=0.45\linewidth]{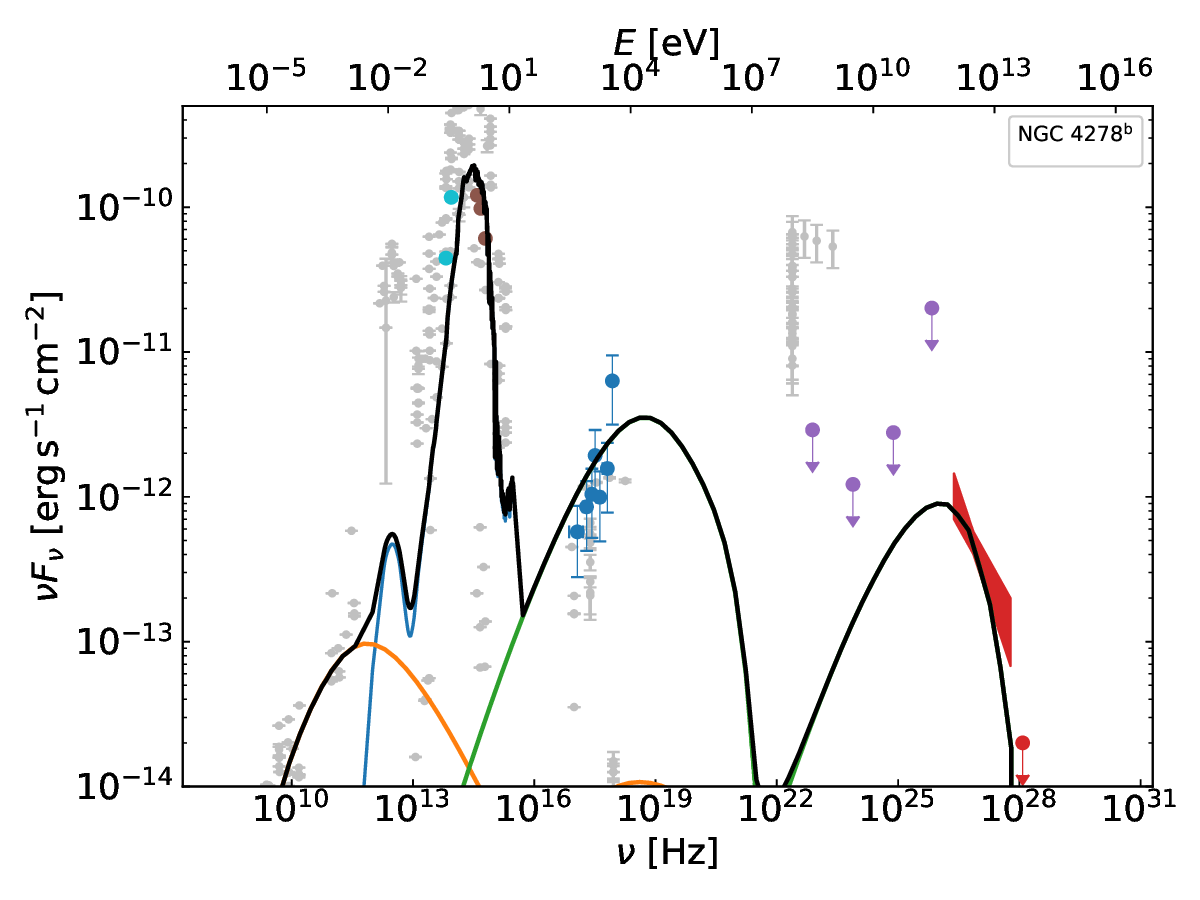}}
  \caption{The first fitting strategy of the spine-layer model. The multi-wavelength data is explained by the emission from one component. The meanings of symbols and line styles are given in the legend of Mrk 421.\label{fig:sl_ec}}
\end{figure*}

\begin{figure*}[htbp]
  \centering
  \subfloat{\includegraphics[width=0.45\linewidth]{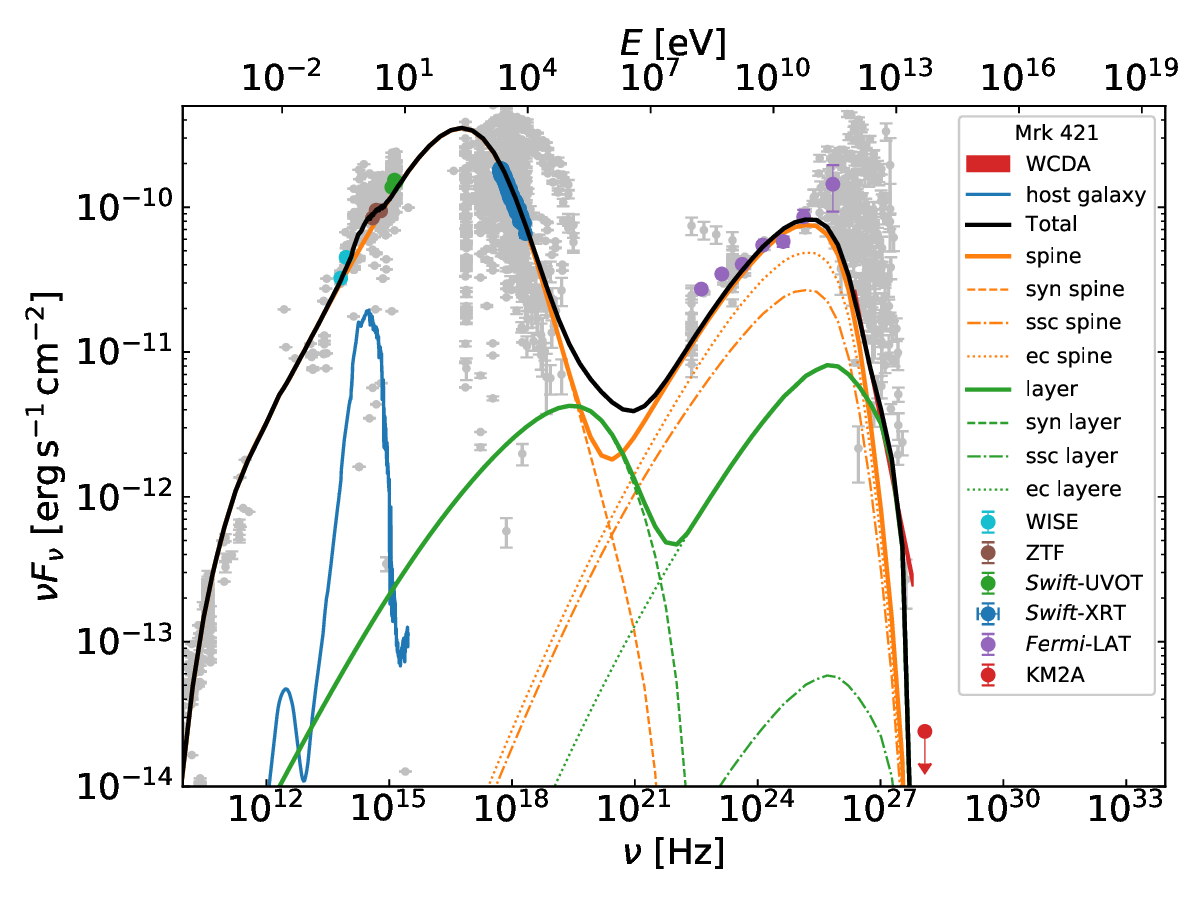}}
  \subfloat{\includegraphics[width=0.45\linewidth]{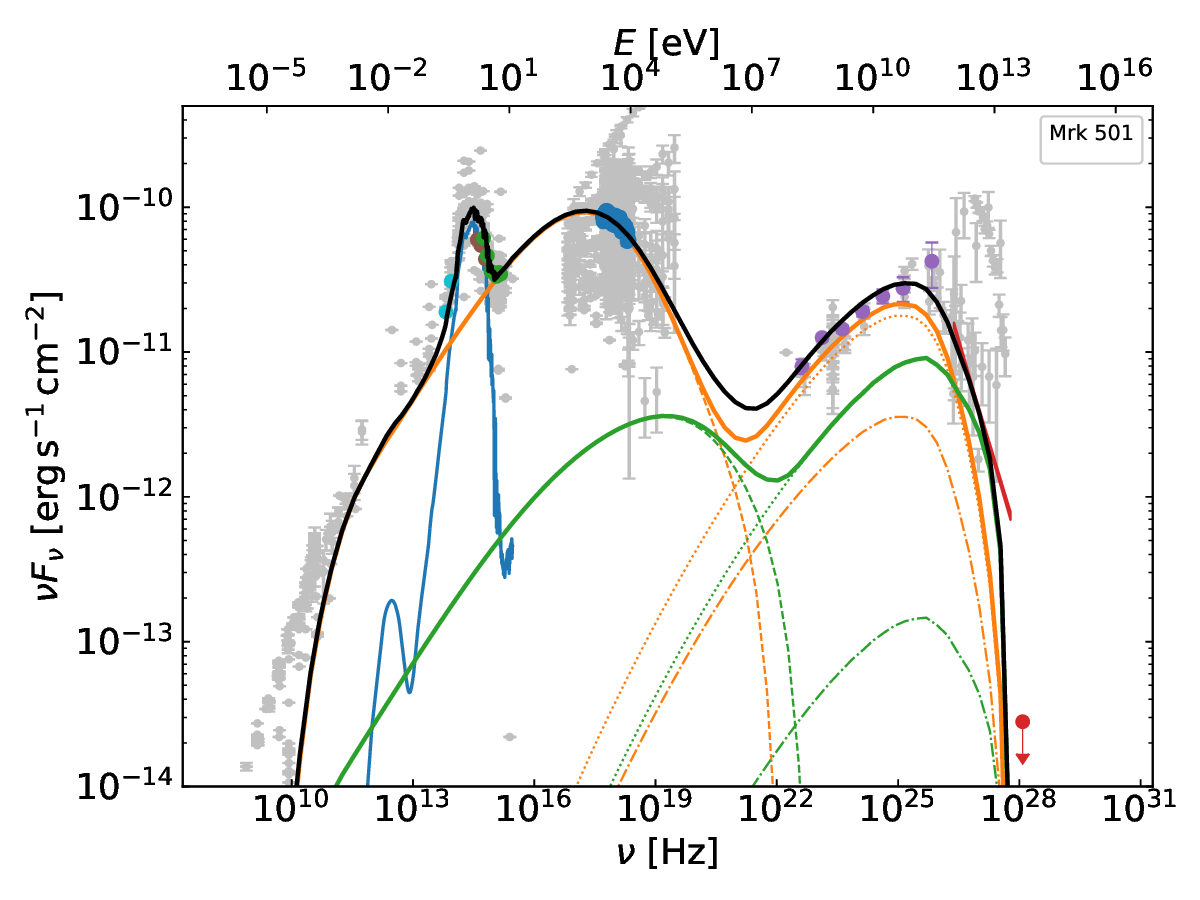}}
  \hfill
  \subfloat{\includegraphics[width=0.45\linewidth]{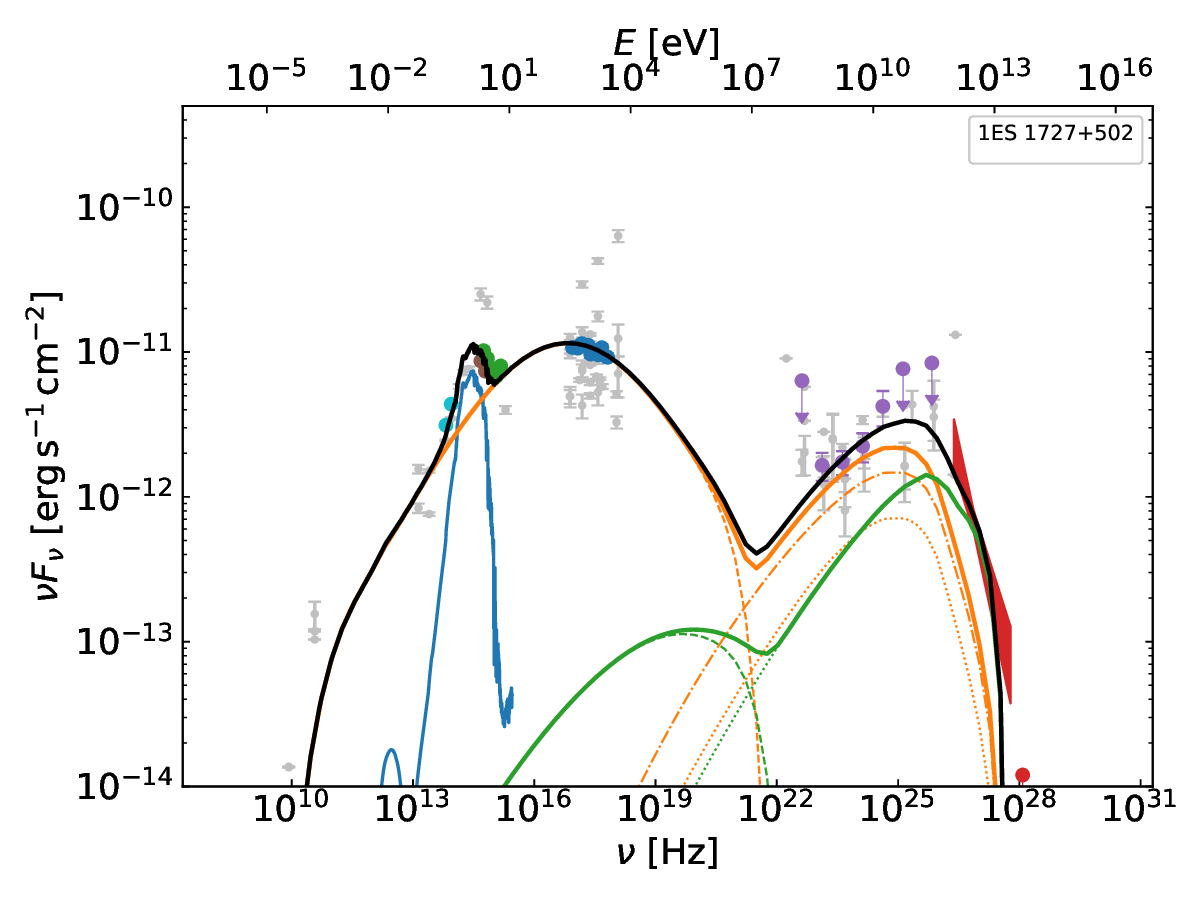}}
  \subfloat{\includegraphics[width=0.45\linewidth]{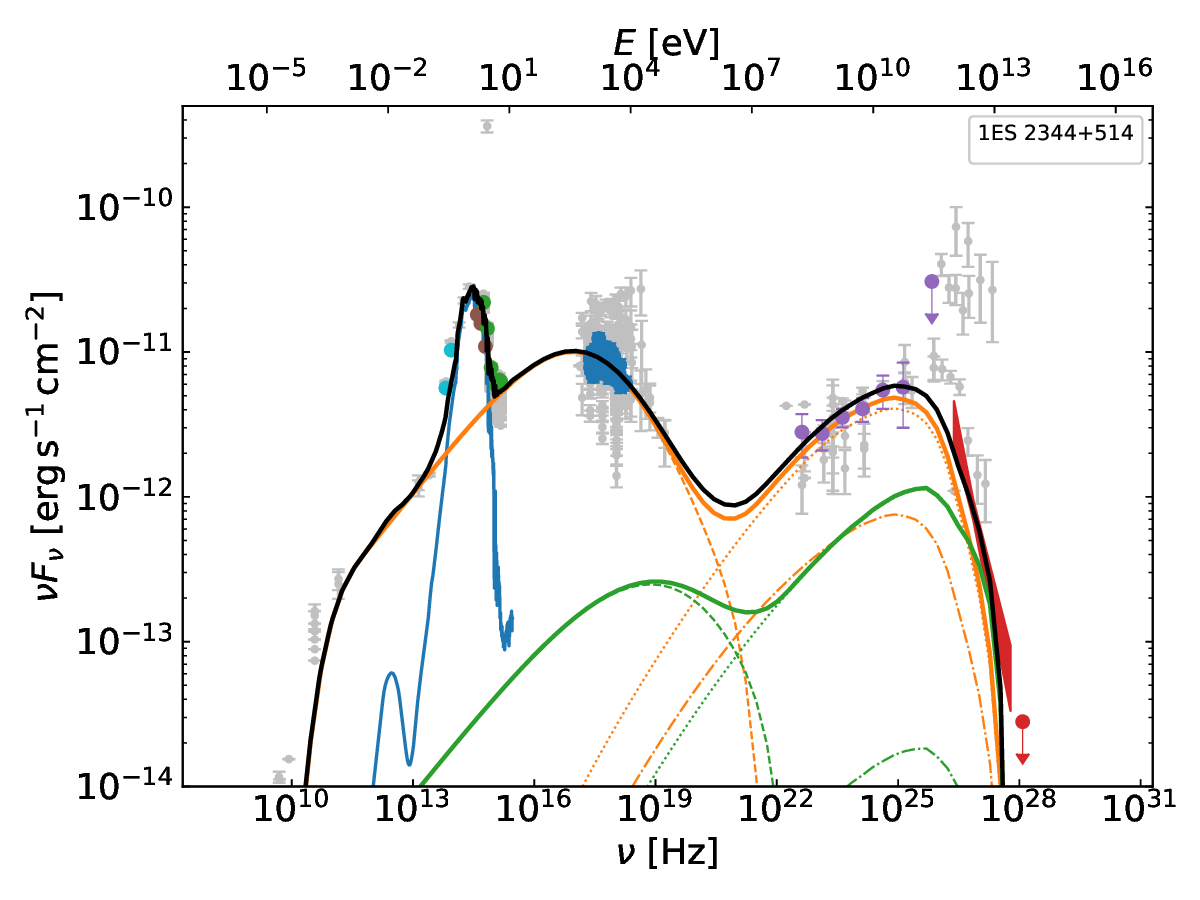}}
  \hfill
  \subfloat{\includegraphics[width=0.45\linewidth]{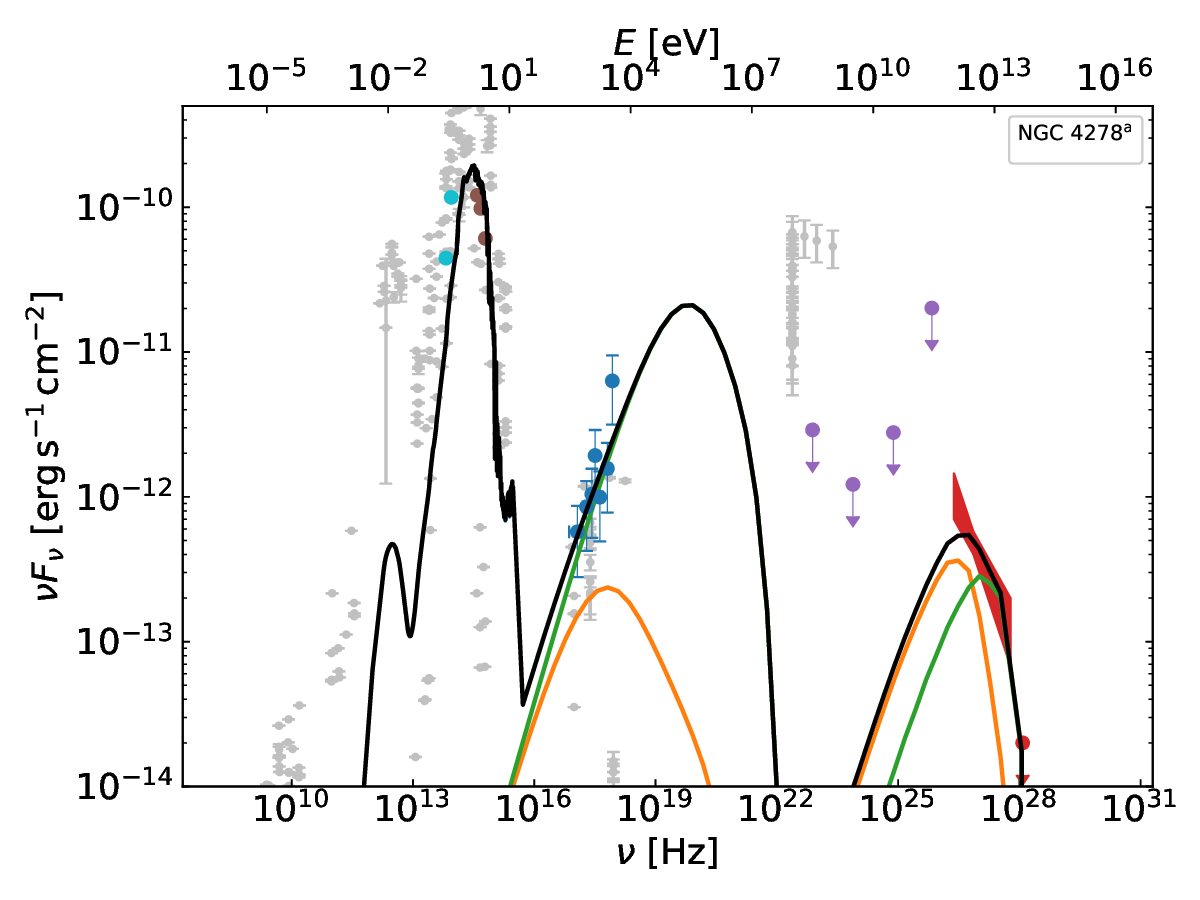}}
  \subfloat{\includegraphics[width=0.45\linewidth]{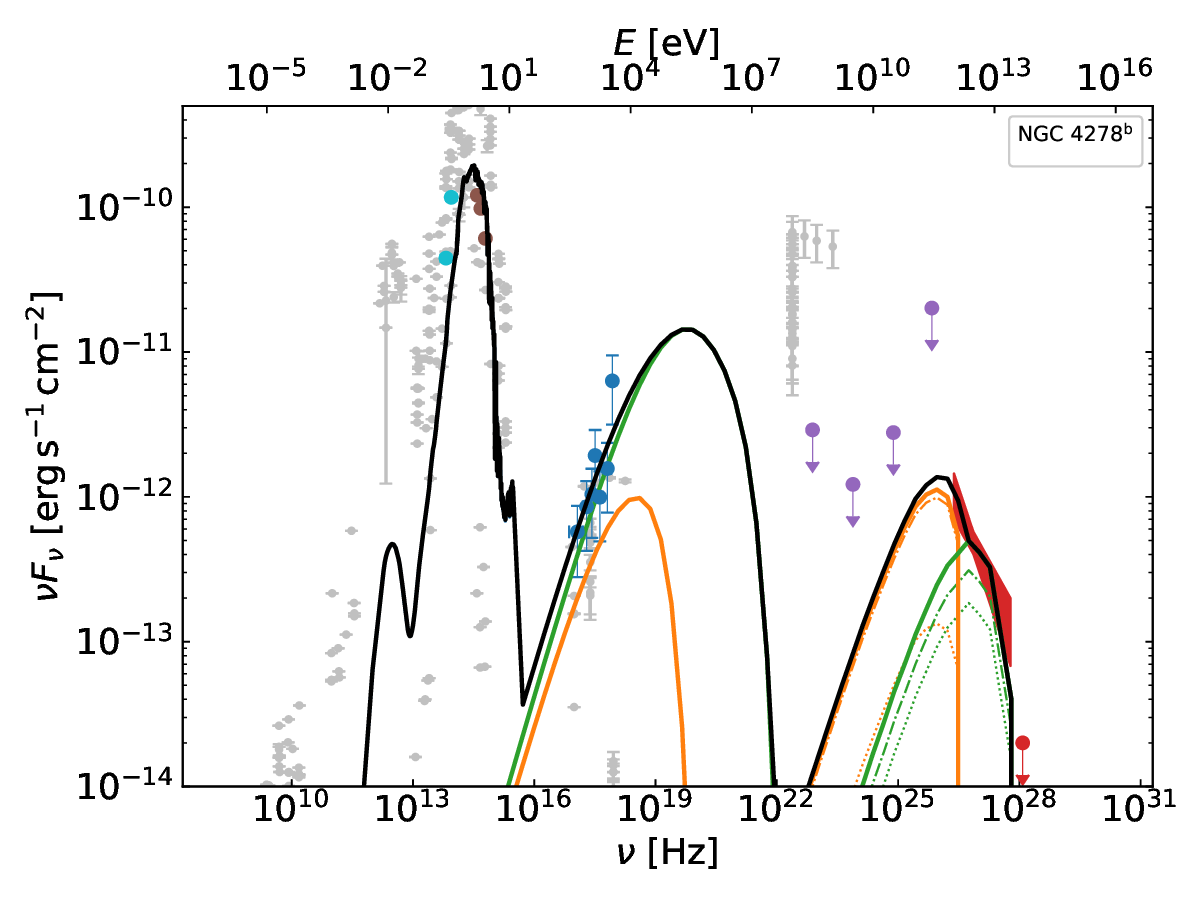}}
  \caption{The second fitting strategy (two zone) of the spine-layer model, the SEDs are reproduced by the superposition of emissions from two components. The meanings of symbols and line styles are given in the legend of Mrk 421. \label{fig:sl}}
\end{figure*}

The observed limb-brightening at the parsec \citep[e,g.,][]{2004ApJ...600..127G,2006ApJ...646..801G,2010ApJ...723.1150P} and the kiloparsec scales \citep[e.g.,][]{1989ApJ...340..698O,2011MNRAS.417.2789L} suggest that the jet could be structured with a fast spine surrounded by a slower layer. Based on this observation, the spine-layer (or structured) jet model is proposed and applied to account for the rapidly variable VHE emission \citep[e.g.,][]{2005A&A...432..401G} and to reproduce the quiescent state SED \citep{2016MNRAS.456.2374T}. In this subsection, we apply the spine-layer model to fit the SEDs.

This model consists of two components, a relatively small cylinder that is the spine (denoted by the subscript `s') and another hollow cylinder wraps around the spine as the layer (denoted by the subscript `l'). Similar to the conventional two-zone model, this model also requires two sets of parameters. The difference is that the spine and the layer influence each other and there is a relationship between these two sets of parameters. We basically use the same settings as the one-zone SSC model for each component. There are three differences from before:
\begin{enumerate}
\item The radiation zone changes from spherical to a cylinder (spine) or hollow cylinder (layer), so the radius of the radiation zone $R$ is changed to the cross-section radius $R_{\rm c}$, and we add a parameter of the length of the cylinder $L$. In addition, all calculations in relation to the shape of the radiation zone must also be replaced. For example, the volume of a sphere in Eq. (\ref{eq:EED}) must be replaced by that for a cylinder or hollow cylinder.
\item We set $R_{\rm l}=1.2R_{\rm s}$ to reduce the number of free parameters, which follows \cite{2005A&A...432..401G}. The spectral indexes of the electron spectrum in the layer are set to be the same as that in the spine, because they cannot be constrained in fitting.
\item Photons produced in one component can enter another component and, as soft photons, enhance the IC emission in both components. And this process of scattering the soft photons coming from another component is commonly known as the external Compton (EC) process. The energy density of the soft photons from another component is calculated as suggested in \cite{2005A&A...432..401G}.
\end{enumerate}

Finally, there are fifteen free parameters, which can be found in Table \ref{tab:par}. In this spine-layer model, we consider three strategies for modeling:
\begin{enumerate}
\item We consider using the EC process to fit the high-energy hump independently. As discussed in Sect. \ref{sec:ssc}, it is difficult to reproduce the entire VHE data using the one-zone SSC model due to the KN effect. In the spine-layer model, if soft photons come from another component, Eq. \ref{eq:KN} is rewritten as
\begin{equation} \label{eq:KN_sl}
        E_{\rm KN}^{\rm obs}\approx \frac{\delta_{\rm s}\delta_{\rm l}m_{\rm e}^{2}c^4}{\Gamma'\left(1+z\right)^{2}}\frac{1}{E_{0}^{\rm obs}},
\end{equation}
where $\Gamma'=\Gamma_{\rm s}\Gamma_{\rm l}\left(1-\beta_{\rm s}\beta_{\rm l}\right)$ is the relative Lorentz factor between the spine and the layer, $\beta_{\rm s}$ and $\beta_{\rm l}$ are velocities for the spine and the layer, respectively. In the EC process, the soft photons provided by another component would not be constrained by the fitting of the low-energy hump as in the SSC process. Therefore, the KN effect could be weakened as long as another component can provide low energy soft photons, as shown in Eq. \ref{eq:KN_sl}. This indicates that the EC process might improve the fitting result of the SSC process on VHE observations. We then fit the SEDs where all the observed radiation is produced in one component and the other component provides only soft photons. The fitting results are shown in Fig.$\,$\ref{fig:sl_ec}. It can be seen that the EC process with slight/no KN effect still cannot effectively reproduce the high-energy tail of LHAASO spectrum of blazars. This is because the radiation spectrum produced by the IC process is not in the standard power-law form. The morphology of the spectrum produced by the IC process approaches a smooth curve because it is influenced by both electron and soft photon spectra, the KN effect, and the absorption of photon-photon interactions. This is evident in the fitting result of Mrk 421, where the EC radiation spectrum is curved compared to the power-law spectrum observed by LHAASO. In the case of NGC 4278$^{\rm a}$, the multi-wavelength radiation is explained by emission produced in the spine, while for NGC 4278$^{\rm b}$ it is explained by an emission from the layer. Because the relativistic beaming effect reduces the flux in the case with a larger viewing angle and a larger bulk Lorentz factor. The Doppler factor $\delta=0.30$ is small for the spine of NGC 4278$^{\rm b}$.


\item The superposition of radiation from two components seems to be a plausible strategy to explain the VHE spectra, as it has minimal parameter constraints. The fitting results are shown in Fig.$\,$\ref{fig:sl}. In the cases of four blazars, the X-ray spectra are explained by synchrotron emission produced in the spine, and the $\gamma$-ray radiation is from the superposition of emission from the spine and the layer. As shown in Fig.$\,$\ref{fig:sl}, the EC emission spectrum produced in the spine rapidly decreases near TeV or sub-TeV due to the KN effect. The EC radiation spectrum produced in the layer can be extended to higher energies, because a larger break electron Lorentz factor is set in the layer. Although the spectral index of the radiation spectrum for each component is different from the observed spectral index in the VHE band due to the influence of the KN effect, the VHE spectrum can still be reproduced by superimposing the radiation of the two regions. In the two cases of NGC 4278, a very hard electron spectrum is still required in fitting. The low-energy and the high-energy humps are both explained by the radiation superposition from two components. 

\item We consider fitting the SED with a superposition of multi-radiation processes from one emitting region. To be specific, the low-energy hump is fitted by synchrotron radiation, and the high-energy hump is fitted by the radiation superposition of the SSC and the EC processes. 
To reproduce the high-energy hump by the radiation superposition of the SSC and the EC processes, we consider a scenario similar to that shown in Fig.$\,$\ref{twozone2344}. In this strategy, the peak energy of SSC or EC radiation is required to reach $\sim 14\,{\rm TeV}$. If the high-energy hump originates from the IC process, the threshold peak energy is
\begin{equation} \label{eq:IC_peak_th}
    E_{\rm IC,peak}^{\rm obs}<\gamma_{\rm e,b}m_{\rm e}c^{2}\frac{\delta}{1+z}.
\end{equation}
Substituting $E_{\rm IC,peak}^{\rm obs}=14\,{\rm TeV}$ into Eq. \ref{eq:IC_peak_th}, we obtain,
\begin{equation} \label{eq:gamma_eb}
    \gamma_{\rm e,b}>2.74\times10^{7}\frac{1+z}{\delta}.
\end{equation}
The characteristic photon energy in the observer frame produced by the electron-synchrotron process can be calculated by
\begin{equation}\label{eq:esyn}
    E_{\rm e,c}^{\rm syn}=\frac{3heB\gamma_{\rm e}^{2}}{4\pi m_{\rm e}c}\frac{\delta}{1+z}\approx1.74\times10^{-8}\gamma_{\rm e}^{2}\frac{B}{1\,{\rm G}}\frac{\delta}{1+z}\,{\rm eV},
\end{equation}
where $h$ is the Planck constant. Based on the observed peak energy of the low-energy hump, we can then derive parameter constraints for the magnetic field strength, the break electron Lorentz factor and the Doppler factor. Substituting the peak energy of low-energy hump $E_{\rm e,peak}^{\rm syn}\lesssim 1\,{\rm keV}$ of Mrk 421 and Mrk 501 and Eq. \ref{eq:gamma_eb} into Eq. \ref{eq:esyn}, we obtain,
\begin{equation}
    B<7.66\times10^{-5}\frac{1+z}{\delta}\,{\rm G},
\end{equation}
which deviates strongly from the median ($\sim0.4\,{\rm G}$) of the magnetic field strength estimated from the VLBI core shift-measurements for BL Lacs \citep{2012A&A...545A.113P}.

\end{enumerate}

The first fitting strategy shows no advantage over the one-zone SSC model, either from the fitting results or from the chi-square results. The second fitting strategy is not recommended either. Despite its seemingly superior reproduction of SEDs to the naked eye, particularly for the LHAASO spectra of Mrk 421 and Mrk 501, it often results in larger $\chi^2/{\rm d.o.f}$. This is primarily due to the fact that it employs nearly twice as many fitting parameters ($n$) as the one-zone SSC model. When considering different models to fit the same SED, the number of observed data points ($m$) remains constant. According to the formula in footnote c of Table 3, an increase in the number of fitting parameters ($n$) leads to a decrease in the denominator ($m-n$), which ultimately results in an increase in $\chi^2/{\rm d.o.f}$.


\section{Discussion and Conclusion}\label{DC}

\subsection{Can emission from $p\gamma$ interactions interpret the LHAASO spectra?}\label{sec:pg}

As shown in Sect.~\ref{sec:ssc}, the one-zone SSC model can fit most of the multi-wavelength spectra, except for the high-energy tail of the LHAASO spectra. To obtain a better fit, we comprehensively test the contributions from $pp$ interactions, proton-synchrotron emission, and the spine-layer model. On the other hand, since various soft photon fields exist in AGNs environment, many works dedicate themselves to studying the electromagnetic and neutrino emissions from $p\gamma$ interactions. As suggested by many recent studies \citep[e.g.,][]{2021ApJ...906...91S,2022MNRAS.515.5235S,2022ApJ...934..158A}, here we analytically discuss if emission from $\pi^0$ decay in the $p\gamma$ interactions can improve the fitting of LHAASO spectra.

Using the $\delta-$approximation, the relation between the energy of $\pi^0$ decay VHE photons $E_{\rm VHE}^{\rm obs}$ and the energy of target photons $E_{\rm tar}^{\rm obs}$ both in the observer' frame can be obtained, 
\begin{equation}
    E_{\rm tar}^{\rm obs} \simeq 0.9~\rm MeV \big(\frac{\delta}{20} \big)^2 \big(\frac{14~\rm TeV}{E_{\rm VHE}^{\rm obs}}\big),
\end{equation}
when considering the peak cross section of photopion interactions due to the $\bigtriangleup^+(1232)$ resonance. By taking $L_{\rm Edd}$ as the maximum proton injection luminosity, $10^{-28}~\rm cm^2$ as the photopion cross section weighted by inelasticity, the lower limit of flux of the target photons can be estimated by
\begin{equation}
\begin{split}
    {\nu F_{\nu}}_{\rm tar}^{\rm obs}\simeq 2.2\times10^{-10}~\rm erg~s^{-1}~cm^{-2}~\big(\frac{\it{R}}{10^{16}~\rm cm} \big) \big(\frac{\Gamma}{10} \big)^2 \\ \big(\frac{\delta}{20} \big)\big(\frac{14~\rm TeV}{E_{\rm VHE}^{\rm obs}} \big)
    \big(\frac{{\nu F_{\nu}}_{\rm 14~TeV}^{\rm obs}}{10^{-13}~\rm erg~s^{-1}~cm^{-2}} \big)\big(\frac{10^9~M_{\odot}}{M_{\rm BH}} \big).
\end{split}
\end{equation}
As shown in Fig.~\ref{fig:ssc}, the model predicted fluxes at $\sim 1~\rm MeV$ for these LHAASO AGNs are $\sim 10^{-12}~\rm erg~s^{-1}~cm^{-2}$. If considering all emission processes occur in one single region, it can be seen that even when the emission from $p\gamma$ interactions is only used to account for the high-energy tail of the LHAASO spectra, the model-predicted flux is still two orders of magnitude lower than the flux required. If one attempts to use the emission from $p\gamma$ interactions to account for the whole LHAASO spectrum, the required flux of target photons would be more than three orders of magnitude higher than the model-predicted flux. Even taking into account that the $p\gamma$ interactions and the leptonic processes may occur in different regions, i.e., a multi-zone case, the required flux at MeV band ($>\sim 10^{-10}~\rm erg~s^{-1}~cm^{-2}$) still far exceeds all the existing AGN observations.Therefore, using the emission from the $p\gamma$ interactions to interpret the LHAASO spectra can be confidently ruled out.

\subsection{The influence of different EBL models}\label{sec:EBL}

\begin{figure*}[htbp]
  \centering
  \subfloat{\includegraphics[width=0.33\linewidth]{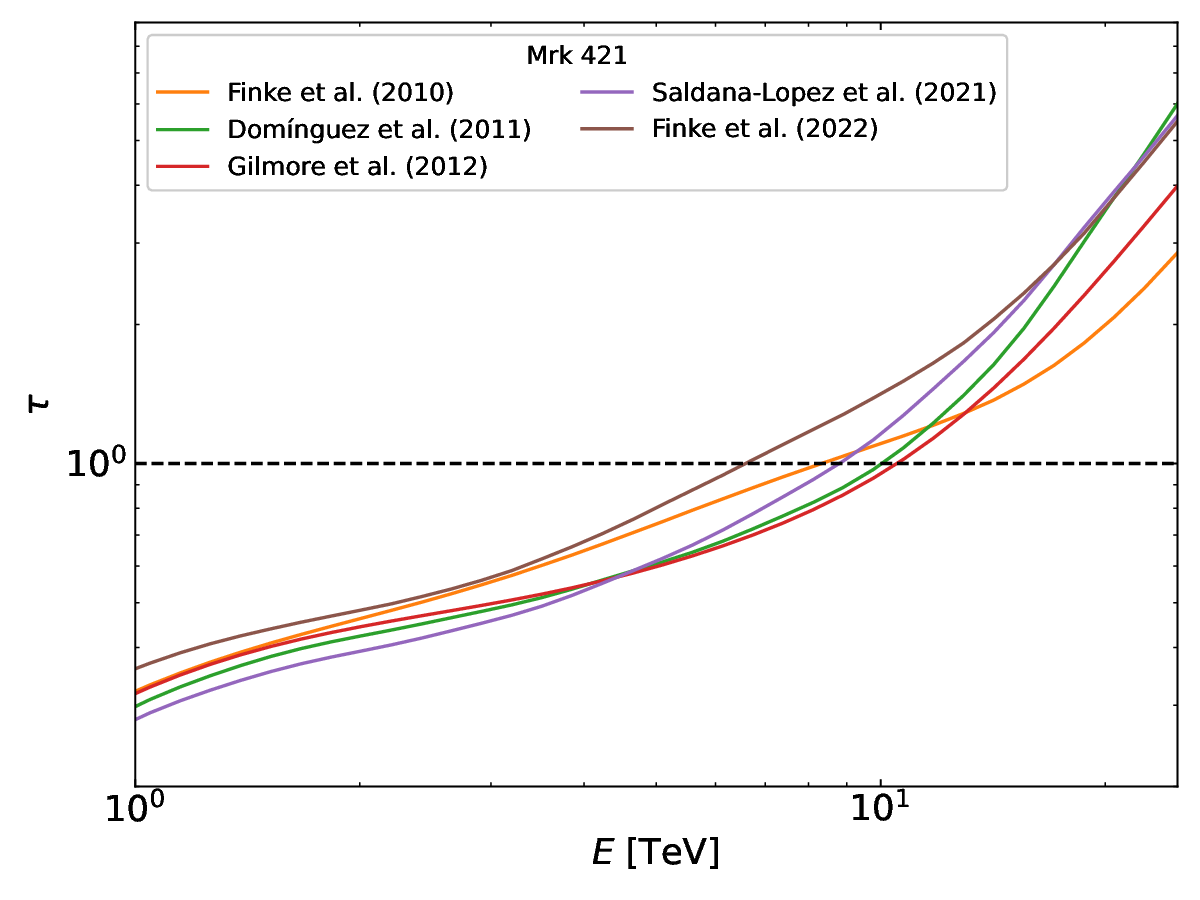} }
  \subfloat{\includegraphics[width=0.33\linewidth]{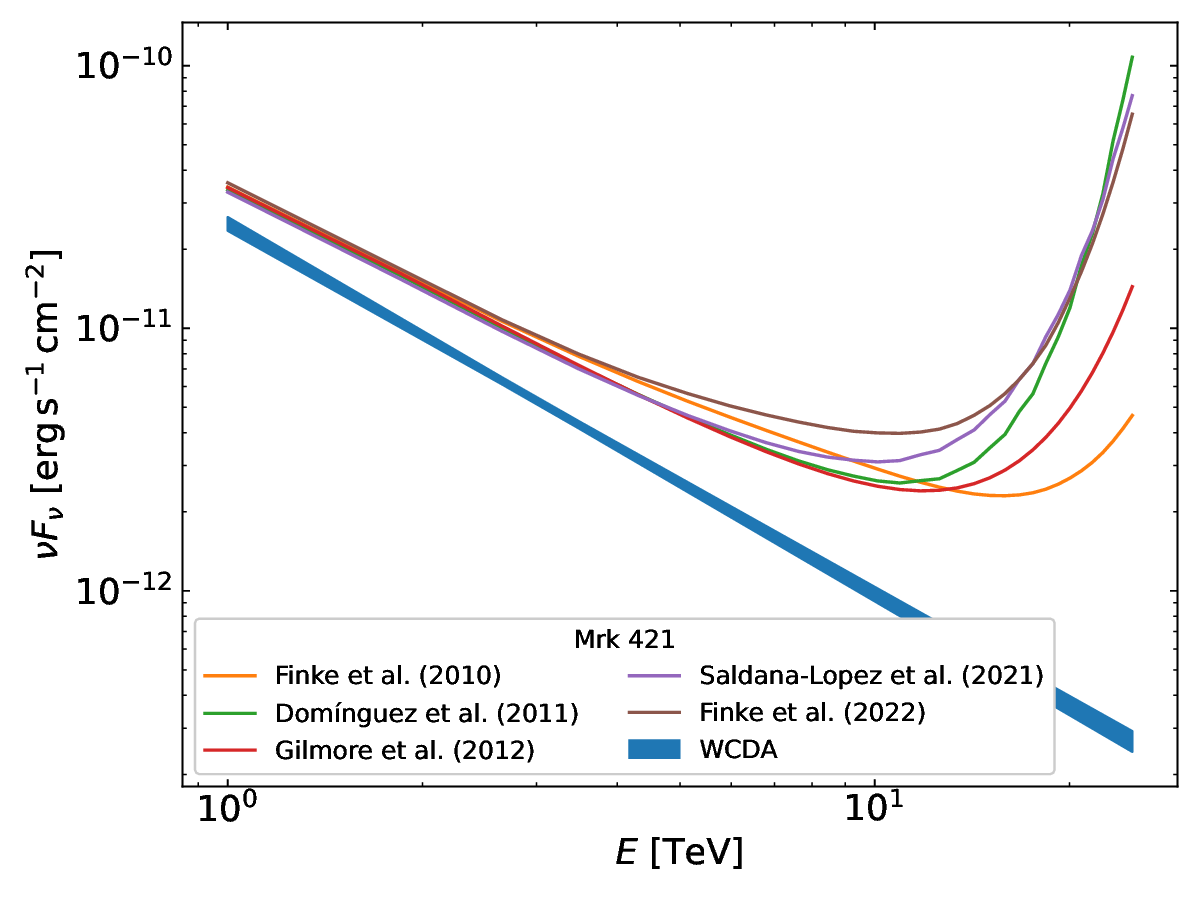}} 
  \subfloat{\includegraphics[width=0.33\linewidth]{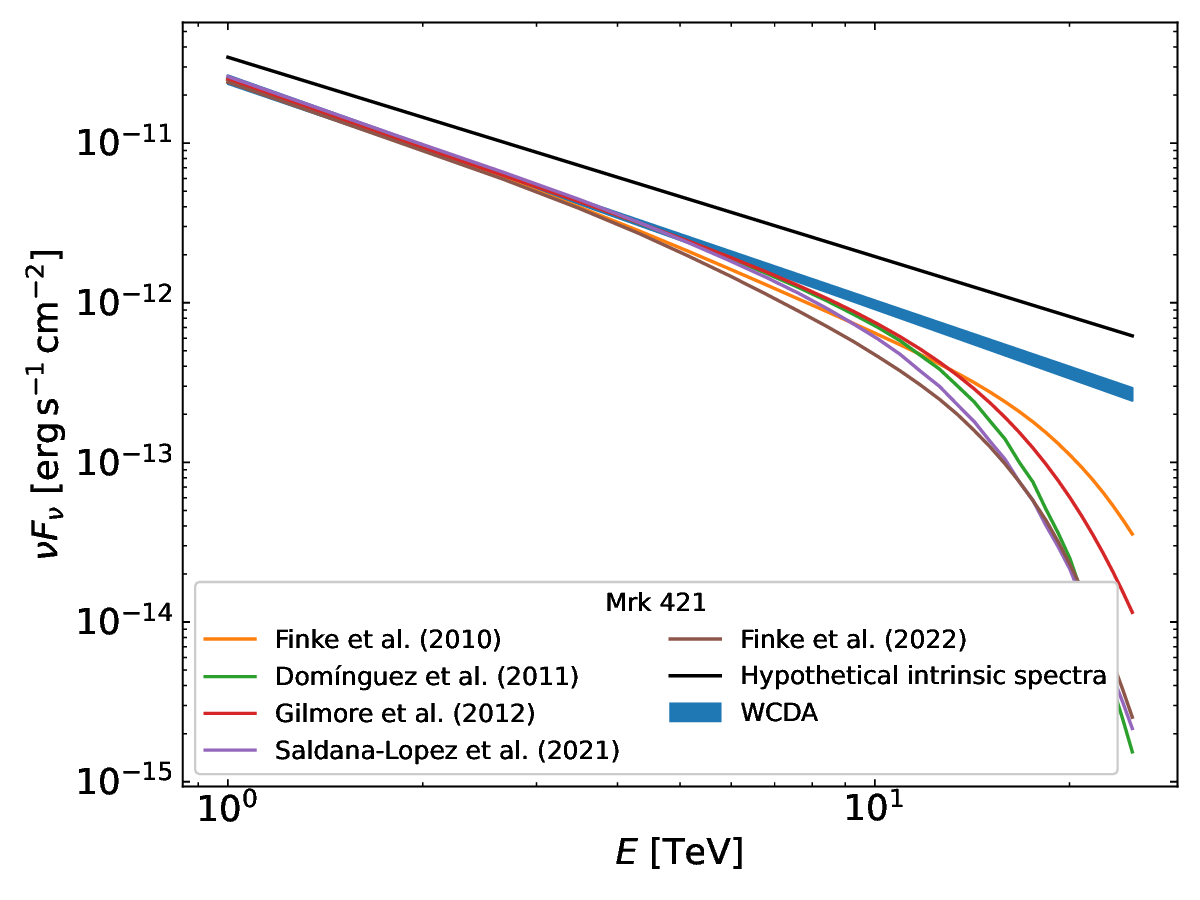}}
  
\caption{The optical depths (left panel), the intrinsic VHE spectra (middle panel) and the hypothetical extended intrinsic spectrum and the expected observed spectrum (right panel) of Mrk 421 (z=0.031) for different EBL models. The optical depths taken from \cite{2010ApJ...712..238F, 2011MNRAS.410.2556D, 2012MNRAS.422.3189G, 2021MNRAS.507.5144S, 2022ApJ...941...33F}. \label{fig:EBL}}
\end{figure*}

The observed $\gamma$-ray spectra of extragalactic sources, especially in the VHE band, are softened by the interactions of the $\gamma$-ray photons with the EBL. The energy spectrum of EBL is difficult to obtain by direct observation, so many researchers have used various methods to estimate it. To evaluate the influence of different EBL models on the fitting results, we apply five EBL models to Mrk 421, showing  the optical depths (left panel in Fig.$\,$\ref{fig:EBL}) and the intrinsic VHE spectra (middle panel in Fig.$\,$\ref{fig:EBL}), respectively. 

The energy of $\gamma$-ray opacity equal to unity varies in different EBL models. Within these models, the energy is focused between the 7-10\,TeV range for Mrk 421. The EBL model (green line in Fig.$\,$\ref{fig:ssc}) used for the calculation in Sect. \ref{model}, shows a relatively moderate optical depth in the energy range of the WCDA. Furthermore, various EBL models are also applied to calculate the corresponding intrinsic VHE spectra, based on the observed spectrum given by WCDA. The middle panel of Fig.$\,$\ref{fig:EBL} shows that all models indicate the presence of a new component beyond $\sim 10~\rm TeV$, which is consistent with our fitting results. From Fig.$\,$\ref{fig:EBL}, it can be seen that all EBL models have an equivalent effect on the VHE spectra below 4\,TeV. Thus, the intrinsic VHE spectrum can be estimated by extrapolating the 1-4\,TeV spectrum up to higher energies, if we assume that the VHE radiation comes from one single component. The right panel of Fig.$\,$\ref{fig:EBL} displays the hypothetical extended intrinsic spectrum (represented by the black line) and the expected observed spectrum after absorption. It appears challenging to reproduce the entire VHE spectrum from the radiation of a single component, even without considering the KN effect in the one-zone SSC model.

The best-fit power-law bow-tie of the LHAASO data cannot be broken into energy bins to evaluate how the EBL impacts it over the entire energy range, so corrections other than a scaling factor would result in break in the power law which are not captured by the best-fit. The right panel of Fig.$\,$\ref{fig:EBL} shows that the power-law spectrum is unlikely to be broken before 7\,TeV by the influence of absorption from the EBL model of \cite{2011MNRAS.410.2556D}, which is used in the calculation of Sect. \ref{model}. From Fig.\,\ref{fig:ssc}, it can be found that the energy spectrum predicted by the one-zone SSC model already deviates from the LHAASO bow-tie at about 2\,TeV, and the model predicted flux is only about half the observed flux at 7\,TeV. This suggests that there are factors other than EBL absorption that cause the power-law spectrum to bend, which is consistent with the conclusion in Sect. \ref{sec:ssc}. In the Sect. \ref{sec:pg}, the theoretical analysis predicts that if the $p\gamma$ interactions are used to explain the observation of the LHAASO data at 14\,TeV, it is necessary to increase the model predicted flux of 1\,MeV by more than 100 times. From the left panel of Fig.$\,$\ref{fig:EBL} it can be seen that the optical depths of different EBL models at 14\,TeV range from 1.2 to 1.7, and the correction factor for the flux is between 0.30 and 0.18. Thus, even if taking into account the effect of EBL absorption on the LHAASO best-fit power-law bow-tie, it is necessary to increase the 1\,MeV flux by more than an order of magnitude.

The above discussion indicates that applying different EBL models does not affect our conclusions. In addition, the observational data from the $\gamma$-ray telescope can help to constrain the absorption optical depth induced by EBL and to constrain the EBL model \citep[e,g.,][]{2007A&A...475L...9A, 2018Sci...362.1031F, 2019ApJ...885..150A, 2019MNRAS.486.4233A}. However, this would require a more abundant or simultaneous set of observation data. As it is beyond the scope of this paper, we will not discuss it further.


\subsection{Comparison with previous TeV AGNs studies}
The full broadband SEDs modeling
has been the main tools for the blazar study. The VHE $\gamma$-ray observation from Imaging Air Cherenkov Telescopes (IACTs) provide a strong constraint to the jet models. The four LHAASO blazars are all known VHE emitters, of which Mrk 421, Mrk 501, and 1ES 2344+514 are the earliest detected extragalactic VHE $\gamma$-ray sources \citep{1992Natur.358..477P, 1996ApJ...456L..83Q, 1999ApJ...513..161C}. The VHE $\gamma$-ray of 1ES 1727+502 has also been observed for more than ten years \citep{2011ApJ...729..115A, 2014A&A...563A..90A}. Therefore, these four blazars have been extensively studied with numerous multi-wavelength SEDs from different periods.

Most previous studies on these four blazars have shown that the one-zone SSC model can reasonably reproduce the SEDs \citep[e.g.,][]{2007ApJ...663..125A, 2009ApJ...705.1624A, 2010MNRAS.401.1570T, 2011ApJ...727..129A, 2011ApJ...734..110B, 2011ApJ...729....2A, 2011ApJ...738...25A, 2011ApJ...738..169A, 2014A&A...563A..90A, 2015ApJ...812...65F, 2016ApJS..222....6B, 2022ApJ...929..125A, 2022MNRAS.515.2633P}. In contrast, some studies suggest that the one-zone SSC model cannot fit the SEDs because it underestimates the TeV $\gamma$-ray flux, e.g. the high-state TeV flux of Mrk 421 observed by the Whipple Observatory between 16 and 20 April 2006 \citep{2005ApJ...630..130B}, the flare-state VHE band (above $6\,{\rm TeV}$) flux of Mrk 501 observed by ARGO-YBJ in October 2011 \citep{2012ApJ...758....2B}, the narrow spectral feature at $\sim 3\,{\rm TeV}$ of Mrk 501 observed by MAGIC in 19 July 2014 \citep{2020A&A...637A..86M}. Some other studies have found that high Doppler factors are necessary for fitting with the one-zone SSC model, e.g. Doppler factors $\gtrsim 60$ obtained by fitting the H.E.S.S. and \textsl{Swift} data of a TeV flare observed from PKS 2155-304 between 28 and 30 July 2006, Doppler factors $\gtrsim 30$ for fitting a TeV flare observed in 2001 from Mrk 421 \citep{2008ApJ...686..181F}. These diverse observations and fitting results suggest that the realized radiation process from these sources may be very complex, and more observations from different time periods and energy bands are key to further research.

In addition to the one-zone SSC model, other models are also used to fit the SEDs of these sources. \cite{2013A&A...556A..67A} found that both one-zone and two-zone SSC models can well reproduce the SED observed by 1ES 2344+514 in late 2008. \cite{2023ApJS..266...37A} suggested that the observed different patterns of variability of Mrk 501 would naturally be expected from the two-zone model. \cite{2011ApJ...736..131A} presented the average SED of Mrk 421 in the low state between 19 January and 1 June 2019, and suggested that both the one-zone SSC model and the hadronic proton-synchrotron Blazar model \citep{2001APh....15..121M} are able to describe the SED well. \cite{2020MNRAS.496.3912M} also found that a flaring state SED of 1ES 2344+514, observed in August 2016, can be successfully described by both the one-zone SSC model and the proton-synchrotron model. Note that a larger magnetic field ($B=50\,{\rm G}$ for Mrk 421 and $B\sim 50\,{\rm G}$ for 1ES 2344+514) is required in their hadronic scenario. \cite{2013MNRAS.434.2684M} found that the observations of Mrk 421 on March 2001 can be naturally reproduced with the leptohadronic model. \cite{2016EPJC...76..127S} concluded that the TeV flaring of Mrk 421 can be well explained by the photohadronic model. \cite{2020A&A...640A.132M} suggested that a co-located two-zone model is a more reasonable explanation for the overall SEDs of five TeV blazers, including 1ES 2344+514 and 1ES 1727+502. \cite{2021A&A...655A..89M} found that the SED of Mrk 421 with a VHE flare observed on 4 February 2017 is better reproduced by a two-zone leptonic model than by a one-zone leptonic model. \citep{2020ApJ...901..132S, 2021ApJ...914..120S, 2022MNRAS.515.5235S} found that the one-zone photohadronic model is inadequate to explain the multi-TeV flaring events from the transient extreme HSP-like sources of Mrk 501, Mrk 421 and 1ES 2344+514, and they proposed a two-zone photohadronic model as an effective methodology. \cite{2023MNRAS.525.3533M} suggested that an additional emission mechanism other than the SSC process is required to explain the TeV observations of Mrk 421 by MAGIC in February 2013, because its VHE spectra are remarkably harder than the X-ray spectra. \cite{2023ApJ...948...82H} included that the SED observed from Mrk 421 in January 2013, in particular the hard X-ray excess, could have been generated as a result of the two-injection scenario.

These sources exhibit a number of observational features, particularly in the flaring state, that cannot be explained by the one-zone SSC model. Higher sensitivity observations at higher energy bands will hopefully verify the above assumptions and models, and LHAASO has great potential in this regard. In this work, we collect multi-wavelength data from five LHAASO AGNs during the same observation period as LHAASO. Based on theoretical and fitting analysis, we suggest that the one-zone SSC model is capable of reproducing most of the SED, with the exception of the VHE tail in the cases of Mrk 421 and Mrk 501. The inability of the VHE tail is mainly due to the collective effect of the KN effect, the EBL absorption and the parameter constraints for other bands observations. This is well demonstrated in the case of NGC 4278, which is very close to us and has almost no EBL absorption. In addition, its multi-wavelength data has very weak parameter constraints. Therefore, when we consider more extreme parameters, the one-zone SSC model can reproduce its SED, especially the LHAASO spectrum. We suggest that the high-energy tail of the LHAASO data of Mrk 421 and Mrk 501 cannot be fitted with the one-zone SSC model, unless very extreme parameters are considered. This is similar to the conclusion of \cite{2005A&A...433..479K}, which suggests that the Thomson scattering into VHE photon energies requires unacceptably large Doppler factors.

To reproduce the SEDs of LHAASO AGNs, we apply the $pp$ model, the proton-synchrotron model and the spine-layer model. The results of fitting and the chi-square test suggest that the one-zone model, upon incorporating $pp$ interactions, effectively accounts for all observations in the SEDs, especially the tail of VHE observation. In addition, a multi-zone model is also feasible if we consider the superposition of radiation generated by different regions to explain VHE observations, as demonstrated in the spine-layer model presented in Sect. \ref{sec:sl}. Despite its seemingly superior reproduction of SEDs to the naked eye, particularly for the LHAASO spectra of Mrk 421 and Mrk 501, it often results in larger $\chi^2/{\rm d.o.f}$.

Our analysis results indicate that the proton-synchrotron model and the $p\gamma$ model are difficult to explain the SEDs without considering very extreme parameters. Of all the sources, only the SEDs of 1ES 2344+514 can be reproduced using the two-zone proton-synchrotron model. A very large magnetic field ($>10\,{\rm G}$) must be introduced to fit the SEDs of the other LHAASO AGNs, whether in one-zone or two-zone proton-synchrotron models. The low interaction efficiency of $p\gamma$ model, brought about by the lack of suitable soft photon fields, prevents it from reproducing the SEDs within reasonable parameters.

NGC 4278 is the most possible association with 1LHAASO J1219+2915. Moreover, it is also found to be positionally consistent with the $\gamma$-ray transient source 1FLT J1219+2907 detected by \textsl{Fermi}-LAT \citep{2021ApJS..256...13B}. Therefore, although \textsl{Fermi}-LAT did not detect high-energy radiation from NGC 4278 during the 500-day period of the LHAASO detection, it remains a candidate with great potential for the VHE source. Unfortunately, the data reduction and SED fitting in this paper do not allow us to determine further whether the VHE radiation comes from NGC 4278.


\subsection{Outlook} \label{sec:OL}
Our results suggest that VHE observations are crucial to constraint the jet model. As mentioned in Sect. \ref{sec:ssc}, the SSC process of HSP enters the KN regime in the VHE band. Detailed observations in the VHE band can verify or rule out the origin of the one-zone SSC model more precisely. Furthermore, Fig.$\,$\ref{fig:EBL} shows that different EBL absorption models have significant different influence on VHE observation beyond 7 TeV. Therefore, by conducting further observations of extragalactic sources exceeding 7 TeV, we can constrain the EBL model better. This method has already been extensively applied in other $\gamma$-ray telescopes \citep[e.g., ][]{2018Sci...362.1031F, 2019ApJ...885..150A, 2019MNRAS.486.4233A}.

Multi-wavelength variability can provide a different perspective to study the emission origin. For example, long-term monitoring is carried out for Mrk 421, as it is one of the closest BL Lac objects. Its VHE variability displays a highly complex behaviour. Most observations have found a strong correlation between flares in the VHE band and the X-ray band \citep{2008ApJ...677..906F, 2021A&A...655A..89M, 2021A&A...647A..88A, 2021MNRAS.504.1427A}. Some observations have reported that variations in the VHE band correlated with X-rays, but not with the optical \citep{2007A&A...462...29G} and the other bands \citep{2015A&A...578A..22A}. Some variability studies indicate that the correlation between the X-ray band and the VHE band shows different behaviour \citep{2020ApJS..248...29A}, and \cite{2020ApJ...890...97A} find that the flux relationship changes from linear to quadratic, to no correlation, and to anti-correlation over the decline epochs. \cite{2005ApJ...630..130B} report the inconsistency of X-ray band and VHE band flare times. Taken together, these phenomena are difficult to explain using the one-zone SSC model. Therefore, the observations of variability in VHE band and the corresponding simultaneously SEDs are very important to investigate the radiation mechanisms and the physical properties of blazars. 

\begin{acknowledgments}
We thank the anonymous referee for insightful comments and constructive suggestions. D. R. X thanks Yao Su and Zhang Guobao for their assistance in X-ray data processing. This work is partially supported by the National Key Research and Development Program of China (grant numbers 2022SKA0130100, 2022SKA0130102 and 2023YFE0101200). J.M. is supported by the NSFC 11673062, CSST grant CMS-CSST-2021-A06, and the Yunnan Revitalization Talent Support Program (YunLing Scholar Project). Z.R.W. acknowledges the support by the NSFC under Grant No. 12203024 and the support by the Department of Science $\&$ Technology of Shandong Province under Grant No. ZR2022QA071. R.X. acknowledges the support by the NSFC under Grant No. 12203043. D.R.X. acknowledge the science research grants from the China Manned Space Project with No. CMS-CSST-2021-A06, Yunnan Province Youth Top Talent Project (YNWR-QNBJ-2020-116) and the CAS “Light of West China” Program. F.K.P. acknowledges support by the National Natural Science Foundation of China (Grants No. 12003002), the University Annual Scientific Research Plan of Anhui Province 2023 (2023AH050146), the Excellent Teacher Training Program of Anhui Province (2023), and the Doctoral Starting up Foundation of Anhui Normal University 2020 (903/752022). L.M.S. acknowledges the support from NSFC grant No. 12103002 and Anhui Provincial Natural Science Fondation grant No. 2108085QA43. 

This work made use of data supplied by the UK \textsl{Swift} Science Data Centre at the University of Leicester. Based on observations obtained with the Samuel Oschin Telescope 48-inch and the 60-inch Telescope at the Palomar Observatory as part of the Zwicky Transient Facility project. ZTF is supported by the National Science Foundation under Grants No. AST-1440341 and AST-2034437 and a collaboration including current partners Caltech, IPAC, the Weizmann Institute for Science, the Oskar Klein Center at Stockholm University, the University of Maryland, Deutsches Elektronen-Synchrotron and Humboldt University, the TANGO Consortium of Taiwan, the University of Wisconsin at Milwaukee, Trinity College Dublin, Lawrence Livermore National Laboratories, IN2P3, University of Warwick, Ruhr University Bochum, Northwestern University and former partners the University of Washington, Los Alamos National Laboratories, and Lawrence Berkeley National Laboratories. Operations are conducted by COO, IPAC, and UW. This research has made use of the NASA/IPAC Extragalactic Database (NED), which is operated by Jet Propulsion Laboratory, California Institute of Technology, under contract with the National Aeronautics and Space Administration.
\end{acknowledgments}

\vspace{5mm}
\facilities{LHAASO, \textsl{Fermi}-LAT, \textsl{Swift}-XRT, \textsl{Swift}-UVOT, ZTF, WISE}

\software{astropy \citep{2013A&A...558A..33A,2018AJ....156..123A, 2022ApJ...935..167A},  
          naima \citep{naima}
          }

\appendix
\section{Proton-synchrotron modeling with strong magnetic fields}\label{Bsyn}
\begin{table*}[]
\caption{\label{tab:par_appendix}The fitting parameters of proton-synchrotron model with strong magnetic fields.}
\begin{tabular}{ccccccccccccc}
\hline\hline
\multicolumn{11}{c}{proton-synchrotron model}                                                                                                                                                                                                          &          &                \\
Source name        & $\theta$ & $\Gamma$ & $L_{\rm e}^{\rm inj}\left({\rm erg}\,{\rm s}^{-1}\right)$ & $\gamma_{\rm e,b}$ & $\gamma_{\rm e,max}$ & $p_{\rm e,1}$ & $p_{\rm e,2}$ & $p_{\rm p,1}$ & $p_{\rm p,2}$ & $L_{\rm p}^{\rm inj}$/$L_{\rm Edd}$ & $\chi^2/{\rm d.o.f}$ & $\chi^2_{\rm WCDA}/{\rm d.o.f}$ \\ \hline
(1)                & (2)      & (3)      & (4)                                                       & (5)                & (6)                  & (7)           & (8)           & (9)           & (10)          & (11)                                & (12)     & (13)           \\
Mrk 421            & 1.8      & 23       & 3.00E+42                                                  & 4.90E+03           & 1.00E+07             & 1.00          & 3.70          & 2.41          & 5.40          & 2.63E-01                            & 3.64     & 44.75          \\
Mrk 501            & 1.8      & 23       & 1.20E+42                                                  & 1.40E+04           & 1.00E+07             & 1.40          & 4.10          & 2.41          & 4.20          & 2.00E-01                            & 3.83     & 34.81          \\
1ES 1727+502       & 1.8      & 23       & 4.50E+41                                                  & 2.00E+04           & 1.00E+07             & 1.80          & 4.20          & 2.20          & 3.00          & 2.17E-01                            & 22.25    & 23.18          \\
1ES 2344+514       & 1.8      & 23       & 2.30E+41                                                  & 2.00E+04           & 1.00E+07             & 1.7e          & 4.20          & 2.60          & 4.20          & 3.70E-01                            & 6.20     & 20.63          \\
NGC 4278$^{\rm a}$ & 1.8      & 5        & 1.30E+39                                                  & 3.00E+05           & 1.00E+07             & 1.00          & 4.90          & 2.10          & 4.90          & 9.09E-07                            & 27.91    & 1.80           \\
NGC 4278$^{\rm b}$ & 30       & 3        & 3.00E+41                                                  & 5.00E+05           & 1.00E+07             & 1.00          & 4.90          & 1.50          & 4.50          & 5.56E-06                            & 32.11    & 4.10          \\ \hline
\end{tabular}
\end{table*}

\begin{figure*}[htbp]
  \centering
  \subfloat{\includegraphics[width=0.45\linewidth]{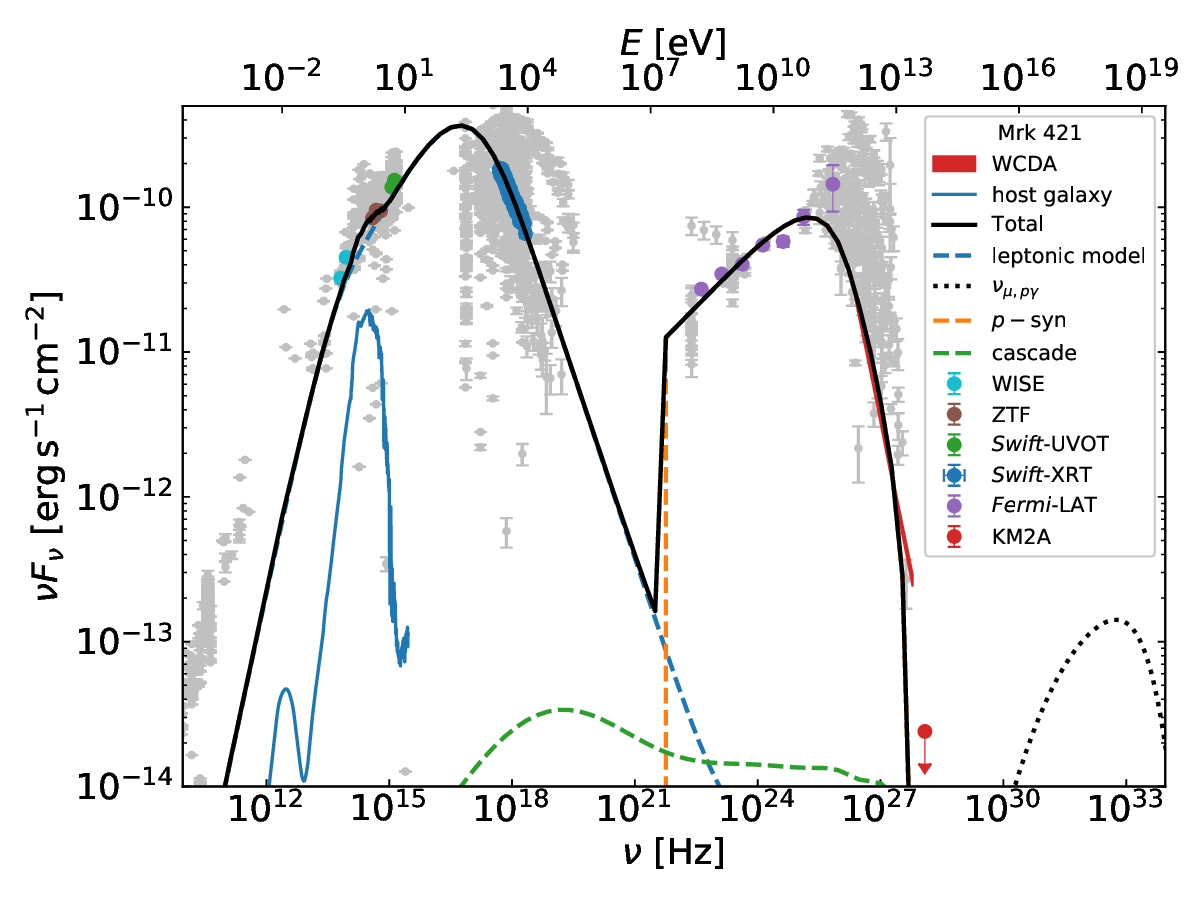}}
  \subfloat{\includegraphics[width=0.45\linewidth]{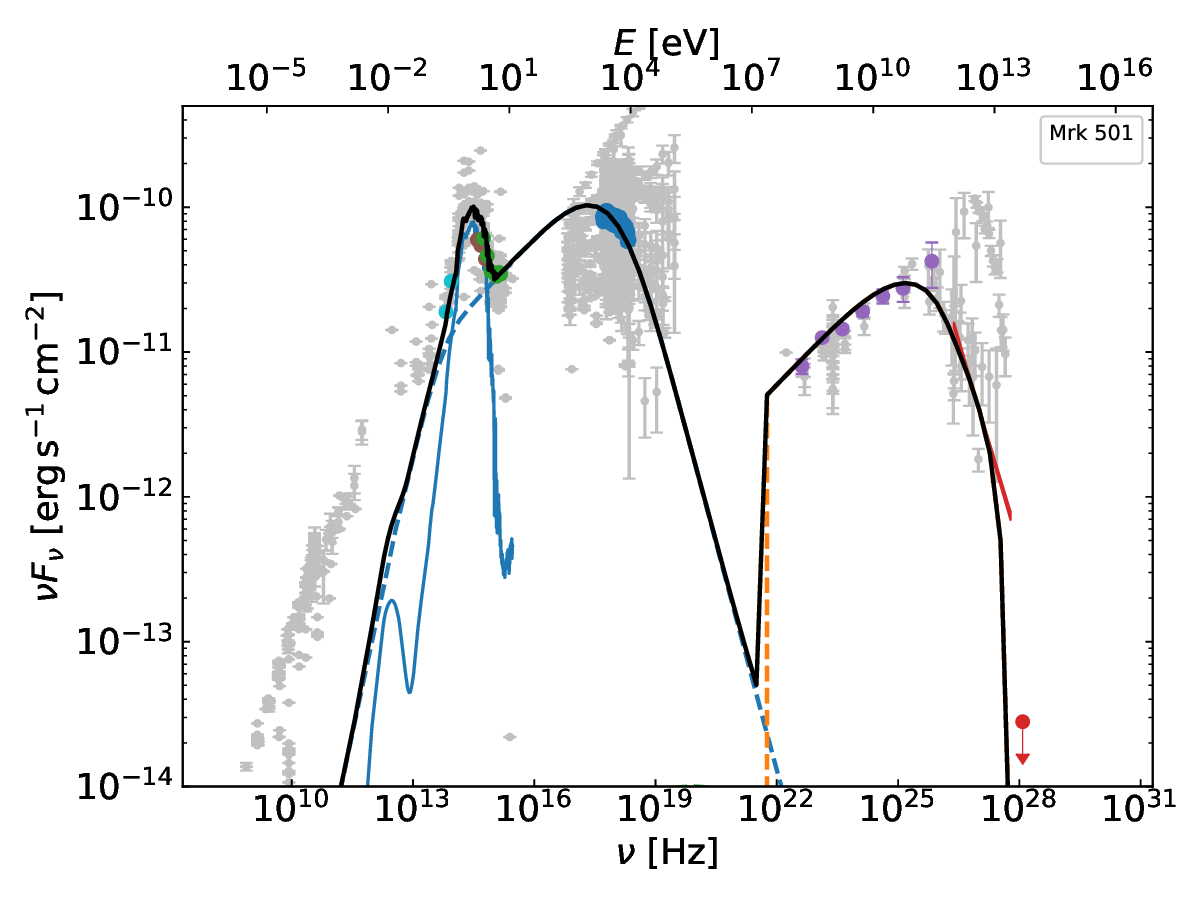}}
  \hfill
  \subfloat{\includegraphics[width=0.45\linewidth]{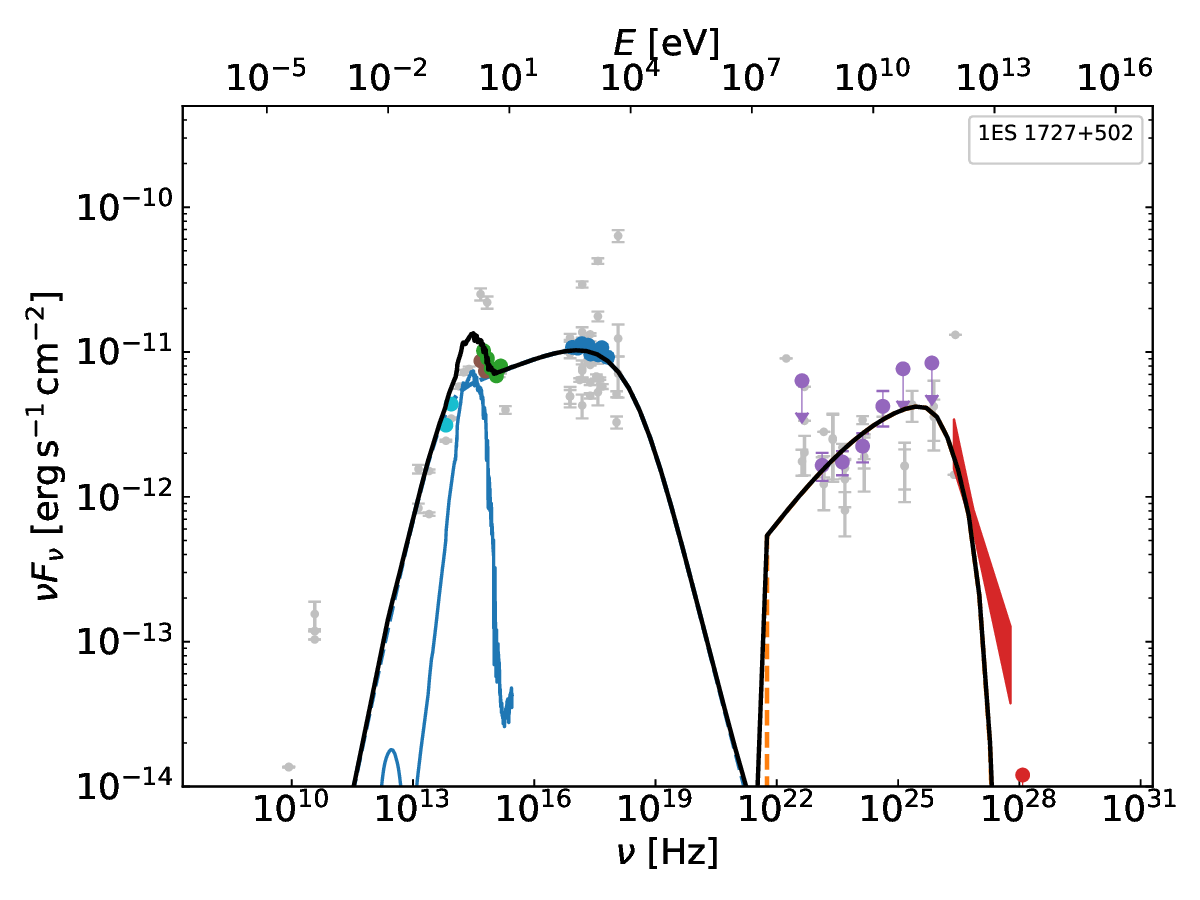}}
  \subfloat{\includegraphics[width=0.45\linewidth]{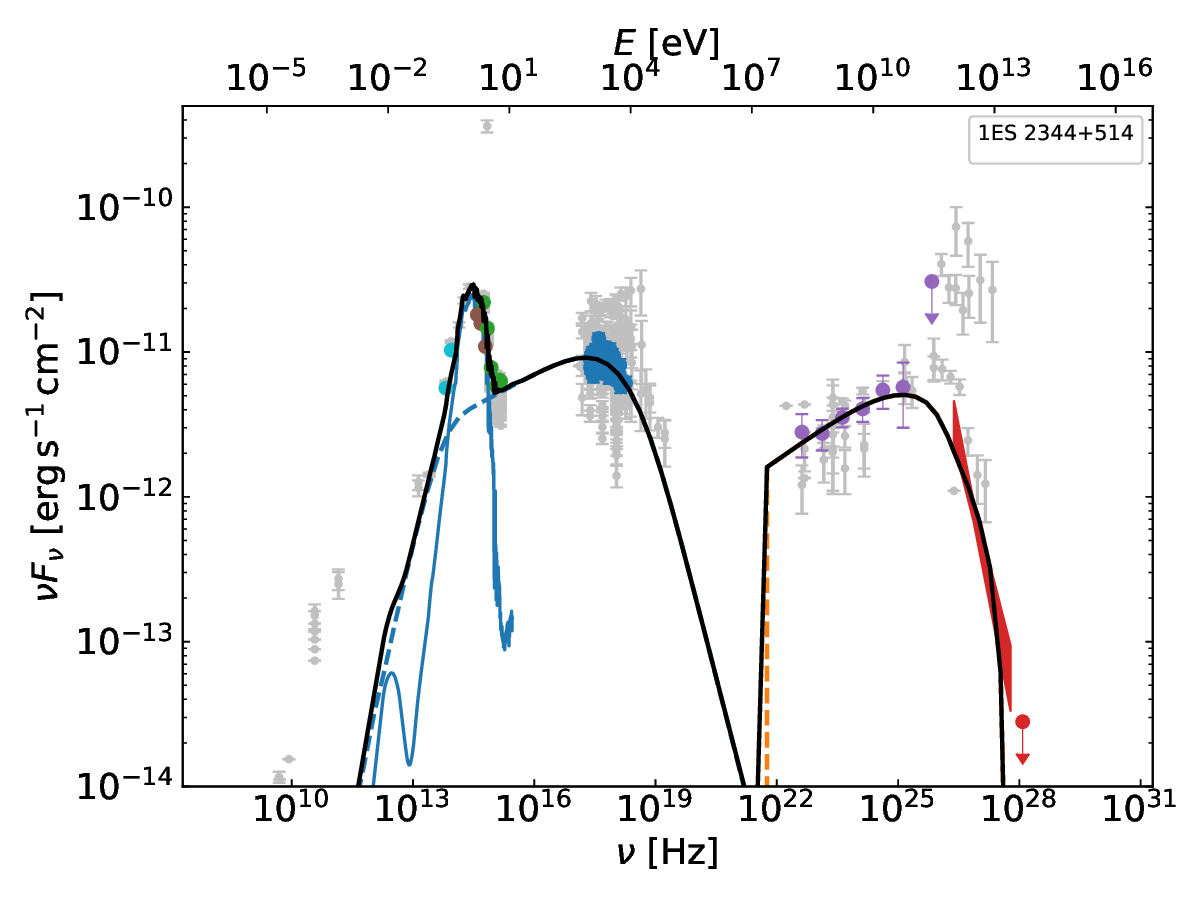}}
  \hfill
  \subfloat{\includegraphics[width=0.45\linewidth]{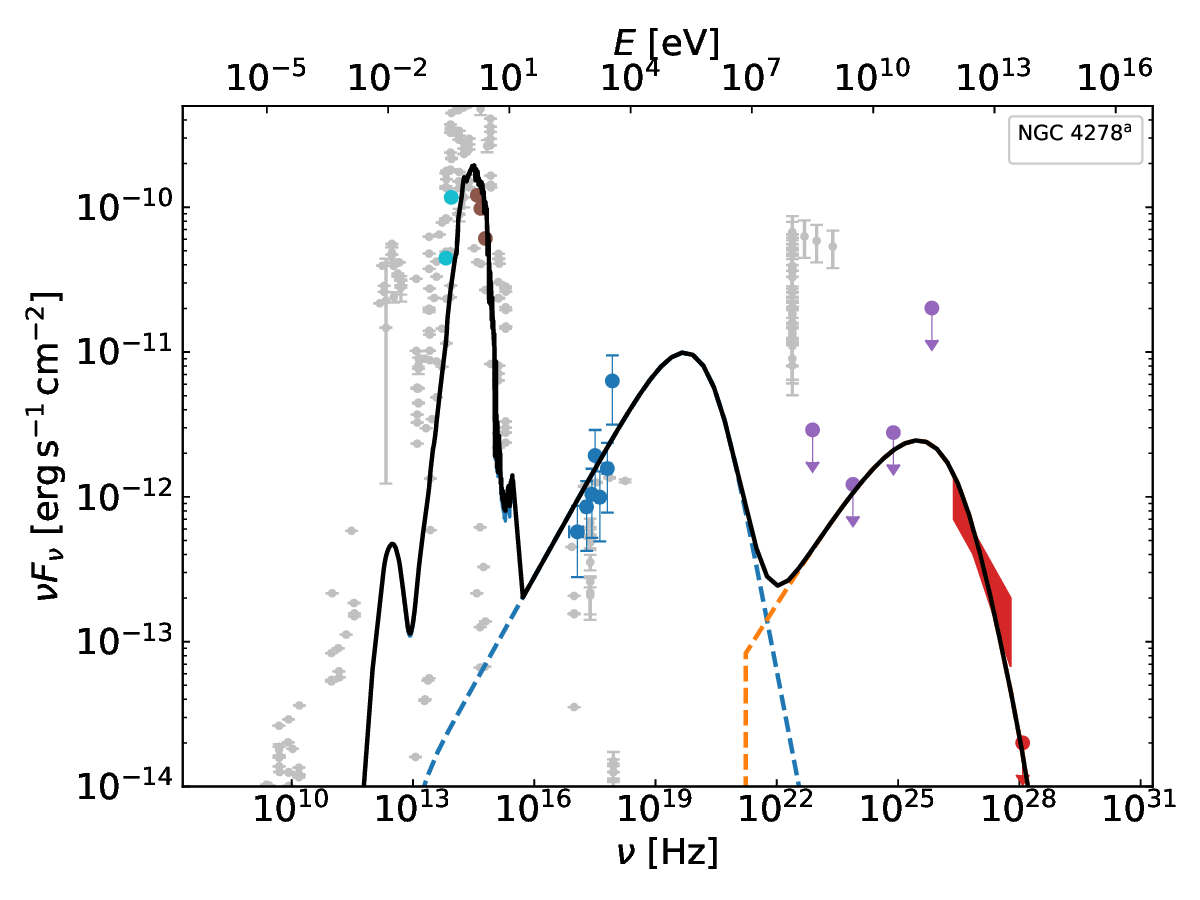}}
  \subfloat{\includegraphics[width=0.45\linewidth]{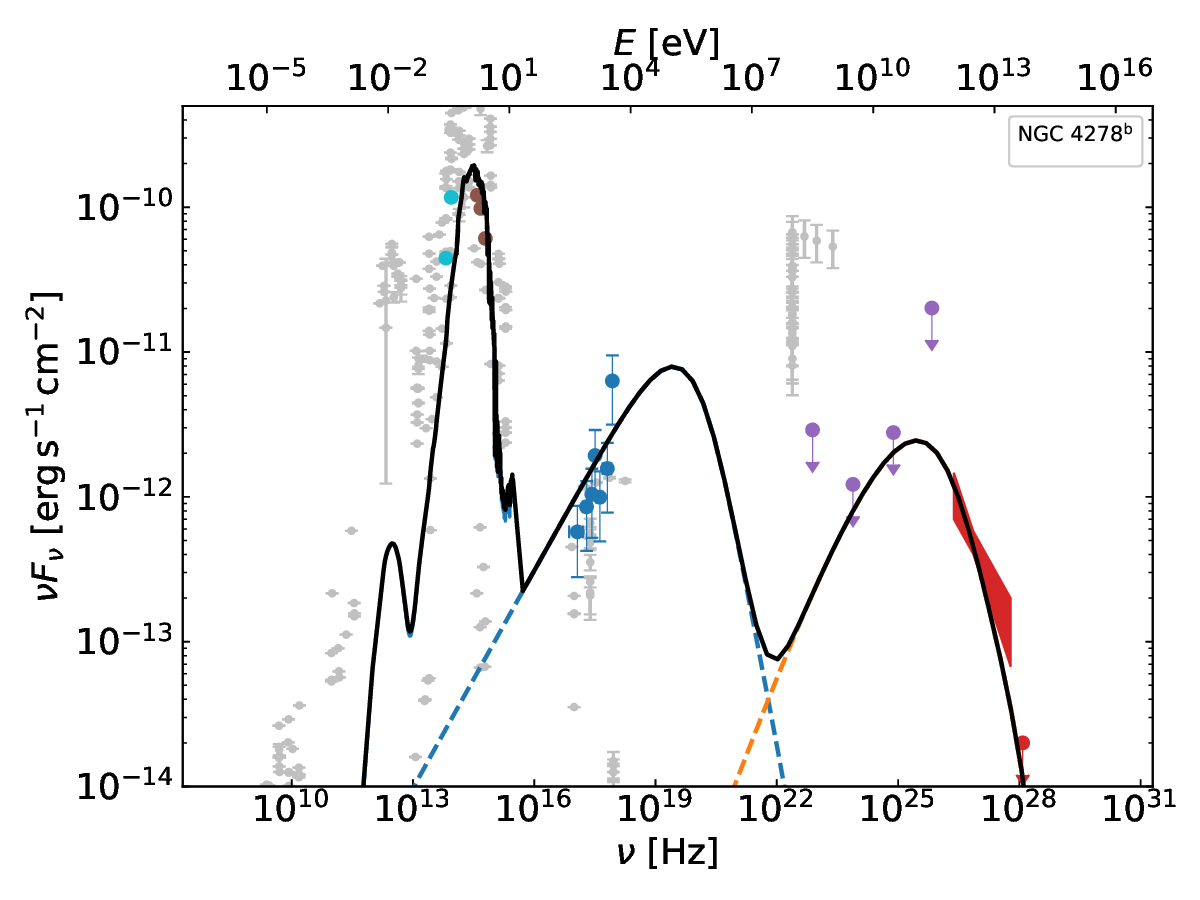}}
  \caption{One-zone proton-synchrotron modeling. The meanings of symbols and line styles are given in the legend of Mrk 421. \label{fig:psyn}}
\end{figure*}

Some studies show that the high-energy hump of SEDs can be fitted by the proton-synchrotron process with a strong magnetic field \citep{2015MNRAS.448..910C, 2023ApJ...942...51W}, although such a large magnetic field strength contradicts current observations. In the following, we will present the fitting results of the proton-synchrotron model, with a strong magnetic field. In this scenario, the leptonic modeling follows that given in Sect. \ref{sec:ssc}, and the hadronic modeling basically follows that given in Sect. \ref{sec:pp}. There are five differences from before: 
\begin{enumerate}
\item The power-law proton spectrum cannot fit the observation of LHAASO and therefore the injection proton density distribution is changed to a broken power-law spectrum, i.e., 
\begin{equation}
    n_{\rm p}^{\rm inj}(\gamma_{\rm p})\propto\left\{\begin{array}{ll}\gamma_{\rm p}^{-s_{\rm p,1}}, & \gamma_{\rm p,min}\leq \gamma_{\rm p}\leq  \gamma_{\rm p,b}~\\\gamma_{\rm p,b}^{s_{\rm p,2}-s_{\rm p,1}}\gamma_{\rm p}^{-s_{\rm p,2}}, & \gamma_{\rm p,b}<\gamma_{\rm p}\leq\gamma_{\rm p,max}\end{array}\right..
\end{equation}
\item In the proton-synchrotron modeling, a higher maximum proton Lorentz factor is required to produce TeV emission than that in the $pp$ interactions. Here we set $\alpha=1$, which implies an extreme acceleration efficiency.
\item In the proton-synchrotron model, a large magnetic field is needed to accelerate protons to higher energies and produce higher energy emissions. We boldly fix the magnetic field $B$ to 35\,G for all of five AGNs.
\item To maximise the efficiency of the proton-synchrotron process within a reasonable parameter space, we assume that the power of the magnetic field equals to half the Eddington luminosity. Then the radius of radiation zone can be written as, 
\begin{equation} \label{eq:R}
    R=\sqrt{L_{\rm Edd}/\left(2\pi\Gamma^{2}cU_{\rm B}\right)}.
\end{equation}
\item During fitting, we find a significant degeneracy between $\gamma_{\rm p,max}$ and $\gamma_{\rm p,b}$. In order to reduce the number of free parameters, we set $\gamma_{\rm p,b}=\gamma_{\rm p,max}/10$.
\end{enumerate}

Finally, there are nine free parameters left, which can be found in Table \ref{tab:par_appendix}. The fitting results are shown in Fig.$\,$\ref{fig:psyn}. It can be seen that the LHAASO observations are well reproduced for Mrk 421, Mrk 501, 1ES 2344+514, NGC 4278$^{\rm a}$ and NGC 4278$^{\rm b}$. In the case of 1ES 1727+502, however, it deviates significantly from the observations, which may be caused by the maximum energy that protons can reach. The characteristic photon energy in the observer's frame produced by the proton-synchrotron process can be calculated by
\begin{equation}\label{eq:psyn}
    E_{\rm p,c}^{\rm syn}=\frac{3heB\gamma_{\rm p}^{2}}{4\pi m_{\rm p}c}\frac{\delta}{1+z}\approx9.46\times10^{-12}\gamma_{\rm p}^{2}\frac{B}{1\,{\rm G}}\frac{\delta}{1+z}\,{\rm eV}.
\end{equation}
To reproduce the VHE spectra, protons with maximum energy should emit at least 20 TeV photons (the energy range of WCDA data is 1-25$\,{\rm TeV}$). Substituting Eq. (\ref{eq:pmax}) and Eq. (\ref{eq:R}) into Eq. (\ref{eq:psyn}) yields
\begin{equation}\label{eq:psynmax}
    E_{\rm p,max}^{\rm syn}=\frac{3he^{3}}{\pi m_{\rm p}^3c^{6}}\frac{B\delta L_{\rm Edd}}{\alpha^{2}\Gamma^{2}\left(1+z\right)}
    \approx 0.16\left(\frac{10}{\alpha}\right)^{2}\frac{B}{1\,{\rm G}}\frac{M_{\rm BH}}{10^{9}M_{\odot}}\frac{\delta}{\Gamma^{2}\left(1+z\right)}\,{\rm TeV}.
\end{equation}
It is clear that $\alpha$, $B$ and $\Gamma$ are the three parameters that will affect the value of $E_{\rm p,max}^{\rm syn}$. To increase $E_{\rm p,max}^{\rm syn}$, $\alpha$ must be lowered, but $\alpha=1$ is already the theoretical minimum value. So to increase $E_{\rm p,max}^{\rm syn}$, alternative acceleration mechanisms with higher efficiency than Fermi first-order are needed.
Similar to $\alpha$, it is also needed to reduce $\Gamma$ to get a larger $E_{\rm p,max}^{\rm syn}$. However, reducing $\Gamma$ will also lead to the observed flux decrease due to the weakening of the beaming effect. Unless we increase $L_{\rm p}^{\rm inj}$ at the same time, but that would cause the jet power to exceed the Eddington luminosity. The proton injection luminosity in fitting of four blazars (shown in Table \ref{tab:par}) is close to half of the Eddington luminosity, and the power of the magnetic field is assumed previously to be half of the Eddington luminosity, so the sum of the two is very close to the Eddington luminosity. Finally, $B$ is the only parameter that can be adjusted to get a larger $E_{\rm p,max}^{\rm syn}$. Substituting $E_{\rm p,max}^{\rm syn}=20\,{\rm TeV}$, $\alpha=1$, $\Gamma$ and $\theta$ used in fitting into Eq. (\ref{eq:psynmax}), we can obtain the minimum required magnetic field strength $B_{\rm min}=421\,{\rm G}$ for 1ES 1727+502. The jet is unlikely to have such a strong magnetic field. 


\bibliography{sample631}{}
\bibliographystyle{aasjournal}

\end{document}